\documentclass[mypaper,10pt,twoside]{CoAst}
\usepackage{epsf,color,graphicx,picinpar,fancyhdr,multicol,multirow,latexsym,fancybox,amssymb,supertabular,longtable,rotating,nicefrac,amsmath}
\usepackage{lscape} 
\usepackage{url} 
\usepackage{sfmath}

\usepackage{ucs}
\usepackage{booktabs}
\usepackage{inputenc}
\usepackage{fontenc}
\usepackage{dcolumn}

\usepackage{natbib}

\renewcommand{\chaptermark}[1]{\markboth{#1}{}}

\renewcommand{\headrulewidth}{0pt}
\def\setpubdate{2010}
\newcommand{\kopf}{\small\itshape 
Comm. in Asteroseismology - Complementary Topics \\ 
Volume 1, \setpubdate \\
\copyright~Austrian Academy of Sciences}

\newcommand{\Authors}[1]{\begin{center}\normalsize\bf\sf #1 \end{center}}

\renewcommand{\author}[1]{\begin{center}\normalsize\bf\sf #1 \end{center}}
\newcommand{\Address}[1]{\begin{center}\small\sf #1 \end{center}}

\renewenvironment{abstract}{\section*{Abstract}\normalsize\sf}{}
\newcommand{\References}[1]{\begin{flushleft}{\large References\\}\vspace*{2mm}\small #1 \end{flushleft}}

\newcommand{\chapterCoAst}[2]
{\chapter[\sf\normalsize #1\\ 
\footnotesize \hspace*{5mm}by #2 \sf\normalsize][]
{#1\\}
\rhead[\fancyplain{}{\sf\footnotesize \center{#1}}]{\fancyplain{}{\sffamily\thepage}}
\lhead[\fancyplain{\kopf}{\sffamily\thepage}]{\fancyplain{\kopf}{\sf\footnotesize \center{#2}}}}

\newcommand{\figureDSSN}[5]{\begin{figure}[#4]
\centering
\includegraphics*[#5]{#1}
\caption{#2}
\label{#3}
\end{figure}}

\renewcommand{\chaptername}{}

\newcommand{\acknowledgments}[1]{\vspace*{5mm}\noindent  \textbf{Acknowledgments.} #1}

\begin{document}

\pagestyle{empty}

\sf


\setcounter{page}{1}
\thispagestyle{empty}
\vspace*{-1cm}
\begin{center}


\huge\sf    Communications in Asteroseismology\\
Complementary Topics\\
\vspace*{1cm}
\large 
   Volume 1 \\ \the\year\\
\vspace*{4cm}
\vspace*{4mm}
\vspace*{5cm}

\begin{figure}[!ht]
\centering
\includegraphics*[width=60mm,clip]{OEAW_englisch_2009_SW}

\end{figure}

\normalsize
\end{center}
\newpage
\thispagestyle{empty}
\normalsize
\vspace*{-1cm}

\begin{center}
\begin{large}Communications in Asteroseismology - Complementary Topics\end{large}\\
Editor-in-Chief: \textbf{Michel Breger}, michel.breger@univie.ac.at\\
Editorial Assistant: \textbf{Isolde M\"uller}, isolde.mueller@univie.ac.at\\
Layout \& Production Manager: \textbf{Isolde M\"uller}, isolde.mueller@univie.ac.at\\

\vspace{2mm}
CoAst and CoAct Editorial and Production Office\\
T\"urkenschanzstra\ss e 17, A - 1180 Wien, Austria\\
\textit{http://www.oeaw.ac.at/CoAst/ \\ Comm.Astro@univie.ac.at\\}
\vspace*{4mm}
\textbf{Editorial Board:} Conny Aerts, Gerald Handler, \\ Don Kurtz, Jaymie Matthews, Ennio Poretti\\

\vspace*{1.5cm}

\begin{large}Cover Illustration\end{large}\\
\end{center}

\vspace*{1.5cm}
\begin{center}
\small
\sf British Library Cataloguing in Publication data.\\
\sf A Catalogue record for this book is available from the British Library.
\end{center}
\vspace*{1.4cm}
\small
\begin{center}
All rights reserved\\
ISBN ???-?-????-????-?\\
ISSN ????-????\\
Copyright~\copyright~2010 by\\
Austrian Academy of Sciences \\
Vienna\\
\vspace*{2mm}
Austrian Academy of Sciences Press\\
A-1011 Wien, Postfach 471, Postgasse 7/4\\
Tel. +43-1-515 81/DW 3402-3406, +43-1-512 9050\\ Fax +43-1-515 81/DW 3400\\
http://verlag.oeaw.ac.at, e-mail: verlag@oeaw.ac.at\\
\end{center}

		\pagestyle{fancyplain}
		\setlength{\columnseprule}{0.2pt}
		\normalsize



\newpage
\thispagestyle{empty}
\normalsize



 


\chapterCoAst{Preface}{J.\,D.~Scargle}

\Authors{Jeffrey D.~Scargle$^1$} 
\Address{$^1$ Space Science and Astrobiology Division, NASA Ames Research Center}

{\sc SigSpec} is a method for detecting and characterizing periodic signals in
noisy data. This is an extremely common problem, not only in astronomy
but in almost every branch of science and engineering. This work will
be of great interest to anyone carrying out harmonic analysis employing
Fourier techniques.

The method is based on the definition of a quantity called
{\em spectral significance} -- a function of Fourier phase and
amplitude. Most data analysts are used to exploring only
the Fourier amplitude, through the power spectrum, ignoring
phase information. The Fourier phase spectrum can be
estimated from data, but its interpretation is usually problematic.
The spectral significance quantity conveys more information
than does the conventional amplitude spectrum alone, and appears
to simplify statistical issues as well as the interpretation of
phase information.



\newpage 

\thispagestyle{empty}
\vspace*{30mm}
\begin{center}

\end{center}


\begin{verbatim}
 
\end{verbatim}

\chapterCoAst{SigSpec User's Manual}{P.~Reegen}

\Authors{P.~Reegen$^1$} 
\Address{$^1$ Institut f\"ur Astronomie, T\"urkenschanzstra\ss e 17, 1180 Vienna, Austria\\
reegen@astro.univie.ac.at}

\noindent
\begin{abstract}
{\sc SigSpec} computes the spectral significance levels for the DFT amplitude spectrum of a time series at arbitrarily given sampling. It is based on the analytical solution for the Probability Density Function (PDF) of an amplitude level, including dependencies on frequency and phase and referring to white noise. Using a time series dataset as input, an iterative procedure including step-by-step prewhitening of the most significant signal components and MultiSine least-squares fitting is provided to determine a whole set of signal components, which makes the program a powerful tool for multi-frequency analysis. Instead of the step-by-step prewhitening of the most significant peaks, the program is also able to take into account several steps of the prewhitening sequence simultaneously and check for the combination associated to a minimum residual scatter. This option is designed to overcome the aliasing problem caused by periodic time gaps in the dataset. {\sc SigSpec} can detect non-sinusoidal periodicities in a dataset by simultaneously taking into account a fundamental frequency plus a set of harmonics. Time-resolved spectral significance analysis using a set of intervals of the time series is supported to investigate the development of eigenfrequencies over the observation time. Furthermore, an extension is available to perform the {\sc SigSpec} analysis for multiple time series input files at once. In this MultiFile mode, time series may be tagged as target and comparison data. Based on this selection, {\sc SigSpec} is capable of determining differential significance spectra for the target datasets with respect to coincidences in the comparison spectra. A built-in simulator to generate and superpose a variety of sinusoids and trends as well as different types of noise completes the software package at the present stage of development.
\end{abstract}

\section{What is {\sc SigSpec}?}

{\sc SigSpec} (abbreviation of `{\bf SIG}nificance {\bf SPEC}trum') is a program that computes a significance spectrum for a time series. It evaluates the {\it Probability Density Function (PDF)} of a given DFT amplitude level analytically, making use of the theoretical concept introduced by Reegen~(2005, 2007). The {\it False-Alarm Probability}, $\Phi_\mathrm{FA}\left( A\right)$, is the probability that an amplitude in the DFT spectrum exceeds a given limit $A$, and is obtained through integration of the PDF (e.\,g.~Scargle 1982). Instead of this frequently used quantity, {\sc SigSpec} calculates the {\it spectral significance} (abbreviated by `sig') of an amplitude $A$ by
\begin{equation}
\mathrm{sig}\left( A\right) := -\log\left[ \Phi_\mathrm{FA}\left( A\right)\right]\, .
\end{equation}

E.\,g., a sig equal to $5$ indicates that the considered amplitude level is due to noise in one out of $10^5$ cases. This value is used as the default threshold for the termination of the prewhitening sequence.

{\sc SigSpec} performs an iterative process consisting of four steps\footnote{The AntiAlC computation (p.\,\pageref{SIGSPEC_AntiAlC}) differs slightly from this procedure.}:
\begin{enumerate}
\item computation of the significance spectrum,
\item exact determination of the peak with maximum sig,
\item a MultiSine least-squares fit of the frequencies, amplitudes and phases of all significant signal components detected so far,
\item prewhitening of the sinusoidal components. The residuals are used as input for the next iteration.
\end{enumerate}

If {\sc SigSpec} is called without any special settings, it produces four files:
\begin{enumerate}
\item the DFT amplitude spectrum {\tt s000000.dat} of the original time series, containing also sig and phase,
\item the DFT amplitude spectrum {\tt resspec.dat} of the residual time series after prewhitening all significant signal components, containing also sig and phase,
\item the residual time series {\tt residuals.dat} after prewhitening all significant signal components,
\item a result file called {\tt result.dat}, which contains a list of significant signal components,
\item MultiSine track files, each of which contains a list of the frequencies, amplitudes and phases for a single sinusoidal component through the prewhitening cascade (pp.\,\pageref{SIGSPEC_MultiSine tracks}, \pageref{SIGSPEC_mstracks}).
\end{enumerate}
Further options may be applied to obtain spectra, residuals, and/or result files (p.\,\pageref{SIGSPEC_spectra}) in the prewhitening sequence. The MultiSine fits, which are performed after each prewhitening step, modify the frequencies, amplitudes and phases of previous components. If the user examines the resulting signal components and decides not to use all of them, the additional result files help to have accurate frequencies, amplitudes and phases in hands also for a shorter list of significant sinusoids without re-running the program.

{\sc SigSpec} can produce additional files containing
\begin{enumerate}
\item a spectral window for the given time series (pp.\,\pageref{SIGSPEC_Spectral window}, \pageref{SIGSPEC_win}),
\item a sampling profile (pp.\,\pageref{SIGSPEC_Sampling profile}, \pageref{SIGSPEC_profile}) containing the parameters $\alpha _0\left(\omega\right)$, $\beta _0\left(\omega\right)$, $\theta _0\left(\omega\right)$ determining the dependency of the sig on the time-domain sampling, as well as on frequency and phase in Fourier space (see Reegen 2007),
\item a preview of the {\sc SigSpec} analysis (pp.\,\pageref{SIGSPEC_Preview}, \pageref{SIGSPEC_preview}),
\item a Sock Diagram (pp.\,\pageref{SIGSPEC_Sock Diagram}, \pageref{SIGSPEC_sock:phases}),
\item a Phase Distribution Diagram (pp.\,\pageref{SIGSPEC_Phase Distribution Diagram}, \pageref{SIGSPEC_phdist:phases}) containing probability densities for the Fourier phases,
\item a correlogram for each step of the prewhitening sequence (pp.\,\pageref{SIGSPEC_Correlograms}, \pageref{SIGSPEC_correlograms}).
\end{enumerate}
These options are deactivated by default.

\label{SIGSPEC_csig}Given a sequence of prewhitenings yielding $N$ significant components with associated sigs $\mathrm{sig}\left( A_n\right)$, it is desirable to additionally know the probability of the entire sequence to be valid. This means that not a single erroneous component is allowed. The False-Alarm Probability $\Phi _{\mathrm{FA}\,n} = 10^{-\mathrm{sig}\left( A_n\right)}$ of an individual peak is the probability that it is generated by noise. The complementary probability that the considered peak is true is $1 - 10^{-\mathrm{sig}\left( A_n\right)}$. If the individual components are statistically independent, the cumulative probability of all components to be real is the product of the individual probabilities,
\begin{equation}
1 - \Phi _{\mathrm{FA}} = \prod _{n=1}^N\left(1 - \Phi _{\mathrm{FA}\,n}\right)\: .
\end{equation}
Consistently, the cumulative sig is introduced as the negative logarithm of this total False-Alarm Probability for all identified signal components, $\Phi _{\mathrm{FA}}$, and in terms of individual sigs, one obtains
\begin{equation}
\mathrm{csig}\left( A_N\right) := - \log\left\lbrace 1 - \prod _{n=1}^N\left[ 1 - 10^{-\mathrm{sig}\left( A_n\right)}\right]\right\rbrace\: .
\end{equation}

In consistency with the definition of the sig associated with an amplitude in the DFT spectrum, a cumulative sig of $3$ means that the prewhitening cascade is entirely true in $999$ out of $1\,000$ cases. Or -- in other words -- in one out of $1\,000$ cases, at least one of the identified components is generated by noise.

Whereas the individual sig of a component in the prewhitening sequence may exceed that of the previously identified maximum, the cumulative sig is a monotone sequence uniquely decreasing with each additional signal component.

The prewhitening loop stops, if no sig level above a pre-defined limit is found. As described in ``Program termination'', p.\,\pageref{SIGSPEC_Program termination}, there are three different criteria that may be applied to determine the conditions for program termination:
\begin{enumerate}
\item the number of iterations in the prewhitening sequence,
\item a lower sig limit for the highest peak in the significance spectrum,
\item a threshold for the cumulative sig related to a combined probability for all detected frequency components.
\end{enumerate}

The program also supports the subdivision of a time series into a set of intervals and the separate analysis of all these parts in order to monitor frequency changes of signal components with time. This method will be called {\it time-resolved analysis}. In this case, the output is somewhat richer, as described in ``Time-resolved Analysis'' (p.\,\pageref{SIGSPEC_Time-resolved Analysis}).

An immanent problem in the analysis of non-equidistantly sampled time series is {\em aliasing}. Due to periodic gaps in the data set, a peak in the amplitude spectrum is accompanied by side peaks. Especially if more than one sinusoidal component is present in the data, the superposition of side peaks may produce a maximum amplitude in the DFT spectrum at a frequency that has nothing in common with the true signal frequencies. Such a misidentification usually damages the complete prewhitening sequence from this point on. As pointed out by Reegen~(2007), {\sc SigSpec} appears less prone to aliasing than the previously used methods, since the noise component is employed into the statistical treatment correctly. However, the superposition mentioned above may also lead to erroneous identifications.

In order to overcome this potential weakness, {\sc SigSpec} supports the simultaneous calculation of more than one signal component simultaneously. Instead of picking only the peak associated to maximum sig, a whole set of highest peaks is examined, searching all possible combinations for several iterations in order to obtain the solution providing a minimum rms residual. This function is called {\em AntiAlC} (ANTI-ALiasing Correction) mode (p.\,\pageref{SIGSPEC_AntiAlC}).

There is a second option to examine multiple peaks simultaneously: a non-sinusoidal periodicity is represented by multiple peaks in the DFT amplitude spectrum. One finds a fundamental frequency, plus one or more harmonics the frequencies of which are integer multiples of the fundamental. In astronomical applications, this may occur if shock waves are present in the stellar pulsation or if surface variations are examined. In such a case, it is desirable to take into account not only the fundamental frequency, but also all available harmonics at once. This analysis of harmonics is described on p.\,\pageref{SIGSPEC_Analysis of Harmonics}).

{\sc SigSpec} is capable of analysing multiple time series input files simultaneously. This MultiFile mode (p.\,\pageref{SIGSPEC_MultiFile Mode}) speeds up the computation considerably for time series with the same sampling.

A further option is the evaluation of differential significance spectra (p.\,\pageref{SIGSPEC_Differential significance spectra}). The user may specify target vs. comparison data among the input files. Then {\sc SigSpec} performs a quantitative comparison of the two groups of time series and returns a measure of the probability that a peak in a target dataset is `true', taking into account amplitudes and phases at the corresponding frequency in the comparison spectra. In this context, the term `true' is used in the sense of `not entirely produced by the same variability as present in the comparison data'.

The examples presented here refer to the sample projects available for download at {\tt http://www.SigSpec.org}.

\section{How to Run {\sc SigSpec}}\label{SIGSPEC_How to Run SigSpec}

\subsection{Projects}\label{SIGSPEC_Projects}

{\sc SigSpec} is called by the command line

\begin{scriptsize}\begin{verbatim}
SigSpec <project>
\end{verbatim}\end{scriptsize}

\noindent where {\tt <project>} is the name (or path, if desired) of the {\sc SigSpec} project. Before running the program, the user has to provide
\begin{enumerate}
\item a directory {\tt <project>} used for the output,
\item a time series input file (see ``The time series input file'', p.\,\pageref{SIGSPEC_The time series input file}).
\end{enumerate}
The project directory and the time series input file have to be located in the same folder. The project directory need not be empty.

\vspace{12pt}
{\bf Caution: {\sc SigSpec} overwrites existing output files!}
\vspace{12pt}

There are two conventions for denominating input files.
\begin{enumerate}
\item The standard method is to pass only one time series input file to the program. {\sc SigSpec} expects the file to be named {\tt <project>.dat}.
\item For an all-in-one analysis of multiple time series input files, i.\,e., for running {\sc SigSpec} in MultiFile mode, a leading six-digit index is expected. In this case, the first file shall be named {\tt 000000.<project>.dat}, the next file is {\tt 000001.<project>.dat}, and so on. For more information on the MultiFile mode, please refer to ``MultiFile mode'', p.\,\pageref{SIGSPEC_MultiFile Mode}.
\end{enumerate}

Furthermore, the user may pass a set of specifications to {\sc SigSpec} by means of a file {\tt <project>.ini} (see ``The {\tt .ini} file'', p.\,\pageref{SIGSPEC_The .ini file}). This file is expected in the same folder as the time series input file and the project directory. For specifications not given by the user, defaults are used.

\vspace{12pt}\noindent{\bf Example.}\label{SIGSPEC_EXnormalrun} {\it The sample project {\tt SigSpecNative} provides a run without any additional options. The command line is {\tt SigSpec SigSpecNative}. The sample input file {\tt SigSpecNative.dat} (381 data points) represents V magnitudes of IC\,4996\,\#\,89 (Zwintz et al.~2004; Zwintz \& Weiss 2006).

The screen output produced by typing {\tt SigSpec SigSpecNative} at runtime starts with a standard header.}

\begin{scriptsize}\begin{verbatim}
 SSSSSS  ii          SSSSSS
SS    SS            SS    SS
SS       ii  gggg g SS       p pppp   eeeee   ccccc
SS       ii gg   gg SS       pp   pp ee   ee cc   cc
 SSSSSS  ii gg   gg  SSSSSS  pp   pp ee   ee cc
      SS ii gg   gg       SS pp   pp eeeeeee cc
      SS ii gg   gg       SS pp   pp ee      cc
SS    SS ii gg   gg SS    SS pp   pp ee   ee cc   cc
 SSSSSS  ii  gggggg  SSSSSS  pppppp   eeeee   ccccc
                 gg          pp
            gg   gg          pp
             ggggg           pp


SIGnificance SPECtrum
Version 2.0
************************************************************
by Piet Reegen
Institute of Astronomy
University of Vienna
Tuerkenschanzstrasse 17
1180 Vienna, Austria
Release date: August 18, 2009
\end{verbatim}\end{scriptsize}

{\it {\sc SigSpec} processes the command line, checks whether a project directory {\tt SigSpecNative} is present, and searches for a file {\tt SigSpecNative.ini} (see ``The {\tt .ini} file'', p.\,\pageref{SIGSPEC_The .ini file}). Since there is no such file present, four warning messages are produced.}

\begin{scriptsize}\begin{verbatim}
*** start **************************************************

command line interface
Checking availability of project directory SigSpecNative...
project directory SigSpecNative ok.
loading .ini file

Warning: IniFile_SSCols 001
         Failed to open .ini file.


Warning: IniFile_WCols 001
         Failed to open .ini file.


Warning: IniFile_LoadIni 001
         Failed to open .ini file.


Warning: IniFile_Cind 001
         Failed to open .ini file.
\end{verbatim}\end{scriptsize}

{\it The next task is to load the input file {\tt SigSpecNative.dat}. {\sc SigSpec} provides the number of rows, the time interval width, and the standard deviation of the observable.}

\begin{scriptsize}\begin{verbatim}
*** loading time series input file(s) **********************

SigSpecNative.dat

*** time series properties *********************************

points 381, time base 9.17532, rms dev 0.00449592
\end{verbatim}\end{scriptsize}

{\it The next section contains the specifications for the DFT and significance spectra to be calculated.}

\begin{scriptsize}\begin{verbatim}
*** preparing to run SigSpec *******************************

Rayleigh frequency resolution             0.1089880382935977
oversampling ratio                       20.0000000000000000
frequency spacing                         0.0054494019146799
lower frequency limit                     0.0000000000000000
upper frequency limit                   100.4651736990383739
Nyquist coefficient                       0.5000000000000000
number of frequencies                 18437
\end{verbatim}\end{scriptsize}

{\it As {\sc SigSpec} performs the prewhitening sequence, a list of detected signal components is displayed. The screen output contains the index of the identified component (a line number), the sig, the time-domain rms deviation before prewhitening the corresponding signal, and the csig. The last line contains an insignificant component that meets the breakup condition. In the present example, the default breakup condition (the sig to drop below 5) is applied, which is satisfied in the fourth iteration, where the maximum sig is 4.10802.}

\label{SIGSPEC_EXnormalrunfreqs}\begin{scriptsize}\begin{verbatim}
*** running SigSpec ****************************************

   1 freq 3.13205  sig 9.54539  rms 0.00449592  csig 9.54539
   2 freq 3.98471  sig 7.43085  rms 0.00422861  csig 7.42753
   3 freq 5.40684  sig 5.30164  rms 0.0040257  csig 5.2984
   4 freq 17.3677  sig 4.13698  rms 0.00388775  csig 4.10802
\end{verbatim}\end{scriptsize}

{\it On exit, {\sc SigSpec} displays a good-bye message.}

\begin{scriptsize}\begin{verbatim}
Finished.

************************************************************

Thank you for using SigSpec!
Questions or comments?
Please contact Piet Reegen (reegen@astro.univie.ac.at)
Bye!
\end{verbatim}\end{scriptsize}

\figureDSSN{f01.eps}{{\em Black circles:} light curve for the sample project {\tt SigSpecNative}. {\em Line:} fit formed by three significant signal components (as listed in the file {\tt SigSpecNative/result.dat}). {\em Grey dots:} residuals after prewhitening of three significant signal components (file {\tt SigSpecNative/residuals.dat}).}{SIGSPEC_normalrun.dat}{!htb}{clip,angle=0,width=110mm}

\figureDSSN{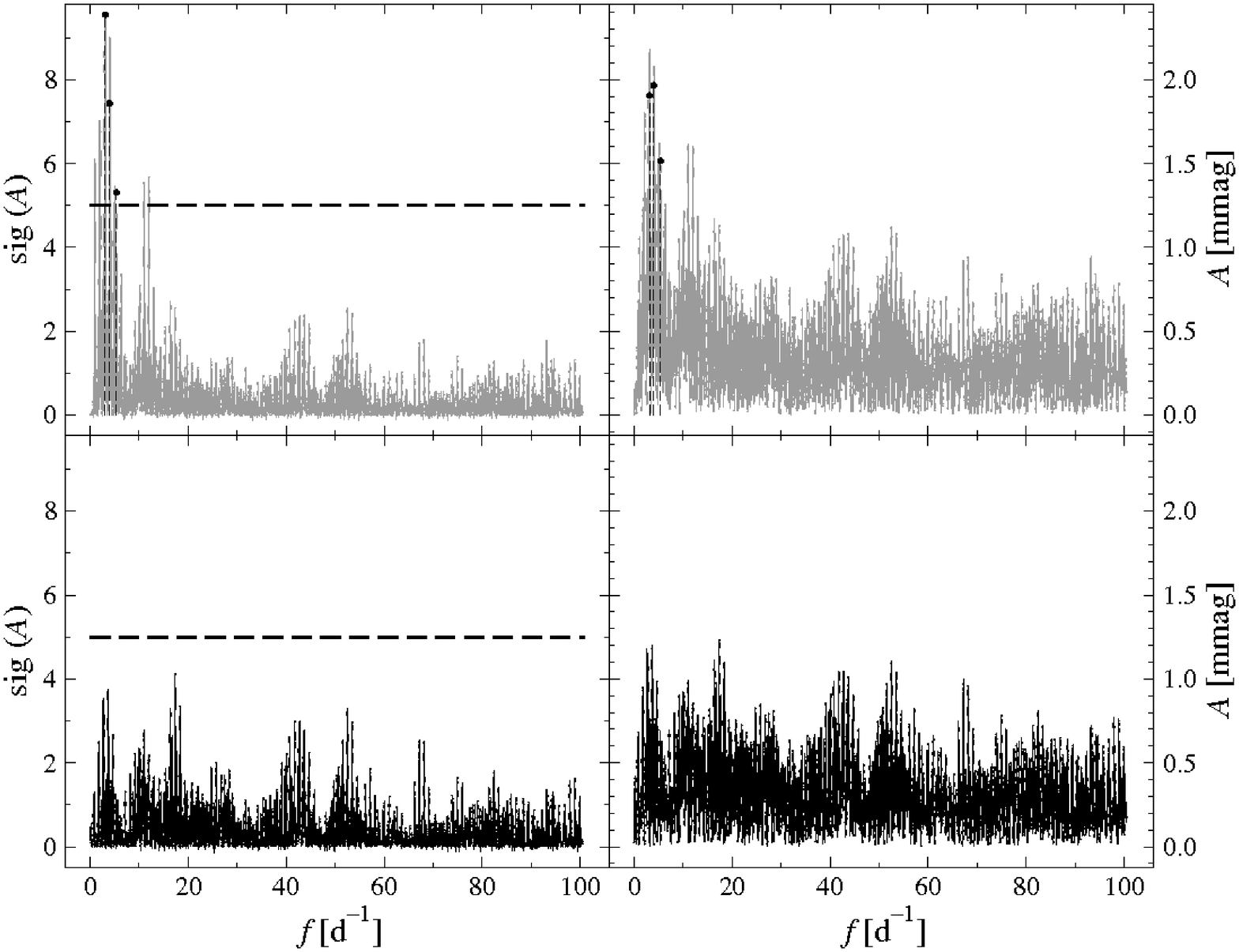}{{\em Grey:} Fourier spectra for the sample project {\tt SigSpecNative}. {\em Left:} significance spectra. {\em Right:} DFT amplitudes. {\em Top:} original spectra, without prewhitening (file {\tt SigSpecNative/s000000.dat}). {\em Bottom:} residual spectra, with three significant signal components prewhitened (file {\tt SigSpecNative/resspec.dat}). In the top panels, the significant components are indicated by {\em dots} with dashed drop lines (file {\tt SigSpecNative/result.dat}). The default sig threshold of 5 is represented by a horizontal {\em dashed line} in the left panels.}{SIGSPEC_normalrun.s}{!htb}{clip,angle=0,width=110mm}

{\it If no special output is selected in a file {\tt SigSpecNative.ini}, {\sc SigSpec} produces the following output files in the project directory {\tt SigSpecNative}:
\begin{itemize}
\item {\tt s000000.dat}: DFT and significance spectrum of the original time series (without any prewhitening),
\item {\tt result.dat}: list of significant signal components detected in the time series,
\item {\tt residuals.dat}: residual time series after prewhitening all significant signal components listed in {\tt result.dat},
\item {\tt resspec.dat}: DFT and significance spectrum of the residual time series {\tt residuals.dat}.
\end{itemize}}
{\it Fig.\,\ref{SIGSPEC_normalrun.dat}} {\it contains the sample input {\tt SigSpecNative.dat}, the multisine fit to the time series according to the list of significant signal components in {\tt SigSpecNative/result.dat}, and the residuals after subtracting the fit \linebreak[4] from the input time series (file {\tt SigSpecNative/residuals.dat}). Fig.\,\ref{SIGSPEC_normalrun.s}} \linebreak[4]{\it refers to the frequency domain: the DFT spectrum of the initial time series {\tt SigSpecNative/s000000.dat}, the three significant signal components contained in {\tt SigSpecNative/result.dat}, and the residual spectrum in the file {\tt SigSpecNative/resspec.dat}. For detailed information on the contents of the output files, please refer to ``Default Output'', p.\,\pageref{SIGSPEC_Default Output}.}

\vspace{12pt}Furthermore, the user may pass a set of specifications to {\sc SigSpec} in a file {\tt <project>.ini} (see ``The {\tt .ini} file'', p.\,\pageref{SIGSPEC_The .ini file}). For specifications not given by the user, defaults are used.

\subsection{Quiet mode}\label{SIGSPEC_Quiet mode}

If the command line is followed by the letter `q', i.\,e.

\begin{scriptsize}\begin{verbatim}
SigSpec <project> q
\end{verbatim}\end{scriptsize}

all screen output is suppressed.

\section{Input}\label{SIGSPEC_Input}

\subsection{The time series input file}\label{SIGSPEC_The time series input file}

The input file for {\sc SigSpec} is a time series. The corresponding file has to be located in the same folder as the project directory. The only restrictions to the format are that the number of items per row has to be constant for all rows in the file and that columns have to be separated by at least one whitespace character or tab. Dataset entries need not be numeric, except for the columns specified as time, observable, and weights (p.\,\pageref{SIGSPEC_Time series columns representing time and observable}).

\subsection{The {\tt .ini} file}\label{SIGSPEC_The .ini file}

An optional file {\tt <project>.ini} consists of a set of keywords and arguments defining project-specific parameters for {\sc SigSpec}. If this file is not present in the same folder as the time series input file(s), {\sc SigSpec} uses a set of default parameters. A complete list of keywords is given in ``Keywords Reference'', p.\,\pageref{SIGSPEC_Keywords Reference}.

Multiple use of the same keyword or the specification of contradictory keywords causes the software to take into account only the last declaration. There are only three exceptions:

\begin{enumerate}
\item {\sc SigSpec} accepts multiple weights columns specified by {\tt col:weights} (p.\,\pageref{SIGSPEC_Time series columns containing statistical weights}),
\item multiple subset identifier columns may be specified by {\tt col:ssid} (p.\,\pageref{SIGSPEC_Time series columns containing subset identifiers}),
\item the simulator may be used to synthesize multiperiodic signal plus various types of noise upon the given sampling, where the keywords {\tt sim:signal}, {\tt sim:poly}, {\tt sim:exp}, {\tt sim:zeromean}, {\tt sim:serial}, {\tt sim:temporal}, and {\tt sim:rndsteps} may be used multiply (see ``The simulator mode'', p.\,\pageref{SIGSPEC_The simulator mode}).
\end{enumerate}

\vspace{12pt}
{\bf Caution: {\sc SigSpec} expects a carriage-return character at the end of the file {\tt <project>.ini}, otherwise the program may hang!}
\vspace{12pt}

Lines in the {\tt .ini} file starting with a {\tt \#} character are ignored by {\sc SigSpec}. This provides the possibility to write comments into the file. Furthermore, additional characters beyond what is expected in a line (keyword plus required number of parameters) is ignored. Thus it is allowed to add comments also at the end of the lines containing relevant information for {\sc SigSpec}.

\subsection{Time series columns representing time and observable} \label{SIGSPEC_Time series columns representing time and observable}

The keywords {\tt col:time}\label{SIGSPEC_keyword.col:time} and {\tt col:obs}\label{SIGSPEC_keyword.col:obs} determine those columns in the time series input file which contain time values and the observable monitored over time, respectively. These columns are required and have to be uniquely specified. Column indices start with 1.

If {\tt col:time} is not specified, the default value $1$ is used. If {\tt col:obs} is not specified, the default value $2$ is used.

\vspace{12pt}\noindent{\bf Example.} {\it The sample project {\tt coltimecolobs} contains a dataset where the time and observable values are found in columns 2 and 3, respectively. The input time series represents the V photometry of IC\,4996\,\#\,89 (see Example {\tt SigSpecNative}, p.\,\pageref{SIGSPEC_EXnormalrun}). The file {\tt coltimecolobs.ini} contains the two lines}

\begin{scriptsize}\begin{verbatim}
col:time 2
col:obs 3
\end{verbatim}\end{scriptsize}

\subsection{Time series columns containing statistical weights}\label{SIGSPEC_Time series columns containing statistical weights}

Furthermore, one or more columns with statistical weights may be specified using the keyword {\tt col:weights}\label{SIGSPEC_keyword.col:weights}. The keyword accepts two arguments: the first is the column index, the second is a floating-point value, say $p_n$ for the $n$th weights column. Given $N$ weights columns indexed according to $n = 1,\, ...,\, N$, the total weight for the $m$th row is evaluated using the weights $w_{nm}$ in the individual columns according to
\begin{equation}
\Gamma _m := \prod _{n=1}^{N}\gamma _{nm}^{p_n}\: .
\end{equation}

Weights need not be normalised, this is performed by {\sc SigSpec}.

Time, observable, and weights columns have to consist of floating-point numbers only. {\sc SigSpec} checks these columns before starting the computations. If a non-numeric entry is found in one of the special columns, the program terminates with an error message.

\vspace{12pt}
{\bf Caution: {\sc SigSpec} does not support the exponential annotation (e.\,g.~{\tt 4.234E03} or {\tt 1.0385e-03})!}
\vspace{12pt}

\vspace{12pt}\noindent{\bf Example.} {\it The sample project {\tt weights} contains a dataset with statistical weights in column 3 the squares of which are used by {\sc SigSpec}, as specified by the {\tt .ini} file entry

\begin{scriptsize}\begin{verbatim}
col:weights 3 2
\end{verbatim}\end{scriptsize}

\noindent The input time series {\tt weights.dat} represents the sampling of IC\,4996\,\#\,89 (V), and the magnitudes were synthesized by
\begin{enumerate}
\item a sinusoid with frequency 4.68573 cycles per day, amplitude 17.27 mmag,
\item Gaussian noise with 25 mmag rms deviation only for the measurements between HJD 2452524 and HJD 2452525,
\item Gaussian noise with 2.5 mmag rms deviation for all other nights.
\end{enumerate}

\figureDSSN{f03.eps}{Light curve for the sample project {\tt weights}.}{SIGSPEC_weights.dat}{!htb}{clip,angle=0,width=110mm}

The resulting light curve is displayed in Fig.\,\ref{SIGSPEC_weights.dat}\it . Fig.\,\ref{SIGSPEC_weights.s} \it compares the frequency domain output (a closeup for frequencies below 10 cycles per day) with and without employing the weights. Without weights, the peak at 4.7 cycles per day visible, but not the most significant one. Moreover, there is no signal that reaches the sig threshold of 5.\footnote{The result without weights is found in the project directory {\tt noweights}.}

\begin{scriptsize}\begin{verbatim}
   1 freq 5.68136  sig 3.75547  rms 0.088716  csig 3.75547
\end{verbatim}\end{scriptsize}

\figureDSSN{f04.eps}{{\em Grey:} Fourier spectra for the sample project {\tt weights}. {\em Left:} significance spectra. {\em Right:} DFT amplitudes. {\em Top:} spectra of the unweighted time series. {\em Bottom:} spectra employing statistical weights. The significant components are indicated by {\em dots} with dashed drop lines (file {\tt weights/result.dat}). The default sig threshold of 5 is represented by a horizontal {\em dashed line} in the left panels.}{SIGSPEC_weights.s}{!htb}{clip,angle=0,width=110mm}

Column 3 in the time series input file {\tt weights.dat} contains zeroes for the measurements between HJD\,2452524 and HJD\,2452525 and values of 1 for the rest. Consequently, in this example, the exponent 2 assigned to the keyword {\tt col:weights} in the file {\tt weights.ini} does not affect the weighting: the result would be the same if, e.\,g., 

\begin{scriptsize}\begin{verbatim}
col:weights 3 1
\end{verbatim}\end{scriptsize}

\noindent were given instead of

\begin{scriptsize}\begin{verbatim}
col:weights 3 2
\end{verbatim}\end{scriptsize}

\noindent Employing the weights column, {\sc SigSpec} provides the component at 4.7 cycles per day as the only significant signal:

\begin{scriptsize}\begin{verbatim}
   1 freq 4.67968  sig 20.395  rms 0.029129  csig 20.395
   2 freq 30.5489  sig 4.47468  rms 0.0252866  csig 4.47468
\end{verbatim}\end{scriptsize}\sf

\subsection{Time series columns containing subset identifiers}\label{SIGSPEC_Time series columns containing subset identifiers}

If the mean magnitude of a light curve is desired to be adjusted to zero for each night, or if the data are obtained from more than one site, one may perform an individual zero-mean correction for subsets of the total time series. This is achieved by the keyword {\tt col:ssid}\label{SIGSPEC_keyword.col:ssid} in the {.ini} file. This keyword is followed by the index of the column which shall be assigned to subset identifiers and may be multiply defined, if more than one subset identifier column is given. Subset identifiers may be arbitrary alpha-numeric strings.

If {\tt col:ssid} is specified, {\sc SigSpec} treats all lines in the dataset with equal subset identifiers as individual subsets, for each of which a zero-mean correction is performed. Subsequently, {\sc SigSpec} performs the appropriate statistical calculations, taking into account that the zero-mean correction for subsets diminishes the degrees of freedom for noise.

If more than one subset column is specified, data points are considered to belong to the same subset, if all corresponding subset identifiers are equal.

\figureDSSN{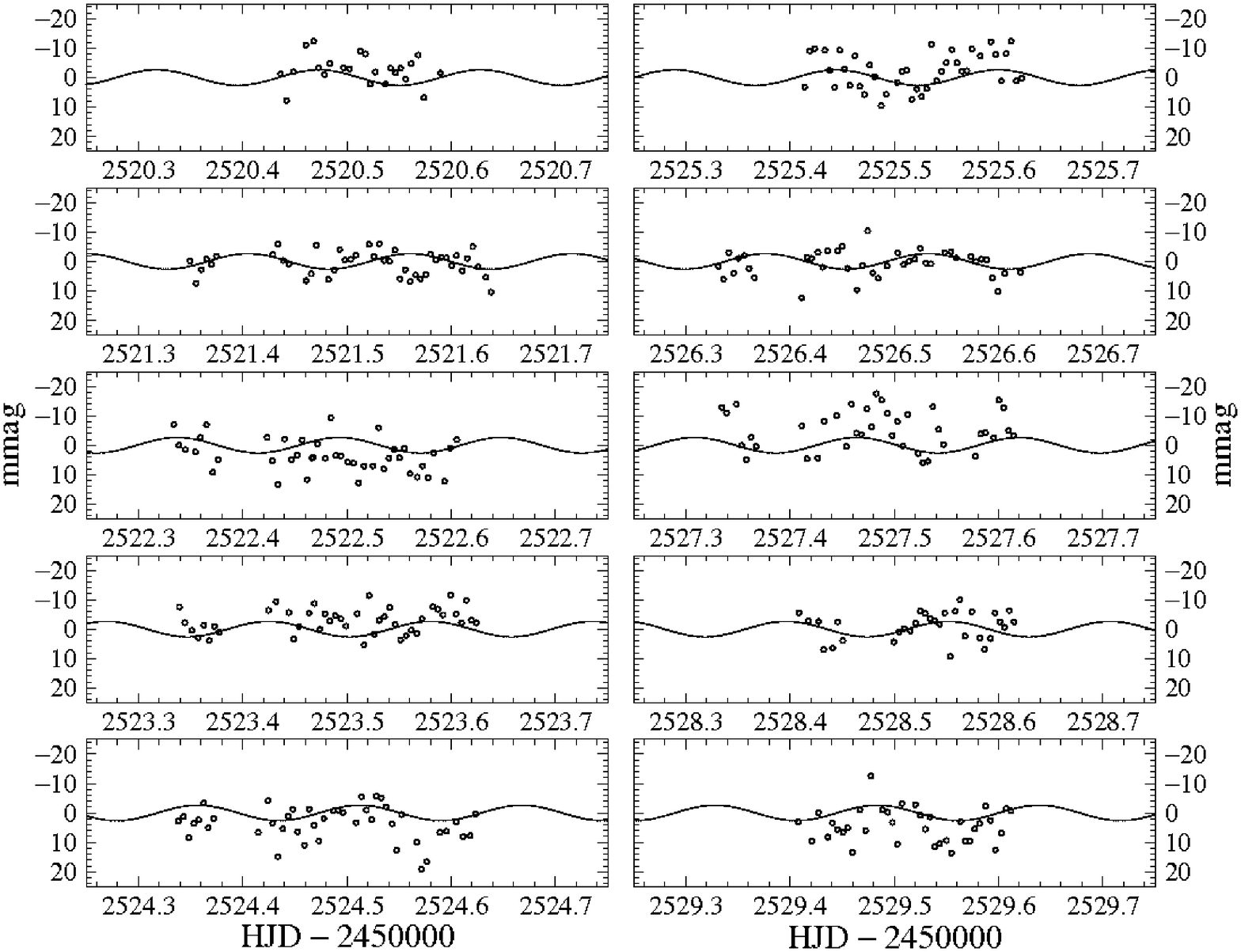}{Light curve for the sample project {\tt subsets}. {\em Solid line:} Sinusoidal signal used as input.}{SIGSPEC_subsets.dat}{!htb}{clip,angle=0,width=110mm}

\vspace{12pt}\noindent{\bf Example.} {\it The sample project {\tt subsets} contains a dataset with subset identifiers in column 3. The input time series {\tt subsets.dat} represents the sampling of IC\,4996\,\#\,89 (V), and the magnitudes were synthesized by adding
\begin{enumerate}
\item Gaussian noise with 5 mmag rms deviation,
\item a sinusoid with frequency 6.43682 cycles per day and amplitude 2.62 mmag,
\item individual constant zeropoint offsets on a millimag range for each night.
\end{enumerate}

The resulting light curve is displayed in Fig.\,\ref{SIGSPEC_subsets.dat}\it , displaying the input signal as a solid line and the data points including the nightly offsets as open dots. Fig.\,\ref{SIGSPEC_subsets.s} \it compares the resulting frequency domain output (a closeup for frequencies below 10 cycles per day) with and without employing the weights. If no subdivision according to the subset identifiers is performed, the spectra show a peak at 6.4 cycles per day plus several spurious peaks at frequencies below 2 cycles per day, which are due to the interpretation of the nightly shifts as signal in the 1-cycle-per-day domain and also harmonics.\footnote{The result without subsets is found in the project directory {\tt nosubsets}.} Consequently, {\sc SigSpec} identifies two additional significant signal components at low frequencies:\pagebreak

\begin{scriptsize}\begin{verbatim}
   1 freq 0.575256  sig 14.5784  rms 6.0813  csig 14.5784
   2 freq 7.43176  sig 6.39232  rms 5.51956  csig 6.39232
   3 freq 0.286066  sig 5.21585  rms 5.29531  csig 5.18785
   4 freq 75.1664  sig 3.39587  rms 5.12278  csig 3.38892
\end{verbatim}\end{scriptsize}

\it Column 3 in the time series input file {\tt weights.dat} contains characters A to J uniquely assigned to each night. Employing the subsets column eliminates the low-frequency artefacts, and {\sc SigSpec} provides the component at 6.4 cycles per day as the only significant signal:}

\begin{scriptsize}\begin{verbatim}
   1 freq 6.4376  sig 8.20485  rms 5.2253  csig 8.20485
   2 freq 75.1661  sig 3.70924  rms 4.95954  csig 3.70922
\end{verbatim}\end{scriptsize}

\figureDSSN{f06.eps}{{\em Grey:} Fourier spectra for the sample project {\tt subsets}. {\em Left:} significance spectra. {\em Right:} DFT amplitudes. {\em Top:} spectra of the total time series. {\em Bottom:} spectra of the subdivided time series. The significant components are indicated by {\em dots} with dashed drop lines (file {\tt subsets/result.dat}). The default sig threshold of 5 is represented by a horizontal {\em dashed line} in the left panels.}{SIGSPEC_subsets.s}{!htb}{clip,angle=0,width=110mm}

\subsection{Lower frequency limit}

The frequency where the computation of spectra starts is specified by the keyword {\tt lfreq}\label{SIGSPEC_keyword.lfreq}. By default, the lower frequency limit is zero.

\vspace{12pt}\noindent{\bf Example.} \it The sample project {\tt limits} illustrates the use of the keyword {\tt lfreq}. It uses the V photometry of IC\,4996\,\#\,89 as input file {\tt limits.dat}, and the file {\tt limits.ini} contains the line

\begin{scriptsize}\begin{verbatim}
lfreq 1
\end{verbatim}\end{scriptsize}

which forces {\sc SigSpec} to perform all computations for frequencies $\ge$ 1 cycle per day. The spectrum {\tt limits/s000000.dat} is displayed in Fig.\,\ref{SIGSPEC_limits.s}\it .\sf

\figureDSSN{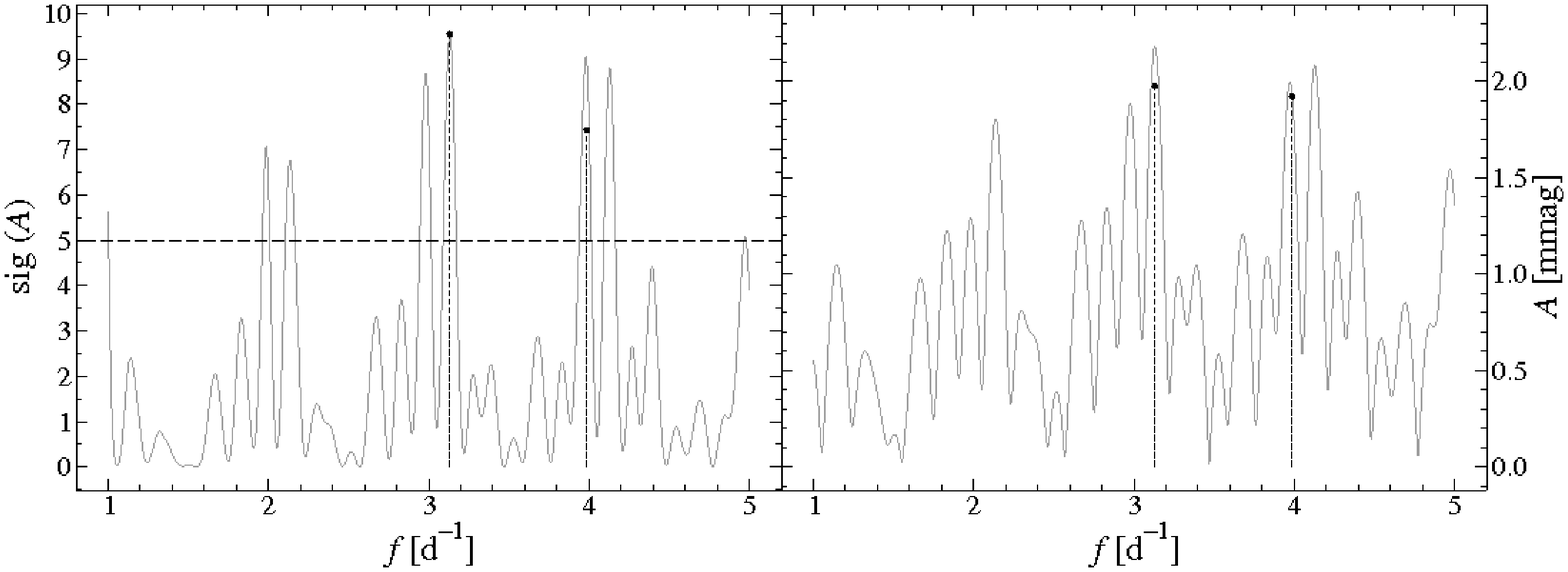}{{\em Grey:} Fourier spectra for the sample project {\tt limits}. {\em Left:} significance spectrum. {\em Right:} DFT amplitudes. The significant components are indicated by {\em dots} with dashed drop lines (file {\tt limits/result.dat}). The default sig threshold of 5 is represented by a horizontal {\em dashed line} in the left panel. The frequency range is set from 1 to 5 cycles per day using the keywords {\tt lfreq} and {\tt ufreq}.}{SIGSPEC_limits.s}{!htb}{clip,angle=0,width=110mm}

\subsection{Upper frequency limit and Nyquist Coefficient}\label{SIGSPEC_Upper frequency limit and Nyquist Coefficient}

The keyword {\tt ufreq}\label{SIGSPEC_keyword.ufreq} allows to determine the upper limit of the frequency interval to be considered. 

An alternative method is the automatic determination of this limit by means of the Nyquist Coefficient (keyword {\tt nycoef}\label{SIGSPEC_keyword.nycoef}). For equidistantly sampled time series with sampling interval width $\delta t$, there is a uniquely defined {\it Nyquist Frequency}
\begin{equation}
f_\nu := \frac{1}{2\,\delta t}\, .
\end{equation}
In case of non-equidistant sampling, each sampling interval between two consecutive time values may be considered to produce its individual Nyquist Frequency, whence this limit is ambiguous. In this case, the Nyquist Coefficient for an arbitrarily given frequency is introduced as the fraction of sampling intervals in the time domain the individual Nyquist Frequency of which is higher than the frequency under consideration. This provides to select an upper frequency limit by specifying a minimum Nyquist Coefficient. E.\,g., specifying a Nyquist Coefficient of 0.5 (which is the default value) guarantees that at least half of the information contained by the spectrum in the considered frequency range is below the Nyquist limit.

Additional information is available by setting the keyword {\tt nyscan}\label{SIGSPEC_keyword.nyscan} in the {\tt .ini} file. If this keyword is specified, {\sc SigSpec} creates a file {\tt nyscan.dat} in the project directory containing the Nyquist Coefficients over the specified frequency range.

\vspace{12pt}\noindent{\bf Example.} {\it The sample project {\tt limits} illustrates the use of the keyword {\tt ufreq}. The line

\begin{scriptsize}\begin{verbatim}
ufreq 5
\end{verbatim}\end{scriptsize}

\noindent in the file {\tt limits.ini} restricts all computations performed by {\sc SigSpec} to frequencies below 5 cycles per day. The spectrum {\tt limits/s000000.dat} (sig and amplitude) is displayed in Fig.\,\ref{SIGSPEC_limits.s}\it . A comparison with the screen output in Example {\tt SigSpecNative}, p.\,\pageref{SIGSPEC_EXnormalrunfreqs}, where no restrictions to the frequency range apply, shows that the screen output in this example contains one line less:

\begin{scriptsize}\begin{verbatim}
   1 freq 3.13205  sig 9.54539  rms 0.00449592  csig 9.54539
   2 freq 3.98471  sig 7.43085  rms 0.00422861  csig 7.42753
   3 freq 2.664  sig 4.60182  rms 0.0040257  csig 4.60117
\end{verbatim}\end{scriptsize}

\noindent The signal component at 5.4 cycles per day is not detected, because it is outside the specified frequency range.}

\vspace{12pt}\noindent{\bf Example.} \it The sample project {\tt nyos} illustrates the use of the keywords {\tt nycoef} and {\tt nyscan} for the V photometry of IC\,4996\,\#\,89. The line

\begin{scriptsize}\begin{verbatim}
nycoef 0.99
\end{verbatim}\end{scriptsize}

\noindent in the file {\tt nyos.ini} provides an upper frequency limit of 110.77 cycles per day. The keyword {\tt nyscan} is given, and the file {\tt nyos/nyscan.dat} contains the Nyquist Coefficients for frequencies from 0 to 110.77 cycles per day, as displayed in Fig.\,\ref{SIGSPEC_nyscan.s}.\sf

\figureDSSN{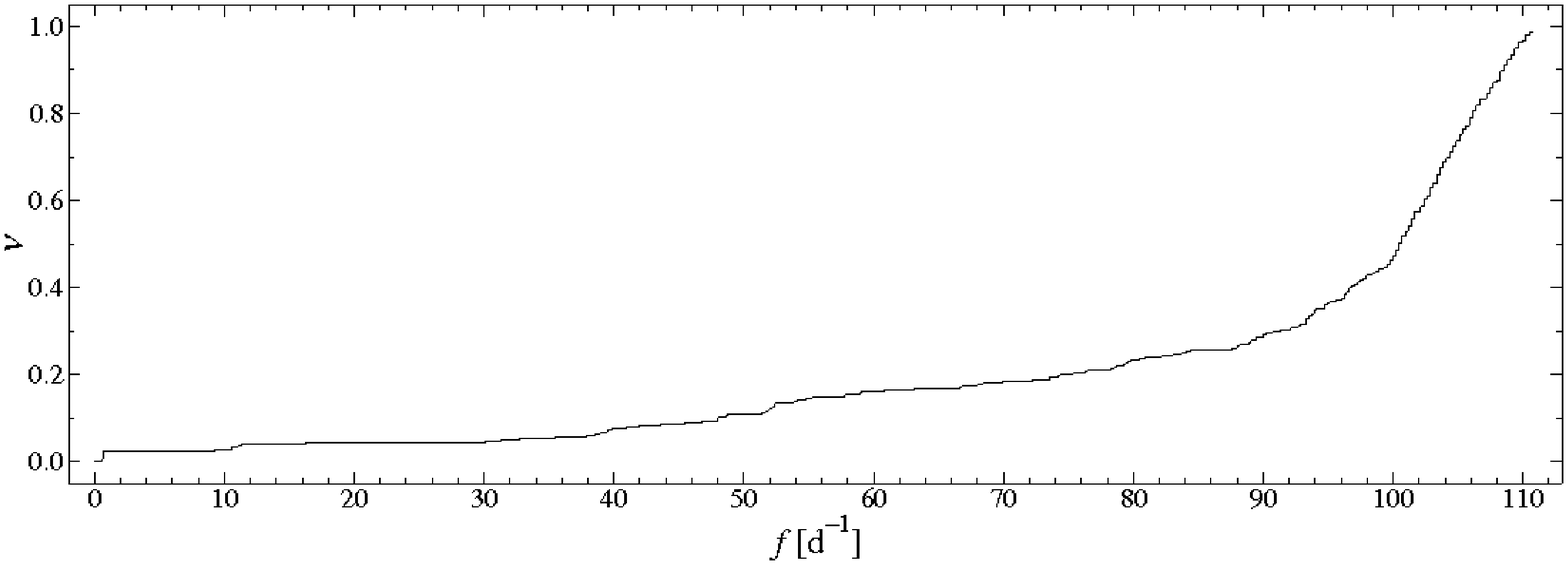}{The file {\tt nyos/nyscan.dat} contains the Nyquist coefficients depending on frequency for the sample project {\tt nyos}.}{SIGSPEC_nyscan.s}{!htb}{clip,angle=0,width=110mm}

\subsection{Frequency spacing and oversampling ratio}

The width of the interval between consecutive frequencies may be specified by the keyword {\tt freqspacing}\label{SIGSPEC_keyword.freqspacing}.

An alternative method is the automatic determination of the spacing by means of the oversampling ratio. In case of equidistantly sampled time series, the frequency spacing is defined by
\begin{equation}
\delta f := \frac{1}{T}\, ,
\end{equation}
where $T$ denotes the width of the total time interval. For non-equidistant time series, it is advisable to use a denser frequency sampling,
\begin{equation}
\delta f := \frac{1}{\Omega T}\, ,
\end{equation}
where $\Omega$ is the oversampling ratio. This quantity may be specified using the keyword {\tt osratio}\label{SIGSPEC_keyword.osratio}. The default value is 20, which is -- in most cases -- sufficient for practical use.

\vspace{12pt}\noindent{\bf Example.} \it The sample project {\tt limits} illustrates the use of the keyword {\tt freqspacing}, an example for the keyword {\tt osratio} is provided in the sample project {\tt nyos}. Both samples use the V photometry of IC\,4996\,\#\,89 as input time series. The line

\begin{scriptsize}\begin{verbatim}
freqspacing 0.001
\end{verbatim}\end{scriptsize}

\noindent in the file {\tt limits.ini} forces {\sc SigSpec} to calculate Fourier amplitudes and sigs for every 0.001 cycles per day. In the file {\tt nyos.ini}, there is a line

\begin{scriptsize}\begin{verbatim}
osratio 12
\end{verbatim}\end{scriptsize}

\noindent which overrides the default oversampling ratio of 20. Fig.\,\ref{SIGSPEC_os.s} \it compares the standard spacing from Example {\tt SigSpecNative}, p.\,\pageref{SIGSPEC_EXnormalrun}), with the spacings obtained applying the two above modifications.\sf

\figureDSSN{f09.eps}{Close-up for the significance spectra generated by the projects {\tt SigSpecNative}, {\tt limits} and {\tt nyos} around the main peak for the V photometry of IC\,4996\,\#\,89. Different settings for frequency spacing and oversampling ratio are applied.}{SIGSPEC_os.s}{!htb}{clip,angle=0,width=110mm}

\subsection{Accuracy of MultiSine fits}\label{SIGSPEC_Accuracy of MultiSine fits}

By default, {\sc SigSpec} performs a MultiSine least-squares fit each time a new significant signal component is detected. The fitting procedure improves the frequencies, amplitudes, and phases of all previously detected signal components. The algorithm applies Newton's root finding scheme to the first derivatives of the residual variance.

The precision of computed frequencies via MultiSine least-sqares fits is defined according to
\begin{equation}
\delta f:=\frac{\mu}{T\,\mathrm{sig}^\frac{\kappa}{2}}\: ,
\end{equation}
where $\mu$ and $\kappa$ are the accuracy parameters for MultiSine fitting. The default value of $\mu$ is $10^{-6}$, that of $\kappa$ is $1$. They may be adjusted by the keyword {\tt multisine:newton}\label{SIGSPEC_keyword.multisine:newton}, followed by $\mu$, $\kappa$ and a third parameter determining the relative tolerance of the time-domain rms error between consecutive iterations (see next paragraph). To reduce the potential time consumption of the procedure, $\mu$ can be adjusted to achieve an overall scaling of the frequency accuracy. The value of $\kappa$ determines the dependence of the demanded frequency precision on the sig of the peak under consideration. Setting $\kappa = 0$ yields the Rayleigh frequency resolution, for $\kappa = 1$ one obtains the Kallinger resolution (Kallinger, Reegen \& Weiss 2008).

The criterion on which MultiSine fitting is based is the minimisation of rms residual. Thus the rms residual is demanded to decrease from one iteration to the next. Otherwise the fitting procedure is terminated. To speed up the computation, the MultiSine fit can be terminated, if the relative improvement of rms residual drops below a positive number. The default value $10^{-6}$. This value may be re-adjusted by the third parameter to the keyword {\tt multisine:newton}.

The two termination conditions are linked by a logical `and', i.\,e. the MultiSine fitting procedure stops if both the desired frequency accuracy is reached for all signal components and the improvement of residual rms drops below the specified threshold.

There is an optional keyword, {\tt multisine:lock}\label{SIGSPEC_keyword.multisine:lock}, that forces the prewhitening cascade to rely on the ``raw'' frequencies, amplitudes and phases (i.\,e. those without MultiSine fitting). Resulting signal components are improved to obtain a least-squares fit in each iteration, but this improvement is ignored in the prewhitening sequence. The default setting is that the improved parameters are used for the subsequent analysis (as also obtained by the keyword {\tt multisine:unlock}\label{SIGSPEC_keyword.multisine:unlock}.

\vspace{12pt}\noindent{\bf Example.} \it The sample project {\tt multisine} illustrates the application of the keyword {\tt multisine:newton} to the IC\,4996\,\#\,89 photometry (V) as input file {\tt multisine.dat}. The file {\tt multisine.ini} contains the line

\begin{scriptsize}\begin{verbatim}
multisine 0.001 0 0.01
\end{verbatim}\end{scriptsize}

\noindent which reduces the accuracy of the MultiSine fit, compared to the default values 0.000001, 1, 0.000001, respectively. The second parameter refers to the Rayleigh frequency resolution rather than the (default) Kallinger frequency resolution. The screen output provides four entries:

\begin{scriptsize}\begin{verbatim}
   1 freq 3.13205  sig 9.54539  rms 0.00449592  csig 9.54539
   2 freq 3.98472  sig 7.43087  rms 0.00422861  csig 7.42755
   3 freq 5.40686  sig 5.29838  rms 0.0040257  csig 5.29516
   4 freq 17.3677  sig 4.13727  rms 0.00388775  csig 4.10809
\end{verbatim}\end{scriptsize}

For comparison, the project {\tt SigSpecNative}, p.\,\pageref{SIGSPEC_EXnormalrun}, employs the default settings.

For the first entry, there is no difference between the two results, but due to propagation of uncertainties, the following entries show slight and increasing deviations. As expected, the rms errors of residuals are higher if the accuracy is reduced.\sf

\subsection{Program termination}\label{SIGSPEC_Program termination}

There are three possibilities to specify a criterion for program termination:
\begin{enumerate}
\item the number of iterations (keyword {\tt iterations}\label{SIGSPEC_keyword.iterations}),
\item a lower sig limit (keyword {\tt siglimit}\label{SIGSPEC_keyword.siglimit}),
\item the reliability of the entire analysis is determined by the {\it cumulative sig}. It is based on the probability that at least one of the frequency components detected so far is due to noise. A threshold in terms of cumulative sig may be defined using the keyword {\tt csiglimit}\label{SIGSPEC_keyword.csiglimit} For an introduction to the cumulative sig, please refer to p.\,\pageref{SIGSPEC_csig}.
\end{enumerate}

Multiple specifications in terms of these keywords cause the prewhitening cascade to terminate if one of the limits is reached.

The default assignment for {\tt siglimit} is $5$. This pre-definition may be deactivated by defining

\begin{scriptsize}\begin{verbatim}
siglimit 0
\end{verbatim}\end{scriptsize}

\noindent in the {\tt .ini} file. The limits {\tt iterations} and {\tt csiglimit} are switched off by default.

\vspace{12pt}\noindent{\bf Example.} \it The sample project {\tt terminate} contains a combination of the keywords {\tt siglimit}, {\tt csiglimit} and {\tt iterations}, applied to the V photometry of IC\,4996\,\#\,89 as input file. For a comparison to the standard output, please refer to Example {\tt SigSpecNative}, p.\,\pageref{SIGSPEC_EXnormalrun}. The file {\tt terminate.ini} contains a combination of three keywords:

\begin{scriptsize}\begin{verbatim}
siglimit 0
csiglimit 3
iterations 10
\end{verbatim}\end{scriptsize}

The first line deactivates the default setting of 5 for the sig limit. The combination of the second and third line forces {\sc SigSpec} to terminate after 10 iterations, or earlier, if the cumulative sig drops below 3. The screen output provides seven lines, corresponding to six significant signal components:

\begin{scriptsize}\begin{verbatim}
   1 freq 3.13205  sig 9.54539  rms 0.00449592  csig 9.54539
   2 freq 3.98471  sig 7.43085  rms 0.00422861  csig 7.42753
   3 freq 5.40684  sig 5.30164  rms 0.0040257  csig 5.2984
   4 freq 17.3677  sig 4.13698  rms 0.00388775  csig 4.10802
   5 freq 3.67101  sig 3.73187  rms 0.00378701  csig 3.57943
   6 freq 52.5182  sig 3.41319  rms 0.00369756  csig 3.18744
   7 freq 41.7372  sig 3.02872  rms 0.00361981  csig 2.80001
\end{verbatim}\end{scriptsize}

The cumulative sig of 2.8 for component 7 is responsible for program termination before the limit of 10 iterations is reached.\sf

\section{Default Output}\label{SIGSPEC_Default Output}

All output files are written into the project directory. A six-digit index denotes the iteration in the prewhitening cascade. E.\,g., an index {\tt 000000} represents a file obtained from the input data without any prewhitening, {\tt 000001} denotes a file after prewhitening of the first sinusoidal component. The general annotation {\tt \#iteration\#} will be used for this six-digit identifier.

\vspace{12pt}\noindent{\bf Example.}\footnote{The sample project {\tt output} is the most time consuming sample of all. The computation takes 90 minutes on an Intel Core2 CPU T5500 (1.66GHz) under Linux 2.6.18.8-0.9-default i686. This is mostly due to the calculations of the Sock and Phase Distribution Diagrams. In order to speed up the program, the user may switch off these operations by placing a {\tt \#} character at the beginning of all lines containing keywords {\tt sock:...} and {\tt phdist:...} in the file {\tt output.ini}.} \it The sample project {\tt output} illustrates how to adjust the output of {\sc SigSpec}. The input file {\tt output.dat} represents 16 nights (992 data points) of Str\o mgren y photometry (Vienna University APT, T6; Strassmeier et al.~1997) of the Delta Scuti star EE\,Cam (Breger, Rucinski \& Reegen 2007). The light curve is displayed in Fig.\,\ref{SIGSPEC_output.dat}\it .

\figureDSSN{f10.eps}{Light curve for the sample project {\tt output}.}{SIGSPEC_output.dat}{!htb}{clip,angle=0,width=110mm}
\pagebreak The vast amount of output provided by Sock Diagrams and Phase Distribution Diagrams makes it necessary to restrict the frequency interval in the file {\tt output.ini}. Especially close to zero frequency, the output may be tremendous. Thus the very low frequencies are avoided:

\begin{scriptsize}\begin{verbatim}
lfreq 1
ufreq 16
\end{verbatim}\end{scriptsize}

\noindent The frequency spacing is adjusted to speed up the computations of Sock and Phase Distribution Diagrams.

\begin{scriptsize}\begin{verbatim}
freqspacing 0.005
\end{verbatim}\end{scriptsize}

All other entries in the file {\tt output.ini} apply to output files and are discussed in the subsequent sections.\sf

\subsection{Spectra}\label{SIGSPEC_Spectra}

By default, two spectra (files {\tt s000000.dat} and {\tt resspec.dat}) are generated. The file {\tt s000000.dat} contains the spectrum of the original time series, and the file {\tt resspec.dat} represents the residual spectrum after finishing the prewhitening sequence. 

The columns are
\begin{enumerate}
\item frequency [inverse time units],
\item sig,
\item DFT amplitude [units of observable],
\item Fourier-space phase angle [rad],
\item Fourier-space phase angle of maximum sig [rad].
\end{enumerate}

To achieve consistency with the output for differential significance spectra (see p.\,\pageref{SIGSPEC_Differential significance spectra}), two further columns are found containing values $-1$ and $0$ only.

The phase angles $\theta$ are given according to a trigonometric fit,
\begin{equation}
F\left( t\right) := A\cos\left( 2\pi ft - \theta\right)\, ,
\end{equation}
with amplitude $A$ and frequency $f$ as given in the file. This convention is compatible to the definition of phase in Fourier space. This definition of phase is consistently used for all types of {\sc SigSpec} output.

If the keyword {\tt spectra}\label{SIGSPEC_keyword.spectra} is provided in the {\tt .ini} file, additional output files {\tt s\#iteration\#.dat} are generated. The index {\tt \#iteration\#} starts with {\tt 000001}, denoting the residual spectrum after the first prewhitening step. 

The keyword {\tt spectra} expects two integer parameters. The first defines the number of iterations for which these files shall be generated. A negative number causes {\sc SigSpec} to generate files for all iterations. The second parameter has to be a positive number and defines a step width. If it is set {\tt 1}, a file is generated after each iteration, if it is set {\tt 2}, after every second iteration (starting with {\tt s000002.dat}), and so on.

\vspace{12pt}\noindent{\bf Example.} \it The sample project {\tt output} uses the keyword {\tt spectra} in the file {\tt output.ini}, namely

\begin{scriptsize}\begin{verbatim}
spectra 10 2
\end{verbatim}\end{scriptsize}

Spectra are written only during the first 10 iterations of the prewhitening sequence. The second parameter provides only every second file to be generated. In this example, the following files are produced:

\begin{scriptsize}\begin{verbatim}
output/s000000.dat
output/s000002.dat
output/s000004.dat
output/s000006.dat
output/s000008.dat
output/s000010.dat
\end{verbatim}\end{scriptsize}

\noindent In addition, the file {\tt resspec.dat} contains the residual spectrum after all iterations.\sf

\subsection{Residual time series}\label{SIGSPEC_Residual time series}

By default, a file {\tt residuals.dat} is generated. It represents the residuals after prewhitening all signal components found significant. The column format is the same as for the time series input file.

If the keyword {\tt residuals}\label{SIGSPEC_keyword.residuals} is provided in the {\tt .ini} file, additional files {\tt t\#iteration\#.dat} are generated, where the index {\tt \#iteration\#} starts with {\tt 000001}, denoting the residuals after the first prewhitening step. 

The keyword {\tt residuals} expects two integer parameters. The first defines the number of iterations for which these files shall be generated. A negative number causes {\sc SigSpec} to generate files for all iterations. The second parameter has to be a positive number and defines a step width. If it is set {\tt 1}, a file is generated after each iteration, if it is set {\tt 2}, after every second iteration (starting with {\tt t000002.dat}), and so on.

\vspace{12pt}\noindent{\bf Example.} \it The sample project {\tt output} uses the keyword {\tt residuals} in the file {\tt output.ini}, namely

\begin{scriptsize}\begin{verbatim}
residuals -1 5
\end{verbatim}\end{scriptsize}

Setting the first parameter $-$1 provides residual time series during the entire prewhitening sequence. The second parameter provides only fifth second file to be generated. Since the number of iterations in this example is 40, the following files are produced:

\begin{scriptsize}\begin{verbatim}
output/t000005.dat
output/t000010.dat
output/t000015.dat
output/t000020.dat
output/t000025.dat
output/t000030.dat
\end{verbatim}\end{scriptsize}

\noindent In addition, the file {\tt residuals.dat} contains the residual time series after all iterations.\sf

\subsection{Result files}\label{SIGSPEC_Result files}

The file {\tt result.dat} contains a list of all identified sig maxima. This file consists of seven columns providing
\begin{enumerate}
\item frequency [inverse time units],
\item sig,
\item amplitude [units of observable],
\item phase [rad],
\item rms scatter of the time series before prewhitening,
\item point-to-point scatter of the time series before prewhitening,
\item the cumulative sig for all frequency components detected so far.
\end{enumerate}

Columns 3 and 4 represent amplitude and phase as the result of a least-squares fit to the time series at the present prewhitening stage (i.\,e.~after subtraction of all previously identified signal components) for the frequency of maximum significance.

The last line in the file contains zeroes for frequency, amplitude, and phase. The non-zero values refer to the (cumulative) sig of the most significant component below the threshold, and to the rms and point-to-point scatter after the last prewhitening step, respectively. This final line is suppressed if the criterion {\tt iterations} is responsible for program termination.

If the keyword {\tt results}\label{SIGSPEC_keyword.results} is provided in the {\tt .ini} file, additional result files {\tt r\#iteration\#.dat} are generated, where the index {\tt \#iteration\#} starts with {\tt 000001}, denoting the result of the first iteration. The files contain the significant components within the prewhitening cascade as preliminary results. The MultiSine least-squares fits which are performed at each step of the prewhitening sequence modify frequencies, amplitudes and phases. Therefore it may be useful to have additional results from earlier iterations in hands, if the user decides not to use all components found by {\sc SigSpec} without re-running the program.

The keyword {\tt results} expects two integer parameters. The first defines the number of iterations for which these files shall be generated. A negative number causes {\sc SigSpec} to generate files for all iterations. The second parameter has to be a positive number and defines a step width. If it is set {\tt 1}, a result file is generated after each iteration, if it is set {\tt 2}, after every second iteration (starting with {\tt r000002.dat}), and so on.

\vspace{12pt}\noindent{\bf Example.} \it The sample project {\tt output} uses the keyword {\tt results} in the file {\tt output.ini}, namely

\begin{scriptsize}\begin{verbatim}
results -1 1
\end{verbatim}\end{scriptsize}

\noindent providing result files {\tt r000001.dat}, {\tt r000002.dat},..., for all iterations of the entire prewhitening sequence. In addition, the final result after all prewhitening iterations is contained in the file {\tt results.dat}.\sf

\section{Analysis of the Time-domain Sampling}\label{SIGSPEC_Analysis of the Time-domain Sampling}

\vspace{12pt}\noindent{\bf Example.} \it The sample project {\tt output} contains the output of a spectral window, a sampling profile, a sock diagram, a phase distribution diagram, a preview, and correlograms.\sf

\subsection{Spectral window}\label{SIGSPEC_Spectral window}

The spectral window is computed, if the keyword {\tt win}\label{SIGSPEC_keyword.win} is given in the {\tt .ini} file. This keyword does not require any parameters. The output is provided in the file {\tt win.dat}. It consists of three columns referring to
\begin{enumerate}
\item frequency [inverse time units],
\item amplitude [units of observable],
\item Fourier-space phase angle [rad].
\end{enumerate}

\vspace{12pt}\noindent{\bf Example.} \it The sample project {\tt output} contains the output of a spectral window. The file {\tt output.ini} contains the keyword {\tt win}, and the corresponding output is found in the file {\tt output/win.dat} and displayed in Fig.\,\ref{SIGSPEC_win.s}\it . The frequency limits determined by the lines

\begin{scriptsize}\begin{verbatim}
lfreq 1
ufreq 16
\end{verbatim}\end{scriptsize}

\noindent also apply to the spectral window.\sf

\figureDSSN{f11.eps}{Spectral window for the sample project {\tt output}.}{SIGSPEC_win.s}{!htb}{clip,angle=0,width=110mm}

\subsection{Sampling profile}\label{SIGSPEC_Sampling profile}

The sampling profile is an essential part of the sig computation. All parameters to describe the influence of the time series sampling in Fourier space is entirely contained in the three parameters $\alpha _0$, $\beta _0$, and $\theta _0$. The values of $\alpha _0$ and $\beta _0$ are measures for the maximum and minimum sig for all phase angles at a given frequency, and the angle $\theta _0$ determines the phase angle where maximum sig is obtained at the frequency under consideration. A detailed description is given by Reegen~(2007). If the keyword {\tt profile}\label{SIGSPEC_keyword.profile} is provided in the {\tt .ini} file, the sampling profile for the given time series is written to the file {\tt profile.dat}. The four columns refer to

\begin{enumerate}
\item frequency [inverse time units],
\item $\alpha _0$,
\item $\beta _0$,
\item $\theta _0$ [rad].
\end{enumerate}

\vspace{12pt}\noindent{\bf Example.} \it In the file {\tt output.ini}, the keyword {\tt profile} is given and forces {\sc SigSpec} to generate an output file {\tt output/profile.dat} representing the sampling profile displayed in Fig.\,\ref{SIGSPEC_profile.s}\it .\sf

\figureDSSN{f12.eps}{Sampling profile for the sample project {\tt output}. The lower curve refers to $\alpha _0$, the upper curve to $\beta _0$. The orientation angle of the rms error ellipse, $\theta _0$ is not plotted.}{SIGSPEC_profile.s}{!htb}{clip,angle=0,width=110mm}

\subsection{Sock Diagram}\label{SIGSPEC_Sock Diagram}

The computation of a Sock Diagram is an optional add-on of {\sc SigSpec}. If the keyword {\tt sock:phases}\label{SIGSPEC_keyword.sock:phases} is given in the {\tt .ini} file, {\sc SigSpec} computes {\it sock significances}, i.\,e.~sig levels for a constant signal-to-noise ratio at a set of different phase angles, and for all frequencies for which spectra are calculated. As described by Reegen~(2007), the expected sig level for a given amplitude signal-to-noise ratio at constant frequency and phase angle is proportional to the squared amplitude signal-to-noise ratio. Sig levels in the Sock Diagram are normalised to an expected value of 1, corresponding to an amplitude signal-to-noise ratio
\begin{equation}
\frac{A}{\left< A\right>} = \frac{2}{\sqrt{\pi\log\mathrm{e}}} \approx 1.712\: .
\end{equation}
The sig level for an arbitrary signal-to-noise ratio may be deduced by multiplying the sig displayed in the Sock Diagram by $\frac{\pi\log\mathrm{e}}{4}\:\left(\frac{A}{\left< A\right>}\right) ^2 \approx 0.341\left(\frac{A}{\left< A\right>}\right) ^2\,$.

Furthermore, the phase angle in the Sock Diagram is given with respect to $\theta _0$, i.\,e.~the phases with maximum sock significance are consistently aligned to zero phase for all frequencies.

The number of phase angles in the interval $\left[ 0,\pi\right[$ to be taken into account for each frequency of the spectrum has to be given as an argument to the keyword {\tt sock:phases} in the {\tt .ini} file. The sig levels in the phase intervals $\left[ 0,\pi\right[$ and $\left[\pi ,2\pi\right[$ are symmetric according to
\begin{equation}
\mathrm{sig}\left( A,\omega ,\phi\right) = \mathrm{sig}\left( A,\omega ,\phi + \pi\right)\,\forall\phi\: ,
\end{equation}
but both given in the output file {\tt sock.dat} for completeness. The result represents a three-dimensional polar diagram of the sampling properties of the time series input file. To enhance the corresponding plot resolution, the number of phases specified with the keyword {\tt sock:phases} in the {\tt .ini} file is scaled by the maximum sig for each frequency. For sig maxima $\le 1$, the specified number is used, for sig maxima between $1$ and $2$, the number is doubled, and so on.

To enhance the quality of Sock Diagrams produced by {\sc SigSpec}, the keyword {\tt sock:fill}\label{SIGSPEC_keyword.sock:fill} can be provided to specify a fill factor (as a floating-point number following the keyword). It is used for adaptive oversampling of frequencies according to the differences of maximum sigs for consecutive frequencies. The fill factor is the (rounded) number of additional frequencies per unit of sig difference. In other words, providing {\tt sock:fill 10} guarantees that the resolution of the resulting Sock Diagram along the sig axis does not exceed 0.1, and an appropriate combination of the keywords {\tt sock:phases} and {\tt sock:fill} produces a Sock Diagram that mimics a continuous surface when plotted in 3D style. The default argument of {\tt sock:fill} is 0, which means that adaptive oversampling is switched off.

\vspace{12pt}
{\bf Caution: the Sock Diagram may become a huge file! Especially for very low frequencies, a tremendous amount of data may be expected. Thus it is advisable either to exclude this frequency region (keyword {\tt lfreq}) or to assign very low values to {\tt sock:phases} and {\tt sock:fill}.}
\vspace{12pt}

The user may choose to obtain the Sock Diagram in three-dimensional cylindrical (default, or keyword {\tt sock:cyl}\label{SIGSPEC_keyword.sock:cyl}) or cartesian coordinates (keyword {\tt sock:cart}\label{SIGSPEC_keyword.sock:cart}).

In any case, the output file {\tt sock.dat} consists of three columns. In cylindrical coordinates, the columns refer to
\begin{enumerate}
\item height coordinate: frequency [inverse time units],
\item azimuthal coordinate: phase with respect to the sig maximum [rad],
\item radial coordinate: sock significance.
\end{enumerate}
In cartesian coordinates, the columns refer to
\begin{enumerate}
\item frequency [inverse time units],
\item sock significance component in the direction of the sig maximum,
\item sock significance component in the direction of the sig minimum.
\end{enumerate}

The keywords {\tt sock:colmodel:lin}\label{SIGSPEC_keyword.sock:colmodel:lin} and {\tt sock:colmodel:rank}\label{SIGSPEC_keyword.sock:colmodel:rank} permit to choose between two different colour models assigning RGB colours to the data points of the Sock Diagram. The linear model ({\tt sock:colmodel:lin}) uses the sock significance as it is for colour scaling, whereas the rank model ({\tt sock:colmodel:rank}) relies on a rank statistics of sock significances.

\vspace{12pt}
{\bf Caution: the computation of ranks may be very time-consum\-ing! The progress control displayed during the calcucation of the rank statistics does not provide linear percentages in time. The percentage values refer to the portion of ranks among the number of data points that are finished.}
\vspace{12pt}

A sequence of keywords {\tt sock:colour}\label{SIGSPEC_keyword.sock:colour} determines a colour path that is assigned to the data points in the Sock Diagram. The keyword is followed by four floating-point arguments. The first three arguments refer to the three RGB channels. Using the linear model ({\tt sock:colmodel:lin}), the fourth argument is the sock significance to which the given colour has to be assigned. For the rank model ({\tt sock:colmodel:rank}), the fourth argument is a floating-point value in the interval $\left[0,1\right]$ and determines the fractile of data points to which the given colour has to be assigned. A value of, e.\,g., 0.5 assigns the specified colour to the median of sock significances. {\sc SigSpec} performs a linear interpolation along this colour path and assigns a fourth column to the output file {\tt sock.dat} containing RGB values. For entries beyond the start or end of the colour path, the start or end colour is used, correspondingly.

\figureDSSN{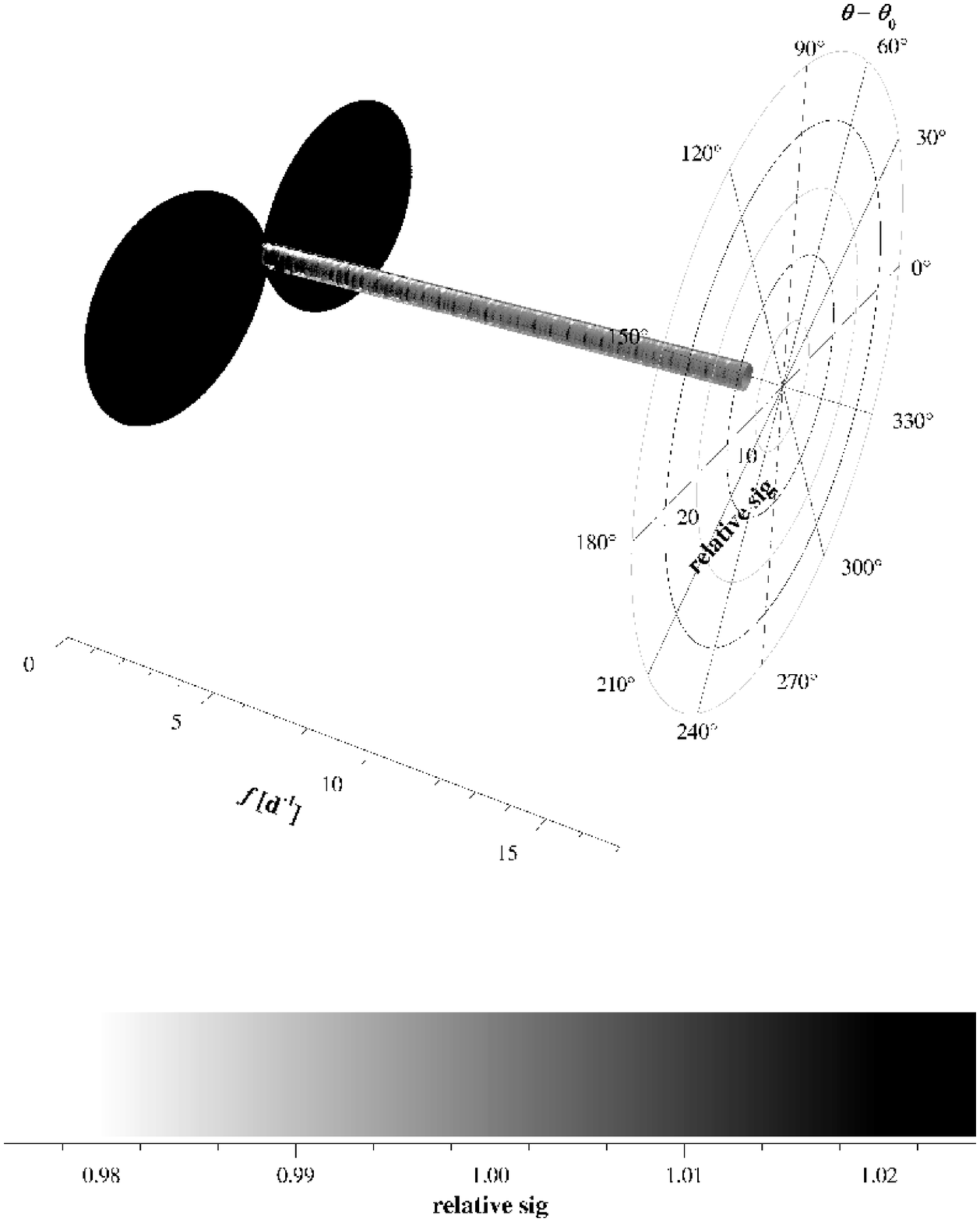}{Sock Diagram for the sample project {\tt output}.}{SIGSPEC_sock.s}{!htb}{clip,angle=0,width=240pt}

\vspace{12pt}\noindent{\bf Example.} \it A linear colour model that produces colours from white via red, yellow, green, cyan, blue, and magenta to black is produced by the following specifications:\sf

\begin{scriptsize}\begin{verbatim}
sock:colmodel:lin
sock:colour 255 255 255 .5
sock:colour 255   0   0 .9
sock:colour 255 255   0 .95
sock:colour   0 255   0 1
sock:colour   0 255 255 1.05
sock:colour   0   0 255 1.1
sock:colour 255   0 255 1.2
sock:colour   0   0   0 2
\end{verbatim}\end{scriptsize}

\vspace{12pt}\noindent{\bf Example.} \it A rank colour model producing greyscale coding is obtained by:\sf

\begin{scriptsize}\begin{verbatim}
sock:colmodel:rank
sock:colour   0   0   0 0
sock:colour 255 255 255 1
\end{verbatim}\end{scriptsize}

\vspace{12pt}\noindent{\bf Example.} \it The Sock Diagram in the sample project {\tt output} is generated according to the following entries in the file {\tt output.ini}:\sf

\begin{scriptsize}\begin{verbatim}
sock:cyl
sock:phases 45
sock:fill 10
sock:colmodel:lin
sock:colour 255 255 255 0.98
sock:colour   0   0   0 1.02
\end{verbatim}\end{scriptsize}

\it The resulting file {\tt output/sock.dat} is displayed in Fig.\,\ref{SIGSPEC_sock.s}\it .\sf

\subsection{Phase Distribution Diagram}\label{SIGSPEC_Phase Distribution Diagram}

In addition to the spectral window and Sock Diagram, {\sc SigSpec} can compute the probability density of phase angles at given frequency as a function of frequency. This is an alternative way to examine the properties of the sampling in the time domain and activated by the keyword {\tt phdist:phases}\label{SIGSPEC_keyword.phdist:phases} in the {\tt .ini} file. The resulting probability densities are normalised in a way that their mean over all phase angles is $\frac{1}{2\pi}$.

The number of phases to be computed is increased according to the eccentricity of the phase distribution at a given frequency.

In perfect analogy to the Sock Diagram (p.\,\pageref{SIGSPEC_Sock Diagram}), there are further keywords available to adjust the contents of the output file {\tt phdist.dat}.

\begin{itemize}
\item {\tt phdist:fill}\label{SIGSPEC_keyword.phdist:fill} determines a filling factor for additional frequencies if the changes between the phase distributions for two adjacent frequencies are too high.
\item {\tt phdist:cyl}\label{SIGSPEC_keyword.phdist:cyl} specifies cylindrical coordinates (height: frequency, azimuth: phase, radial: probability density of phase)
\item {\tt phdist:cart}\label{SIGSPEC_keyword.phdist:cart} specifies cartesian coordinates
\item {\tt phdist:colmodel:lin}\label{SIGSPEC_keyword.phdist:colmodel:lin}
\item {\tt phdist:colmodel:rank}\label{SIGSPEC_keyword.phdist:colmodel:rank}
\item {\tt phdist:colour}\label{SIGSPEC_keyword.phdist:colour}
\end{itemize}

Please refer to ``Sock Diagram'' (p.\,\pageref{SIGSPEC_Sock Diagram}) for a detailed description.

\vspace{12pt}
{\bf Caution: For frequencies close to zero, tremendous output may be expected! Try to avoid the very low frequency region, if possible.}
\vspace{12pt}

\figureDSSN{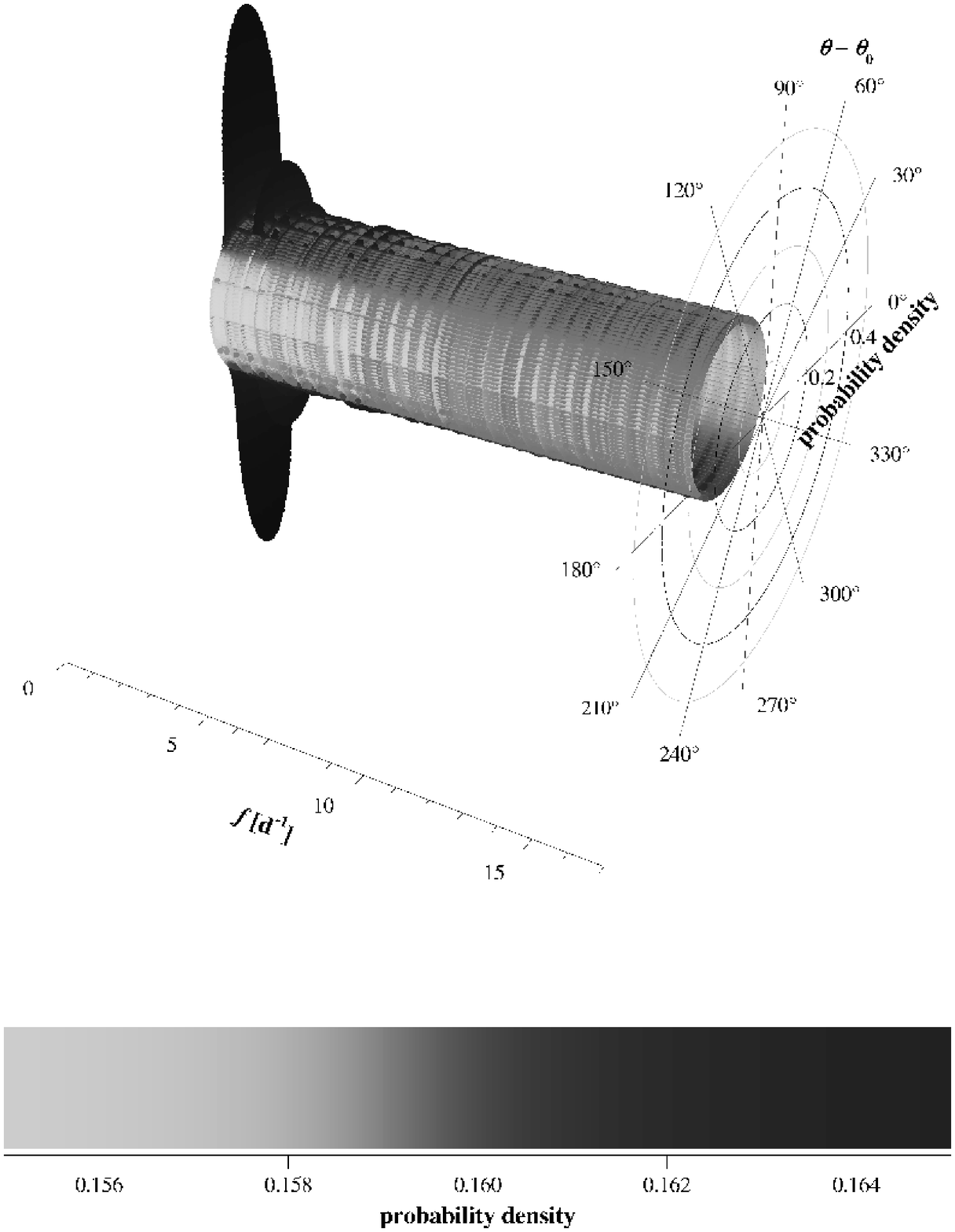}{Phase Distribution Diagram for the sample project {\tt output}.}{SIGSPEC_phdist.s}{!htb}{clip,angle=0,width=240pt}

\vspace{12pt}\noindent{\bf Example.} \it The Phase Distribution Diagram in the sample project {\tt output} is generated according to the following entries in the file {\tt output.ini}:\sf

\begin{scriptsize}\begin{verbatim}
phdist:cart
phdist:phases 30
phdist:fill 50
phdist:colmodel:rank
phdist:colour 223 223 223 0
phdist:colour  31  31  31 1
\end{verbatim}\end{scriptsize}

The resulting file {\tt output/phdist.dat} is displayed in Fig.\,\ref{SIGSPEC_phdist.s}\it .\sf

\section{MultiSine Output}\label{SIGSPEC_MultiSine Output}

After each step of prewhitening, {\sc SigSpec} performs a MultiSine least-squares fit over all significant signal components detected so far. Two optional types of output may help the user comprehend how this procedure performs at runtime.

\subsection{MultiSine tracks}\label{SIGSPEC_MultiSine tracks}

The MultiSine tracks allow to examine the changes in frequency, amplitude and phase of each signal component in the prewhitening cascade and are an alternative representation of the result files. Instead of a file index that refers to the iteration, the file index of the MultiSine track files {\tt m\#index\#.dat} refers to the index of the component in the result files and lists its
\begin{enumerate}
\item frequency [inverse time units],
\item amplitude [units of observable],
\item phase [rad]
\end{enumerate}
for each prewhitening step. In other words, a result file displays all the components for an iteration, whereas the MultiSine track file displays all the iterations for a component. Thus the MultiSine track provides a good estimate for the reliability and accuracy of the components found significant.

If the keyword {\tt mstracks}\label{SIGSPEC_keyword.mstracks} is provided in the {\tt .ini} file, MultiSine track files {\tt m\#index\#.dat} are generated, where the index {\tt \#index\#} starts with {\tt 000001}, denoting the first significant signal component.

The keyword {\tt mstracks} expects two integer parameters. The first defines the number of iterations for which entries in the MultiSine track files shall be generated. A negative number causes {\sc SigSpec} to generate entries for all iterations. The second parameter has to be a positive number and defines a step width. If it is set {\tt 1}, a line in the MultiSine track files is generated for each iteration, if it is set {\tt 2}, for every second iteration (starting with {\tt r000002.dat}), and so on.

\vspace{12pt}\noindent{\bf Example.} \it The sample project {\tt output} uses the keyword {\tt mstracks} in the file {\tt output.ini}, namely

\begin{scriptsize}\begin{verbatim}
mstracks -1 1
\end{verbatim}\end{scriptsize}

\noindent providing MultiSine tracks {\tt m000001.dat}, {\tt m000002.dat},..., for all iterations of the entire prewhitening sequence. The MultiSine track for the primary signal component (file {\tt m000001.dat}) is displayed in Fig.\,\ref{SIGSPEC_mstracks.s}\it .\sf

\figureDSSN{f15.eps}{MultiSine track of the most dominant signal component in the light curve of the sample project {\tt output} (8.59 cycles per day), according to the output file {\tt output/m000001.dat}. {\em Left}: amplitude vs.~frequency. {\em Mid:} phase vs.~frequency. {\em Right:} amplitude vs.~phase.}{SIGSPEC_mstracks.s}{!htb}{clip,angle=0,width=110mm}

\subsection{MultiSine profiles}\label{SIGSPEC_MultiSine profiles}

A closer examination of the accuracy of the MultiSine fitting procedure is provided by the MultiSine profiles. If the user specifies the keyword {\tt msprofs} in the {\tt .ini} file, {\sc SigSpec} produces additional output files {\tt f\#iteration\#.dat}, {\tt a\#iteration\#.dat}, and {\tt p\#iteration\#.dat}. The idea is to evaluate the rms residual through modifying a single parameter of a single signal component, keeping all other parameters constant. Performing this operation for the frequency of each component produces a set of rms-residual-vs.-frequency plots, all written to the file {\tt f\#iteration\#.dat}. Correspondingly, the files {\tt a\#iteration\#.dat} and {\tt p\#iteration\#.dat} contain rms-residual-vs.-amplitu\-de and rms-residual-vs.-phase plots. The frequencies are scanned around the best fit by $\pm\frac{1}{T\sqrt{\mathrm{sig}}}$, $T$ denoting the time interval width of the input time series, and sig referring to the signal component under consideration. The amplitudes are calculated from zero to twice the amplitude of best fit, and the phases in a range of $\pm\pi$ around the phase of best fit.

The keyword {\tt msprofs} is followed by three integer values, the first denoting the number of data points an individual MultiSine profile shall consist of.\footnote{Due to the internal accuracy of the index computation, the actual number of points may differ from this value by $\pm 1$.} The second parameter defines the number of iterations for which MultiSine profiles shall be generated. A negative number causes {\sc SigSpec} to generate profiles for all iterations. The third parameter has to be a positive number and defines a step width. If it is set {\tt 1}, a MultiSine profile is generated after each iteration, if it is set {\tt 2}, after every second iteration (starting with {\tt f000002.dat}, {\tt a000002.dat}, {\tt p000002.dat}), and so on.

The output files consist of seven columns:
\begin{enumerate}
\item frequency, amplitude, or phase, respectively,
\item rms residual,
\item first-order approximation, based on the tangential gradient (which should be zero, so that the deviation from zero is a measure of the accuracy of the MultiSine fitting procedure),
\item second-order approximation, based on the first and second derivatives at the parameter value of best fit,
\item point-to-point scatter,
\item index of the signal component,
\item index of the harmonic ($0$ for fundamental), see ``Analysis of Harmonics'', p.\,\pageref{SIGSPEC_Analysis of Harmonics}.
\end{enumerate}

For each signal component, the first row refers to the parameter value of best fit, as used in the result file.

\vspace{12pt}\noindent{\bf Example.} \it The sample project {\tt output} uses the keyword {\tt msprofs} in the file {\tt output.ini}, namely

\begin{scriptsize}\begin{verbatim}
msprofs 10000 50 3
\end{verbatim}\end{scriptsize}
\noindent providing MultiSine profiles ({\tt f000003.dat}, {\tt a000003.dat}, {\tt p000003.dat}), \linebreak[4] ({\tt f000006.dat}, {\tt a000006.dat}, {\tt p000006.dat}),..., for a maximum of 50 iterations of prewhitening sequence. Each MultiSine profile is specified to contain 10\,000 data points. The number of significant components found in the time series {\tt output.dat} is 33, so that the last set of MultiSine profiles ({\tt f000033.dat}, {\tt a000033.dat}, {\tt p000033.dat}) refers to the final solution contained in the file {\tt result.dat}. The MultiSine ptofiles for the primary signal component at 8.59 cycles per day are displayed in Fig.\,\ref{SIGSPEC_msprofs}\it .\sf

\figureDSSN{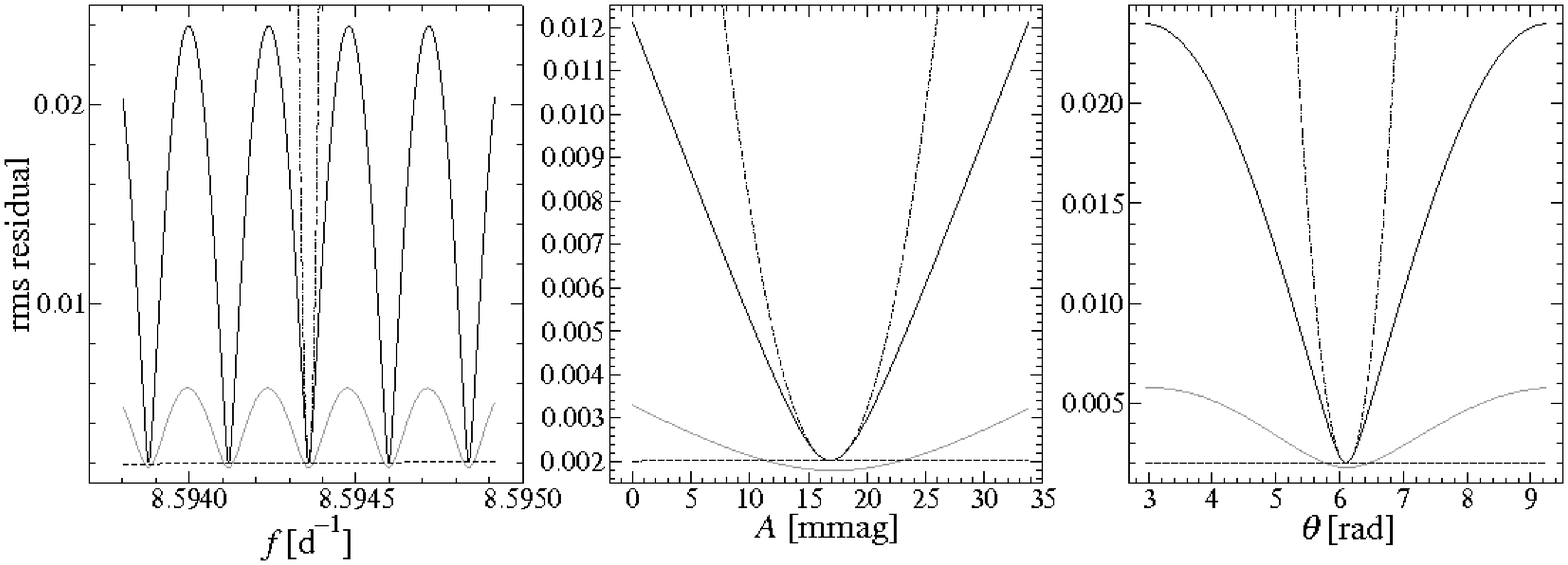}{MultiSine profiles of the most dominant signal component in the light curve of the sample project {\tt output} (8.59 cycles per day), according to the output files {\tt output/f000033.dat}, {\tt output/a000033.dat}, {\tt output/p000033.dat}. {\em Left}: rms residual vs.~frequency. {\em Mid:} rms residual vs.~amplitude. {\em Right:} rms residual vs.~phase. The {\em solid black} line refers to the rms residual, the {\em dashed black} line to the tangential gradient at the value of best fit (which should be zero), and the {\em dashed-dotted black} line to a second-order approximation based on the first two derivatives of rms residual. The {\em solid grey} line represents the point-to-point scatter.}{SIGSPEC_msprofs}{!htb}{clip,angle=0,width=110mm}

\section{Preview}\label{SIGSPEC_Preview}

Since the prewhitening cascade performed by {\sc SigSpec} may be extremely time consuming, the program can compute a preview. This add-on is activated by the keyword {\tt preview}\label{SIGSPEC_keyword.preview} in the {\tt .ini} file.

Whereas the significance spectra rely on the False-Alarm Probability compared to a noise dataset with the same rms error as the given time series (or series of residuals, respectively), the significance spectrum provided in the file {\tt preview.dat} represents a set of identified maxima in the significance spectrum of the original time series, but based on the point-to-point scatter in the time domain rather than on the standard deviation of observables. The lower sig limit for writing a local maximum to the file {\tt preview.dat} is specified as the argument to the keyword {\tt preview} in the {\tt .ini} file. 

The calculation of the sig is based on the assumption that only the point-to-point scatter is random, and everything else contributing to the rms error represents signal that will be prewhitened in the course of the subsequent loop. The preview output is to be considered as a rough estimate for the final result obtained by step-by-step prewhitening and contains not only the intrinsic variations but also all aliases, which will not occur in the following analysis. The file {\tt preview.dat} consists of four columns referring to
\begin{enumerate}
\item frequency [inverse time units],
\item sig,
\item DFT amplitude [units of observable],
\item phase [rad].
\end{enumerate}

\vspace{12pt}\noindent{\bf Example.} \it The sample project {\tt preview} contains a preview file for the V photometry of IC\,4996\,\#\,89. In the file {\tt preview.ini}, the line

\begin{scriptsize}\begin{verbatim}
preview:siglimit 5
\end{verbatim}\end{scriptsize}

\noindent sets the sig threshold to 5. The output file {\tt preview/preview.dat} contains 11 components, sorted by frequency. The frequencies and corresponding sigs in the first two columns are

\begin{scriptsize}\begin{verbatim}
 0.9945158494303480	  6.1674140356166323
 1.9917563998155081	  6.9302735632389876
 2.1388902515119841	  6.7175642729893710
 2.9835475482878717	  8.6773027802854656
 3.1361308018982843	  9.4899187898938777
 3.9862375005883859	  8.9589776551282210
 4.1333713522847031	  8.7492615592402885
 4.9780286490607102	  5.0523760377159039
 5.1360613045861099	  5.3438911274207790
11.0268647743572874	  5.5214237212500406
12.0241053247411784	  5.6674302270769710
\end{verbatim}\end{scriptsize}

\noindent Fig.\,\ref{SIGSPEC_preview.s} \it displays the significance and amplitude spectrum of the original time series. Since the preview does not employ any prewhitenings, aliases are present in the file.
\begin{itemize}
\item The signal at 3.132 cycles per day corresponds to components \# 3, 5, 7, and 9.
\item The signal at 3.986 cycles per day corresponds to components \# 1, 2, 4, 6, and 8.
\item The signal at 5.409 cycles per day is not found in the preview. In the result of the prewhitening sequence, its sig is 5.02. Since the sig in the preview relies on the rms deviation of the original time series, whereas the final sig is based on the rms deviation after the previous prewhitening step, the sig associated to this frequency falls below the pre-selected threshold of 5 in the preview. The significance spectrum (grey line in the left panel of Fig.\,\ref{SIGSPEC_preview.s}\it ) shows a peak at the frequency under consideration the sig of which is $\approx$ 4.8.
\item Components \# 10 and 11 are 1-cycle-per-day aliases of each other, but do not show up in the final result, {\tt preview/result.dat}.
\end{itemize}\sf

\figureDSSN{f17.eps}{{\em Grey:} Fourier spectra for the sample project {\tt preview}. {\em Left:} significance spectrum. {\em Right:} DFT amplitudes. The significant components in the preview are indicated by {\em dots} with dashed drop lines (file {\tt preview/preview.dat}). The default sig threshold of 5 is represented by a horizontal {\em dashed line} in the left panel.}{SIGSPEC_preview.s}{!htb}{clip,angle=0,width=109mm}

\section{Correlograms}\label{SIGSPEC_Correlograms}

{\sc SigSpec} is able to compute correlograms of the time series for each stage of prewhitening. The correlogram files are named {\tt c\#iteration\#.dat}. The calculation of correlograms is activated by the keyword {\tt correlograms}\label{SIGSPEC_keyword.correlograms}, which requires three integer parameters. The first parameter represents the maximum order to which to compute serial correlations, i.\,e.~the limit of index lag for each correlogram. Setting it zero forces {\sc SigSpec} to adjust it to half the number of data points in the time series. The second parameter is the maximum number of iterations for which to compute correlograms. If the number of prewhitening iterations exceeds this value, then no correlogram is generated for the iterations after this limit. If a number $\le 0$ is given, then a correlogram is computed for each prewhitening stage. The third parameter has to be a positive number and defines a step width. If it is set {\tt 1}, a file is generated after each iteration, if it is set {\tt 2}, after every second iteration (starting with {\tt c000002.dat}), and so on. The correlogram computation is switched off by default.

A file {\tt rescorr.dat} is generated, if the keyword {\tt correlograms} is specified, no matter which parameter constellation is chosen.

A correlogram file consists of two columns referring to
\begin{enumerate}
\item index lag,
\item serial correlation coefficient.
\end{enumerate}

\vspace{12pt}\noindent{\bf Example.} \it The sample project {\tt correlograms} illustrates how correlograms are generated with {\sc SigSpec} using the V photometry of IC\,4996\,\#\,89 as time series input file {\tt correlograms.dat}. The file {\tt correlograms.ini} contains the line

\begin{scriptsize}\begin{verbatim}
correlograms 100 -1 1
\end{verbatim}\end{scriptsize}

\noindent which forces {\sc SigSpec} to evaluate correlograms with a maximum index lag of 100 (first parameter) for all iterations (negative value of second parameter). After each iteration, a correlogram is generated (third parameter). The output files

\begin{scriptsize}\begin{verbatim}
correlograms/c000000.dat
correlograms/c000001.dat
correlograms/c000002.dat
correlograms/rescorr.dat
\end{verbatim}\end{scriptsize}

\noindent are generated as displayed in Fig.\,\ref{SIGSPEC_correlograms.s}\it .\sf

\figureDSSN{f18.eps}{Correlograms for the sample project {\tt correlograms}. {\em Solid:} correlogram of the initial time series (file {\tt correlograms/c000000.dat}). {\em Dashed:} correlogram after one prewhitening (file {\tt correlograms/c000001.dat}). {\em Dashed-dotted:} correlogram after two prewhitenings (file {\tt correlograms/c000002.dat}). {\em Dotted:} residual correlogram after three prewhitenings (file {\tt correlograms/rescorr.dat}).}{SIGSPEC_correlograms.s}{!htb}{clip,angle=0,width=109mm}

\section{Time-resolved Analysis}\label{SIGSPEC_Time-resolved Analysis}

\begin{table}
\begin{tabular}{lll}
{\bf keyword}         & {\bf arguments}  & {\bf weight function}\\
\hline
{\tt timeres:w:none}  &                  & $1$ \\
{\tt timeres:w:ipow}  & $\xi$            & $0$ if $t=t_C$, $\left|\, t-t_C\right| ^{-\xi}$ else \\
{\tt timeres:w:gauss} & $\sigma$         & $\mathrm{e}^{-\left(\frac{ t-t_C}{\sigma}\right) ^2}$ \\
{\tt timeres:w:exp}   & $\zeta$          & $\mathrm{e}^{-\frac{\left| t-t_C\right|}{\zeta}}$ \\
{\tt timeres:w:damp}  & $\zeta$          & $\mathrm{e}^{-\frac{t-t_B}{\zeta}}$ \\
{\tt timeres:w:cos}   & $\nu ,\Phi$      & $\cos\left( 2\pi\nu\left|\, t-t_C\right| -\Phi\right)$\\
{\tt timeres:w:cosp}  & $\nu ,\Phi ,\xi$ & $\cos\, ^{\xi}\left( 2\pi\nu\left|\, t-t_C\right| -\Phi\right)$\\
\hline
\vspace{12pt}
\end{tabular}
\caption{Weight functions for time-resolved {\sc SigSpec} analysis. The beginning of the time interval associated with the referring subset is denoted $t_B$, whereas $t_C$ symbolises the centre of the time interval.}\label{SIGSPEC_TABwts}
\end{table}

In time-resolved mode, {\sc SigSpec} performs an analysis for a set of time intervals rather than for the entire time series. An interval of width given by the keyword {\tt timeres:range}\label{SIGSPEC_keyword.timeres:range} is moved in steps the width of which is given by the keyword {\tt timeres:step}\label{SIGSPEC_keyword.timeres:step} from the beginning of the time series to the end.\footnote{In general, the step width is slightly modified by the software to achieve time-resolved analysis over the entire time series.} Consecutive time intervals are free to overlap. Time series data within such an interval are used to form a subset for which the analysis is performed. In addition, statistical weights may be applied to the subset data, all with respect to the centre of the interval, which shall be denoted $t_C$.

The only exception is the keyword {\tt timeres:w:damp}. In this case, the analysis is optimised for signal excited at the beginning of the time interval corresponding to the subset under consideration, $t_B$ and exponentially damped towards the end of the subset.

The weight functions of time are given in Table\,\ref{SIGSPEC_TABwts}. The normalisation of weights is automatically performed by {\sc SigSpec}. Also the combination of a weight function for time-resolved mode with weights columns (keyword {\tt col:weights}) is supported.

In time-resolved mode, the set of output files as given in ``Default Output'', p.\,\pageref{SIGSPEC_Default Output}, is generated for each subset of the time series. This requires the introduction of an additional six-digit index, {\tt \#interval\#}, in addition to {\tt \#iteration\#}, and the annotation for the output files is
\begin{enumerate}
\item {\tt wts.\#interval\#.dat} for the weight function vs.~time in each subset,
\item {\tt s\#iteration\#.\#interval\#.dat} for the spectra,
\item {\tt t\#iteration\#.\#interval\#.dat} for the residuals after each step of prewhitening,
\item {\tt r\#iteration\#.\#interval\#.dat} for the results after each step of pre\-whitening,
\item {\tt m\#index\#.\#interval\#.dat} for the results after each step of prewhitening,
\item {\tt result.\#interval\#.dat} for the result files, each with a list of significant signal components,
\item {\tt residuals.\#interval\#.dat} for the final residuals after the prewhitening of all significant signal components,
\item {\tt resspec.\#interval\#.dat} for the residual spectrum after the prewhitening of all significant signal components,
\end{enumerate}
The column syntax is strictly consistent with the time-unresolved versions (see ``Default Output'', p.\,\pageref{SIGSPEC_Default Output}). The additional files, {\tt wts.\#interval\#.dat}, are in two-column format. The first column represents the time values in the corresponding subset, the second column contains the weight function without normalisation.

Furthermore, {\sc SigSpec} generates a file {\tt t000000.\#interval\#.dat}, which contains the part of the original time series which is actually used as input.

Special functions -- as introduced in ``Analysis of the Time-domain Sampling'' (p.\,\pageref{SIGSPEC_Analysis of the Time-domain Sampling}), ``Preview'' (p.\,\pageref{SIGSPEC_Preview}), and ``Correlograms'' (p.\,\pageref{SIGSPEC_Correlograms}) -- are also supplied with the {\tt \#interval\#} index, i.\,e.
\begin{enumerate}
\item {\tt win.\#interval\#.dat} for the amplitude windows,
\item {\tt profile.\#interval\#.dat} for the sampling profiles,
\item {\tt sock.\#interval\#.dat} for the Sock Diagrams,
\item {\tt phdist.\#interval\#.dat} for the phase distribution diagrams,
\item {\tt preview.\#interval\#.dat} for the previews,
\item {\tt c\#iteration\#.\#interval\#.dat} for the correlograms after each step of prewhitening,
\item {\tt rescorr.\#interval\#.dat} for the final correlograms after the prewhitening of all significant signal components.
\end{enumerate}

\figureDSSN{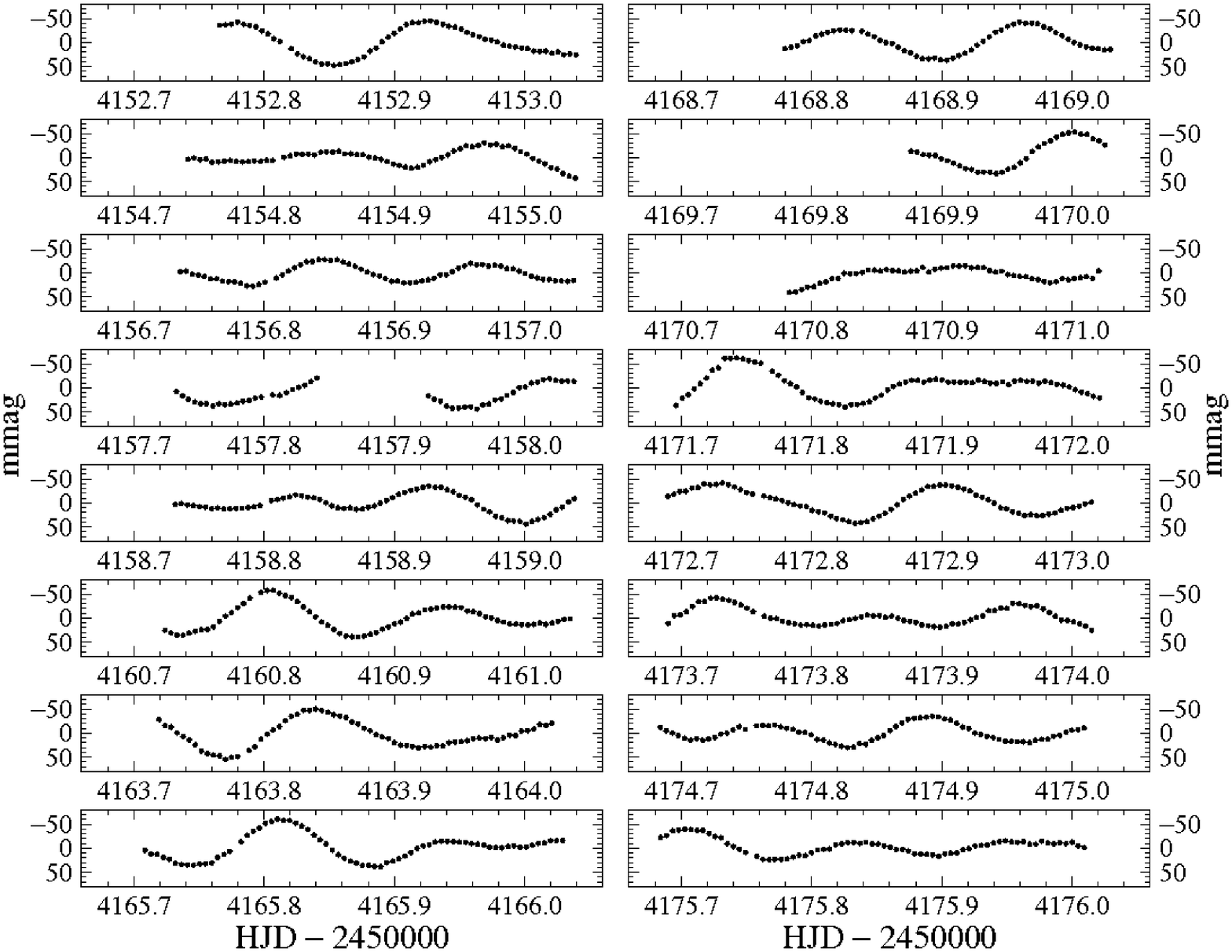}{Time series used for the sample project {\tt timeres}, representing 14 nights of Str\o mgren $y$ photometry of the Delta Scuti star 4\,CVn, acquired in February and March, 2007.}{SIGSPEC_timeres.dat}{!htb}{clip,angle=0,width=109mm}

\vspace{12pt}\noindent{\bf Example.}\label{SIGSPEC_timeres.ex} \it The sample project {\tt timeres} illustrates the time-resolved analysis using Str\o mgren y photometry of the Delta Scuti star 4\,CVn acquired with the Vienna University Automatic Photoelectric Telescope (Strassmeier et al.\,1997). The data represent 16 nights from February 21 to March 16, 2007, and are displayed in Fig.\,\ref{SIGSPEC_timeres.dat}\it .

\figureDSSN{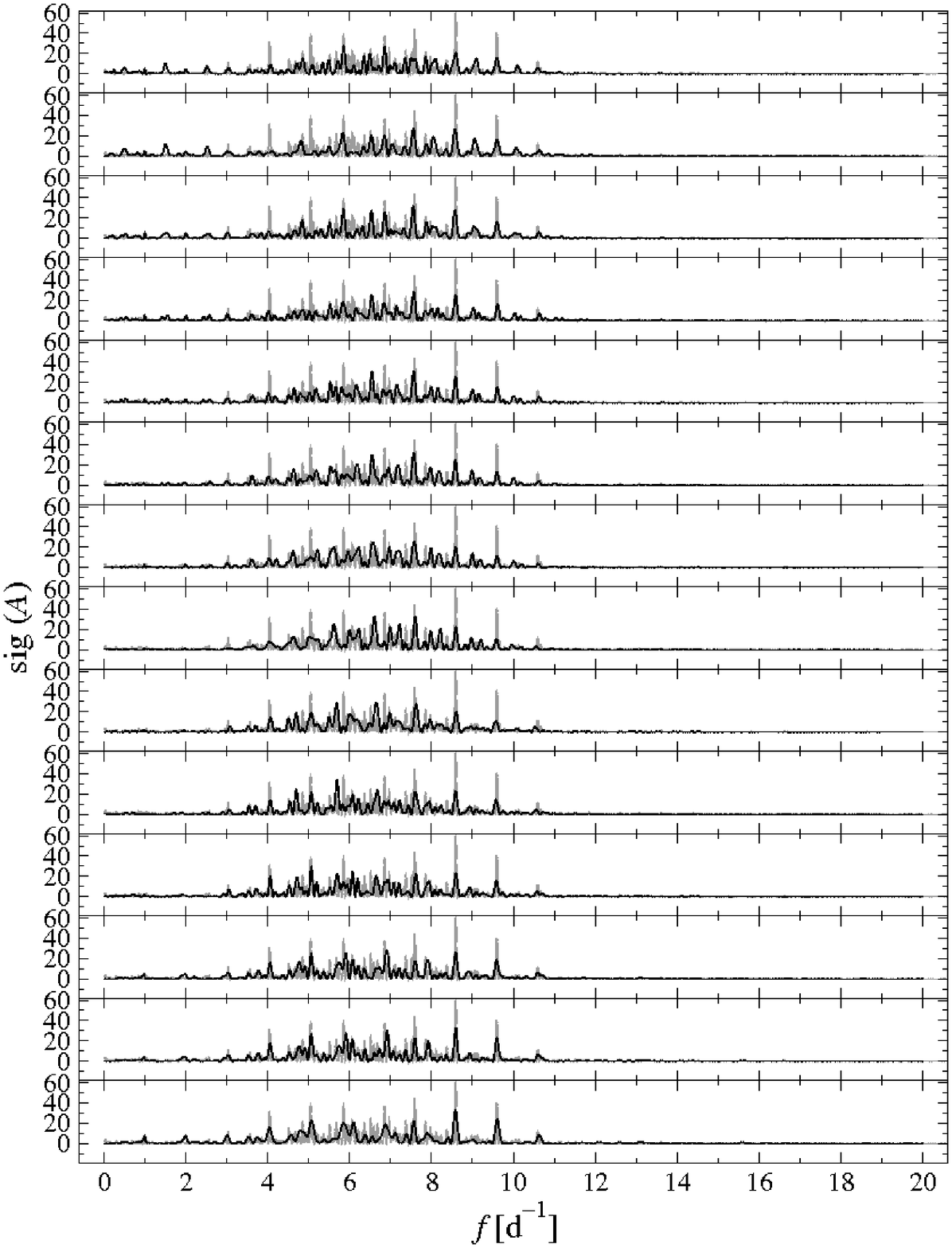}{Time-resolved significance spectra for 14 subsets (from {\em top} to {\em bottom}) automatically generated in the sample project {\tt timeres}. in each panel, the significance spectrum of the full dataset is displayed in {\em grey} colour for comparison.}{SIGSPEC_timeres.s}{!htb}{clip,angle=0,width=96mm}

\pagebreak The file {\tt timeres.ini} contains the specifications 

\begin{scriptsize}\begin{verbatim}
timeres:range 10
timeres:step 1
\end{verbatim}\end{scriptsize}
\noindent which provide a 10-day interval moving over the time base of 24 days, with one-day steps. The resulting 14 subsets are represented by the files \linebreak[4] {\tt timeres/t000000.000000.dat} to {\tt timeres/t000000.000013.dat}. Gaussian weight functions with a standard deviation of 5 days are applied:

\begin{scriptsize}\begin{verbatim}
timeres:w:gauss 5
\end{verbatim}\end{scriptsize}

\noindent The files {\tt timeres/wts.000000.dat} to {\tt timeres/wts.000013.dat} contain the weights applied to each datapoint within each subset. Further output files are

\begin{itemize}
\item {\tt timeres/s000000.\#\#\#\#\#\#.dat} for the significance spectra of the original time series without prewhitening (Fig.\,\ref{SIGSPEC_timeres.s}\it ,
\item {\tt timeres/result.\#\#\#\#\#\#.dat} for the lists of significant signal components,
\item {\tt timeres/residuals.\#\#\#\#\#\#.dat} for the residual time series after all prewhitening steps (divided into subsets according to the time intervals), and
\item {\tt timeres/resspec.\#\#\#\#\#\#.dat} for the significance spectra of residuals.
\end{itemize}
Here {\tt \#\#\#\#\#\#} denotes six-digit numbers ranging from {\tt 000000} to {\tt 000013}.\sf

\section{{\sc SigSpec} AntiAlC: Anti-aliasing Correction Mode}\label{SIGSPEC_AntiAlC}

In AntiAlC mode, {\sc SigSpec} does not follow a strict step-by-step prewhitening sequence. Instead, test runs are performed for a number of candidate peaks in the significance spectrum in order to find the solution that produces a minimum residual rms scatter after a user-given number of prewhitenings.

\begin{enumerate}
\item All peaks above a given sig limit are taken into consideration. The keyword {\tt antialc:par}\label{SIGSPEC_keyword.antialc:par} in the {\tt .ini} file is followed by a floating-point number. This quantity is the AntiAlC parameter $p_\mathrm{al}$, which has to attain a value in the interval $\left] 0,1\right]$. If the highest sig in the considered frequency range is $\max\left[\mathrm{sig}\left( A\right)\right]$, then the sig limit is $p_\mathrm{al}\max\left[\mathrm{sig}\left( A\right)\right]$. I.\,e., the AntiAlC parameter determines the sig limit for the candidate peak selection relative to the highest peak in the spectrum under consideration. Alternatively or in addition, a sig threshold for the AntiAlC candidate selection may be defined using the keyword {\tt antialc:siglimit}\label{SIGSPEC_keyword.antialc:siglimit}. If neither {\tt antialc:par} nor {\tt antialc:siglimit} are present, the sig limit specified by {\tt siglimit} in the {\tt .ini} file (p.\,\pageref{SIGSPEC_Program termination}) is used for the AntiAlC candidate selection also.
\item The candidate selection is performed for each step in the test pre\-whitening sequence.
\item The resulting procedure is the computation of all combinations of candidate peaks above a sig threshold determined by the AntiAlC parameter. The number of iterations for these test prewhitenings is determined by the keyword {\tt antialc:depth}\label{SIGSPEC_keyword.antialc:depth}, followed by an integer value. It specifies the depth of the AntiAlC computation.
\item The successful combination of peaks is selected upon the minimum residual rms deviation out of all examined combinations.
\item {\sc SigSpec} does not necessarily adopt all iterations performed in the test run for the main prewhitening cascade. The integer value following the keyword {\tt antialc:adopt}\label{SIGSPEC_keyword.antialc:adopt} determines how many prewhitening steps shall be adopted. This quantity must not exceed the computation depth provided by the keyword {\tt antialc:depth}. If the limits specified by the keywords {\tt iterations}, {\tt siglimit}, or {\tt csiglimit} are reached, the output may even terminate before the number specified by the keyword {\tt antialc:adopt}.
\end{enumerate}

According to Reegen~(2007), the expected sig is approximately proportional to the squared amplitude, if all influences by the time-domain sampling are neglected. The combination of $n$ sinusoidal signal components interacting via aliasing is expected to produce a maximum amplitude that does not exceed the sum of amplitudes of the sinusoidal components. Consequently, the square root of the sig of such a combination, $\mathrm{sig}_\mathrm{al}$, is very likely below the sum of square roots of individual sigs $\mathrm{sig}_n$,
\begin{equation}
\sqrt{\mathrm{sig}_\mathrm{al}} < \sum _n\sqrt{\mathrm{sig}_n}\: .
\end{equation}
If these all are assumed equal and denoted $\mathrm{sig}_\mathrm{ind}$, then the upper sig limit for the alias is $\mathrm{sig}_\mathrm{ind}\sqrt{n}$. In other words, if a given peak with a sig $\mathrm{sig}_\mathrm{al}$ is an alias of a combination of $n$ signal components with unique sigs $\mathrm{sig}_\mathrm{ind}$, then the individual significances are probably higher than $\frac{\mathrm{sig}_\mathrm{al}}{\sqrt{n}}$. In terms of the AntiAlC parameter, one obtains
\begin{equation}
n \approx \frac{1}{\sqrt{p_\mathrm{al}}}
\end{equation}
for the approximate number of signal components that can be assigned alias-free for a given AntiAlC parameter $p_\mathrm{al}$. Based on these considerations, {\sc SigSpec} evaluates the AntiAlC computation depth using the AntiAlC parameter, if the keyword {\tt antialc:depth} is not provided in the {\tt .ini} file.

The AntiAlC mode produces additional screen output, if a combination of candidate peaks yields a lower residual scatter than the previous minimum, a two-line screen message is returned. The first line is a set of indices. In the example below, the AntiAlC parameter (keyword {\tt antialc:par}) is set $0.5$, and the AntiAlC computation depth (keyword {\tt antialc:depth}) is $3$. Correspondingly, the first line of output applies to the first of altogether ten candidate peaks in the first iteration, the first out of three in the second iteration, and the first out of seven in the third iteration. This peak constellation produces an rms deviation of residuals as displayed in the second line of output (in the example $0.00\,405\,851$). After finishing the test cascade, the number of iterations specified by the keyword {\tt antialc:adopt} (in the present example, this number is $2$) is adopted for the main cascade. The screen output produced by the main cascade is the same as for a normal {\sc SigSpec} prewhitening cascade without AntiAlC. The files containing spectra and residuals, respectively, are updated each time the residual rms deviation improves.

\figureDSSN{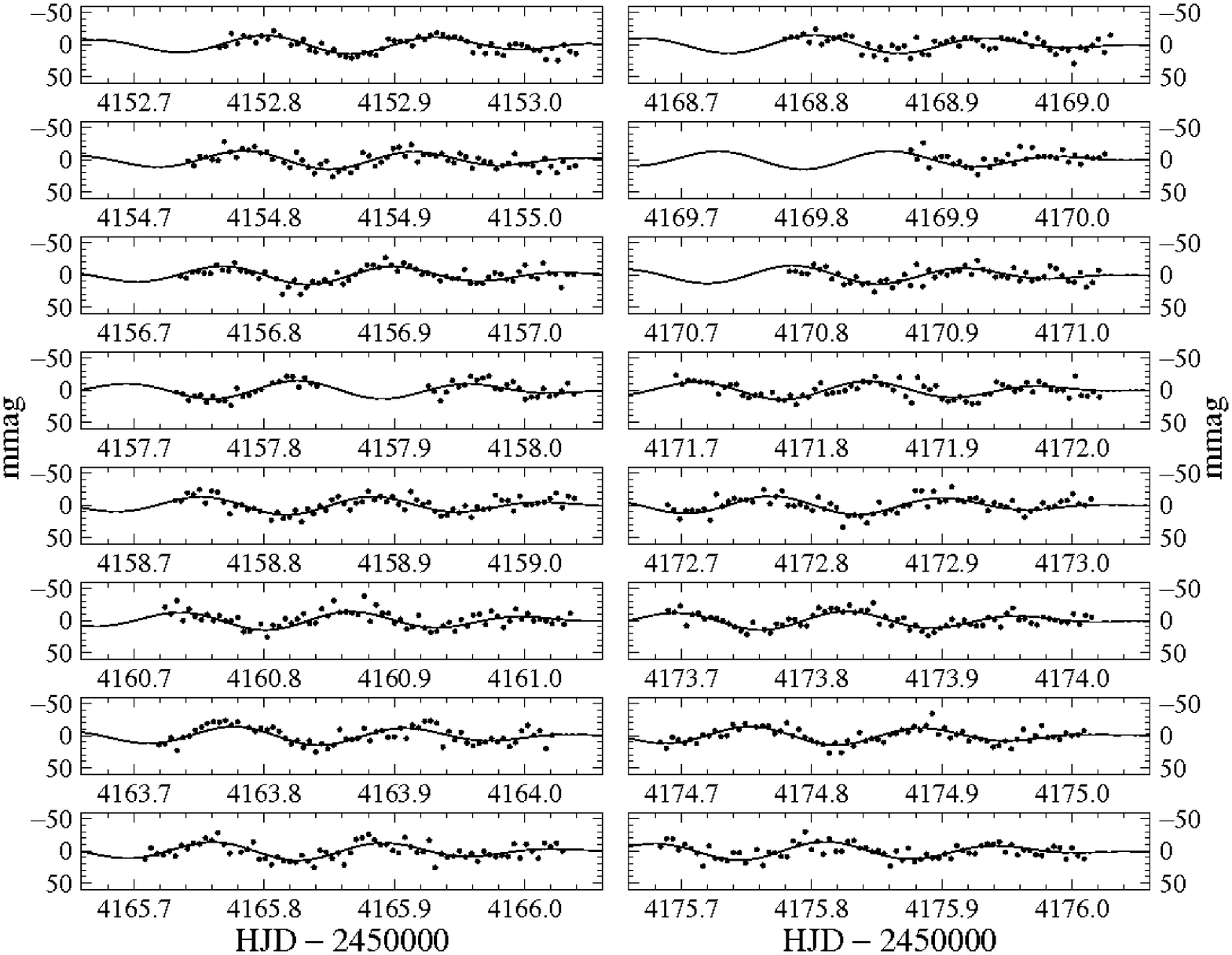}{Time series used for the sample project {\tt antialc} ({\em dots}). The sampling represents 14 nights of Str\o mgren $y$ photometry of the $\delta$\,Sct star 4\,CVn, acquired in February and March, 2007. The magnitude values are synthesized forming two sinusoidal signals ({\em solid line}) plus Gaussian noise. }{SIGSPEC_antialc.dat}{!htb}{clip,angle=0,width=109mm}

\vspace{12pt}\noindent{\bf Example.}\footnote{The computation of the sample project {\tt antialc} takes 7 minutes on an Intel Core2 CPU T5500 (1.66GHz) under Linux 2.6.18.8-0.9-default i686.} \it The sample project {\tt antialc} illustrates the anti-aliasing correction using the same sampling as the data for the sample project {\tt timeres} (p.\,\pageref{SIGSPEC_timeres.ex}), 
\begin{enumerate}
\item a sinusoid with frequency 6.5598 cycles per day, amplitude 7.29 mmag,
\item a sinusoid with frequency 8.5637 cycles per day, amplitude 6.87 mmag,
\item Gaussian noise with 7.36 mmag rms deviation,
\end{enumerate}
as displayed in Fig.\,\ref{SIGSPEC_antialc.dat} \it . The two signal frequencies differ by almost exactly 2 cycles per day and may easily be misidentified as aliases of each other. There are two identical versions of the light curve provided for comparison: {\tt alc.dat} and {\tt antialc.dat}.

\figureDSSN{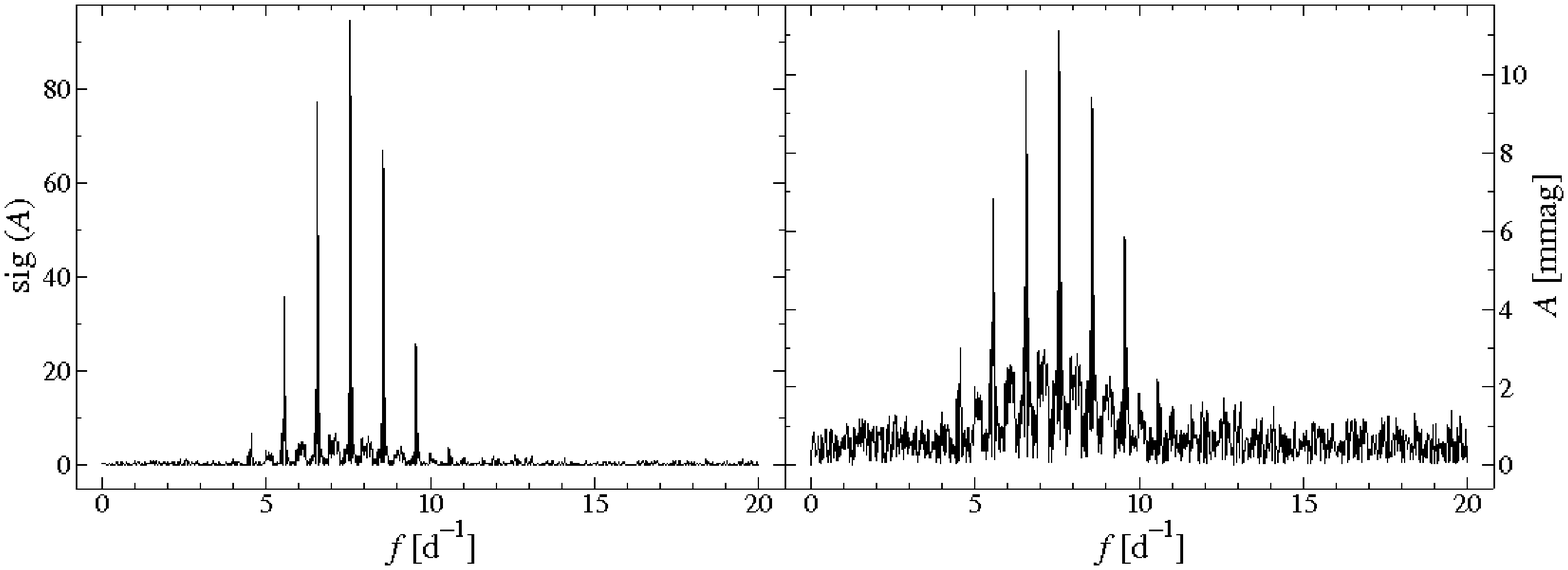}{Fourier spectra for the sample project {\tt antialc}. {\em Left:} significance spectrum. {\em Right:} DFT amplitudes.}{SIGSPEC_antialc.s}{!htb}{clip,angle=0,width=110mm}

The file {\tt alc.dat} corresponds to the project directory {\tt alc}, representing a normal {\sc SigSpec} run without a file {\tt alc.ini}. Running {\tt SigSpec alc}, the resulting frequencies (screen output) are

\begin{scriptsize}\begin{verbatim}
   1 freq 7.55917  sig 55.8792  rms 10.0617  csig 55.8792
   2 freq 5.55706  sig 31.5539  rms 8.65888  csig 31.5539
   3 freq 10.5668  sig 11.011  rms 7.81469  csig 11.011
   4 freq 2.55231  sig 4.9934  rms 7.60001  csig 4.9934
\end{verbatim}\end{scriptsize}

\noindent Instead of the two signal components, 1-cycle-per-day aliases are identified. The significance and Fourier amplitude spectra of the dataset show the highest peak at 7.56 cycles per day, which represents a superposition of the first upper side peak of the signal at 6.56 cycles per day and the first lower side peak of the signal at 8.56 cycles per day (Fig.\,\ref{SIGSPEC_antialc.s}\it . This leads to an imperfect prewhitening of the two components, and the remaining signal is detected as a third component at 9.56 cycles per day.

The alternative AntiAlC analysis is provided by the file {\tt antialc.ini}, which contains the specifications

\begin{scriptsize}\begin{verbatim}
antialc:par 0.5
antialc:depth 2
antialc:adopt 1
antialc:siglimit 4
\end{verbatim}\end{scriptsize}

\noindent All peaks that reach at least 50\,\% of the highest significance in the spectrum are taken into account. {\sc SigSpec} computes two consecutive iterations, but adopts only the first of these two iterations. A sig limit of 4 is assumed for the AntiAlC calculations (contrary to the default sig limit of 5 still valid as a breakup condition for the whole procedure). Running {\tt SigSpec antialc}, the screen output is

\begin{scriptsize}\begin{verbatim}
   1 freq 6.55844  sig 55.0218  rms 10.0617  csig 55.0218                
   2 freq 8.56169  sig 43.6737  rms 8.68212  csig 43.6737                
   3 freq 33.7207  sig 3.97249  rms 7.48075  csig 3.97249                
\end{verbatim}\end{scriptsize}

\noindent Both signals are recovered at a reasonable frequency accuracy. Moreover, according to the file {\tt antialc/result.dat}, the amplitudes of the two signals are recovered to a satisfactory precision (7.22 mmag, 6.47 mmag).\sf

\section{Analysis of Harmonics}\label{SIGSPEC_Analysis of Harmonics}

If a non-sinusoidal, but periodic process is measured, DFT does not only produce the fundamental frequency, which is the repetition rate of the non-sinusoid. The shape of the periodicity is recovered by a number of harmonics (also called overtones) the frequencies of which are integer multiples of the fundamental. In this case it may be considered insufficient to determine the exact frequency of the process by employing only the peak at the fundamental frequency and ignoring the harmonics. The keyword {\tt harmonics}\label{SIGSPEC_keyword.harmonics}, followed by an integer determining the upper limit of the harmonic order, allows to compute the sig of the fundamental plus the desired number of overtones. The specification {\tt harmonics 20} forces {\sc SigSpec} to take into account altogether 21 frequencies.

As pointed out by Reegen~(2007), {\sc SigSpec} treats False-Alarm Probabilities in a statistically clean and unbiased way. In analogy to the comb analysis introduced by Kjeldsen et al.~(1995), but benefitting from the exact statistical treatment of noise, it is possible to extend the method in order to evaluate the probability of a whole set of peaks to be generated by noise simultaneously. This strategy helps to take into account a fundamental frequency plus a set of integer multiples at once and permits to evaluate the most likely solution for a non-sinusoidal signal. In addition, the Fourier Space parameters obtained for the signal components provide a fit to the data in terms of a fundamental frequency plus overtones.

Given a set of amplitude levels $A_h$, $h=0,1,...,H$, at different frequencies with associated False-Alarm Probabilities $\Phi _{\mathrm{FA}}\left( A_h\right)$, the probability that all amplitude levels are due to noise is given by the product of the individual False-Alarm Probabilities,
\begin{equation}\label{EQ.totalFAP}
\Phi _{\mathrm{FA}}\left(\bigwedge _{h=0}^H A_h\right) = \prod _{h=0}^H \Phi _{\mathrm{FA}}\left( A_h\right)\: ,
\end{equation}
if the noise amplitudes at the two frequencies are assumed statistically independent. This is the probability that all amplitude levels are generated by noise.

Since the sig is defined as the negative logarithm of False-Alarm Probability, the above expression leads to
\begin{equation}
\mathrm{sig}\left(\bigwedge _{h=0}^H A_h\right)=\sum _{h=0}^H\mathrm{sig}\left( A_h\right)\: .
\end{equation}
In this context, the sig represents the number of cases in one out of which all amplitude levels $A_h$ are not generated by noise. This logical concept is the representation of an AND operator, as indicated by the argument to sig in the equation.

Reegen~(2007) evaluated the expected value of the sig (ignoring the variations with frequency and phase) to be $\frac{\pi}{4}\log\mathrm{e} \approx 5.4575$. Considering $H$ different amplitude levels simultaneously rescales this expected sig, so that we obtain $\frac{H\pi}{4}\log\mathrm{e}$. This rescaling may cause inconvenience, whence we use the {\em mean sig} of an individual peak out of this sample of fundamental plus harmonics,
\begin{equation}\label{EQ.msig}
\mathrm{msig}\left( A_h\right):=\frac{1}{H+1}\mathrm{sig}\left(\bigwedge _{h=0}^H A_h\right)\: ,
\end{equation}
instead. It is the expected sig obtained for an arbitrarily picked element out of the $H$ peaks: if each of the considered peaks would have $\mathrm{msig}\left( A_h\right)$, then the total sig of the fundamental plus harmonics would be $\mathrm{sig}\left(\bigwedge _{h=0}^H A_h\right)$. The statistical properties of $\mathrm{msig}\left( A_h\right)$ are the same as for the ``normal'' sig analysis. If the keyword {\tt harmonics} is provided in the {\tt .ini} file, the sig levels returned in the second column of the file {\tt result.dat} are mean sigs.

The result files display only the fundamentals of the solution, and information on the harmonics is stored in additional output files. The names are generated from the name of the corresponding result file without the extension {\tt .dat}, plus {\tt -h\#index\#.dat}, where {\tt \#index\#} refers to the index of the item in the result file. For example, the harmonics for the third component in the file {\tt result.dat} are stored in the file {\tt result-h000003.dat}. The files contain the harmonics in ascending order, starting with the fundamental. The three columns are
\begin{enumerate}
\item sig of the individual peak,
\item DFT amplitude [units of observable],
\item Fourier-space phase angle [rad].
\end{enumerate}

\figureDSSN{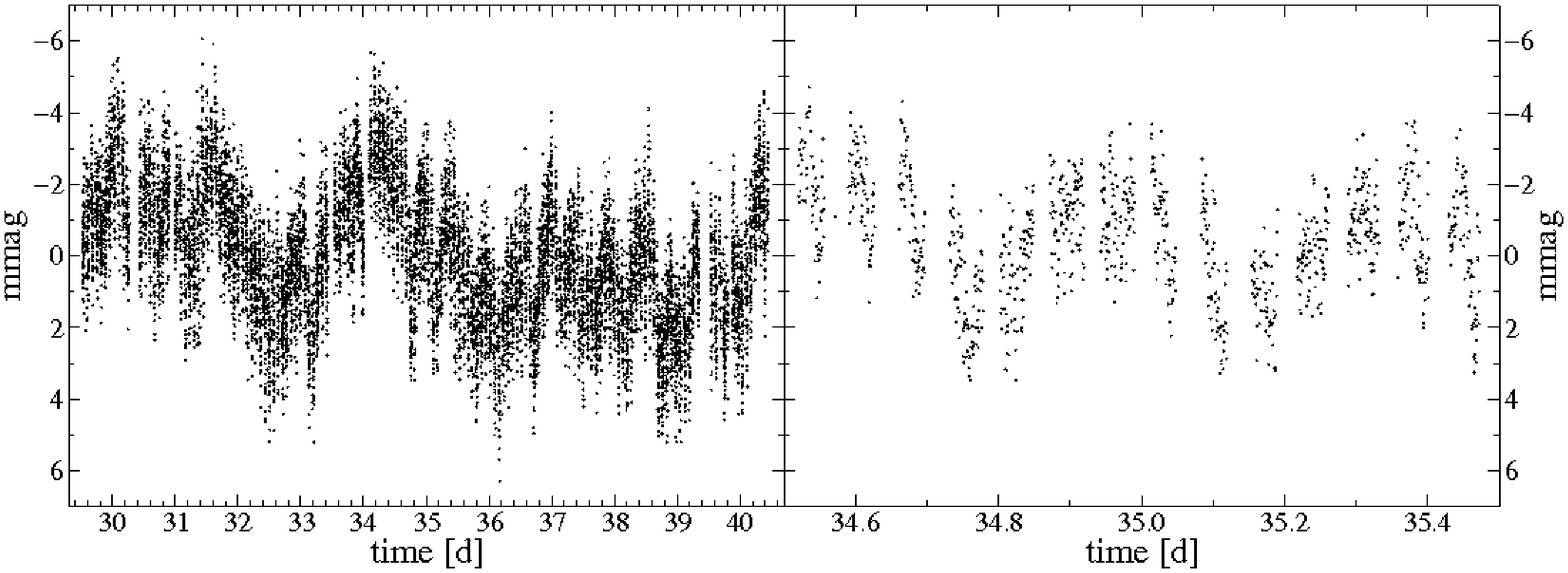}{Time series used for the sample project {\tt harmonics}.}{SIGSPEC_harmonics.dat}{!htb}{clip,angle=0,width=110mm}

\vspace{12pt}\noindent{\bf Example.}\label{SIGSPEC_EXharmonics} \it The sample project {\tt harmonics} illustrates the determination of a non-sinusoidal signal using the analysis of harmonics. The dataset represents (yet unpublished) space photometry of a star that exhibits surface activity. The task is to determine the rotation period of the star. For comparison, two identical versions of the time series are avalable (Fig.\,\ref{SIGSPEC_harmonics.dat}\it ). The file {\tt noharmonics.dat} is used together with the file {\tt noharmonics.ini} to perform a {\sc SigSpec} analysis without harmonics and associated with the project directory {\tt noharmonics} containing the output. It contains four lines:

\begin{scriptsize}\begin{verbatim}
ufreq 13
freqspacing .001
iterations 1
siglimit 0
\end{verbatim}\end{scriptsize}

\noindent In this constellation, {\sc SigSpec} computes the significance spectrum between 0 and 13 cycles per day, with steps of 0.001 cycles per day (Fig.\,\ref{SIGSPEC_harmonics.s}\it , left panel). Only one iteration (i.\,e. no prewhitening) is performed. The highest peak is found at 0.296 cycles per day, which corresponds to a period of 3.38 days.

\figureDSSN{f24.eps}{Fourier spectra for the sample project {\tt harmonics}. {\em Left:} significance spectrum without employing the analysis of harmonics ({\em solid line}). The fundamental and twelve harmonics of the alternative solution are indicated by vertical {\em dashed lines} for comparison. {\em Right:} significance spectrum displaying the mean sig for fundamental plus twelve harmonics ({\em solid line}). Note that the frequency interval differs from the left panel. For comparison, the solution without harmonics is displayed as a {\em dashed line}.}{SIGSPEC_harmonics.s}{!htb}{clip,angle=0,width=110mm}

\pagebreak The file {\tt harmonics.dat} is the same as {\tt noharmonics.dat}, but the associated file {\tt harmonics.ini} specifies a different setup by the lines

\begin{scriptsize}\begin{verbatim}
lfreq 0.125
ufreq 1
freqspacing .001
iterations 1
siglimit 0
harmonics 12
\end{verbatim}\end{scriptsize}

It is advisable not to set the lower frequency limit zero, because below the Rayleigh frequency resolution, consecutive harmonics hit the same peak and produce unreliable results. In the present case, the Rayleigh frequency resolution is 0.091 cycles per day, and to be fairly on the safe side, the lower frequency limit is adjusted to 0.125 cycles per day. Fig.\,\ref{SIGSPEC_harmonics.s}\it (right panel) contains the mean sig of the fundamental plus twelve harmonics vs.~frequency.

The amplitudes of the fundamental and twelve harmonics are displayed vs.~frequency in Fig.\,\ref{SIGSPEC_harmonics.h}\it . The maximum sig is found at 0.155 cycles per day, i.\,e., the rotation period is 6.46 days, indicating that the analysis without harmonics led to a misidentification of the first harmonic as the ``true'' rotational frequency. For comparison, the left panel of Fig.\,\ref{SIGSPEC_harmonics.s} \it contains the fundamental plus harmonics found by this procedure as vertical dashed lines.

Moreover, for the analysis of harmonics, there is additional information in the screen output provided by {\sc SigSpec}. The standard screen output for the project {\tt noharmonics} contains the lines

\figureDSSN{f25.eps}{Frequencies and amplitudes of the harmonics associated to the most significant signal found for the sample project {\tt harmonics} ({\em dots with drop lines}). The DFT amplitudes obtained by {\sc SigSpec} without employing the analysis of harmonics are displayed as a {\em solid line} for comparison.}{SIGSPEC_harmonics.h}{!htb}{clip,angle=0,width=110mm}

\begin{scriptsize}\begin{verbatim}
*** preparing to run SigSpec *******************************

Rayleigh frequency resolution             0.0914470160931467
oversampling ratio                       91.4470160931467433
frequency spacing                         0.0010000000000000
lower frequency limit                     0.0010000000000000
upper frequency limit                    13.0000000000000000
Nyquist coefficient                       0.9993990384615384
number of frequencies                 13000
\end{verbatim}\end{scriptsize}

For the project {\tt harmonics}, the corresponding output is richer.

\begin{scriptsize}\begin{verbatim}
*** preparing to run SigSpec *******************************

Rayleigh frequency resolution             0.0914470160931467
oversampling ratio                       91.4470160931467433
frequency spacing                         0.0010000000000000
lower frequency limit                     0.1250000000000000
upper frequency limit                    13.0000000000000000
Nyquist coefficient                       1.0000000000000000
number of frequencies                 12876
upper fundamental frequency               1.0000000000000000
number of fundamental frequencies       876
\end{verbatim}\end{scriptsize}

Although the upper frequency limit is set 1 cycle per day by the keyword {\tt ufreq}, {\sc SigSpec} has to compute the Fourier spectrum up to a frequency of 13 cycles per day in order to cover also the 12 harmonics. Two additional lines are provided corresponding to the upper limit for the fundamental frequencies, which is related to the specification by {\tt ufreq} in the file {\tt harmonics.ini}, and the number of fundamental frequencies.\sf

\section{MultiFile Mode}\label{SIGSPEC_MultiFile Mode}

\subsection{How to handle multiple time series}\label{SIGSPEC_How to handle multiple time series}

An additional feature of {\sc SigSpec} is the ability to handle multiple time series input files at once. This increases the performance of the program significantly, if the time values in all input files are identical.

\begin{itemize}
\item The user has to provide only one project directory {\tt <project>} -- just as in SingleFile mode (as described in ``Projects'', p.\,\pageref{SIGSPEC_Projects}).
\item Parameter specifications in the file {\tt <project>.ini} are uniquely applied to all time series input files. Thus {\sc SigSpec} expects the same column format for all time series input files and applies the settings specified in the {\tt .ini} file to all input files.
\item Time series files have to be indexed as {\tt \#multifile\#.<project>.dat}, where {\tt \#multifile\#} represents a six-digit index starting with {\tt 000000}. Note that strictly ascending indices are required.
\item All output files are supplied with the leading index {\tt \#multifile\#}. For example, {\tt 000012.s000009.000002} denotes the significance specturm the second iteration for time interval number nine in a time-resolved analysis of the 12th file in MultiFile mode.
\item The MultiFile mode is activated by the keyword {\tt multifile}\label{SIGSPEC_keyword.multifile}, followed by an integer value. This value is interpreted as the maximum index up to which the calculations shall be performed. This permits a restriction for, e.\,g., test runs. If the index limit is assigned a negative value, {\sc SigSpec} analyses as many files as available.
\item The sampling profile of the file {\tt 000000.<project>.dat} is always written to a file. For subsequent and consistent time series, the sampling profile is taken from this file, which saves computation time. Only if the new time values are inconsistent with those of the precursor, the profile is re-calculated and stored in a corresponding output file for later use. The keyword {\tt profile} in the {\tt .ini} file is ignored in MultiFile mode.
\end{itemize}

In MultiFile mode, {\sc SigSpec} terminates, if a {\tt \#multifile\#} index is reached, for which no time series input file is available.

A further keyword to restrict the MultiFile analysis is {\tt mfstart}\label{SIGSPEC_keyword.mfstart}, which permits to specify a MultiFile index to start with (instead of 0).

\vspace{12pt}\noindent{\bf Example.} \it The two lines

\begin{scriptsize}\begin{verbatim}
mfstart 4
multifile 16
\end{verbatim}\end{scriptsize}

\noindent activate the MultiFile mode for input files from {\tt 000004.<project>.dat} to {\tt 000016.<project>.dat}.

The big advantage of the MultiFile mode is that sampling profiles are computed only if necessary. If the time-domain sampling is identical to a previously examined time series, the sampling profile of this time series is used. If the keyword {\tt profile} is set in the {\tt .ini} file, a file {\tt assign.log} is generated. It contains a table of assignments between time series file indices and profile indices.\sf

\vspace{12pt}\noindent{\bf Example.} \it The line

\begin{scriptsize}\begin{verbatim}
           000013                           000002}
\end{verbatim}\end{scriptsize}

in the file {\tt assign.log} means that for {\tt 000013.<project>.dat}, the profile with index {\tt 000002} is used.\sf

\vspace{12pt}\noindent{\bf Example.} \it The sample project {\tt multifile} illustrates the simultaneous analysis of multiple time series. The project contains 10 time series files from {\tt 000457.multifile.dat} to {\tt 000467.multifile.dat}. However, the files do not represent a complete sequence, since {\tt 000466.multifile.dat} is missing.
The lines

\begin{scriptsize}\begin{verbatim}
mfstart 457
multifile 467
\end{verbatim}\end{scriptsize}

\noindent in the file {\tt multifile.ini} would force {\sc SigSpec} to process the complete sequence of time series input files. Indeed, the program starts with the file {\tt 000457.multifile.dat} and proceeds until {\tt 000465.multifile.dat}. Since the next file, {\tt 000466.multifile.dat} is missing, it stops its calculations with {\tt 000465.multifile.dat} and displays a corresponding warning:

\begin{scriptsize}\begin{verbatim}
Warning: MultiFile_Count 002
         MultiFile limit exceeds number of available
         time series input files, limit re-adjusted to 465.
\end{verbatim}\end{scriptsize}

The keyword {\tt profile} in the file {\tt multifile.ini} forces {\sc SigSpec} to generate the following files in the project directory:

\begin{scriptsize}\begin{verbatim}
000457.profile.dat
000458.profile.dat
000460.profile.dat
000463.profile.dat
\end{verbatim}\end{scriptsize}

\noindent The reason why only four profiles are computed for nine time series is found in the file {\tt assign.log}:

\begin{scriptsize}\begin{verbatim}
time series input file           profile and spectral window

          000457                           000457
          000458                           000458
          000459                           000457
          000460                           000460
          000461                           000457
          000462                           000457
          000463                           000463
          000464                           000457
          000465                           000457
\end{verbatim}\end{scriptsize}

\noindent The contents of this file tell the user that the samplings of all time series files are identical, except for those with indices {\tt 000458}, {\tt 000460} and {\tt 000463}. In order to speed up the computations, {\tt SigSpec} generates only one profile for the files with identical sampling and re-uses this profile for all of them. The first file with this sampling is {\tt 000457.multifile.dat}, and the associated profile is also used for

\begin{scriptsize}\begin{verbatim}
000459.multifile.dat
000461.multifile.dat
000462.multifile.dat
000464.multifile.dat
000465.multifile.dat
\end{verbatim}\end{scriptsize}

\noindent If the keyword {\tt win} is added to the file {\tt multifile.ini}, this assignment applies to the files containing the spectral windows as well.\sf

\subsection{Differential significance spectra}\label{SIGSPEC_Differential significance spectra}

Practical astronomical time series analysis occasionally comes along with target and comparison datasets that show coincident peaks in the DFT amplitude spectra. In this case, {\sc SigSpec} provides a possibility to compute the probability that a peak in the target dataset is significant in spite of a given peak in the comparison dataset. Moreover, multiple target and/or comparison datasets may be handled the same way. The idea is to identify common (instrumental and/or environmental) effects and to distinguish them from periodicities exclusively found in a target dataset.

In the {\tt .ini} file, there are three different keywords reserved for the specification of dataset types. Each expects one integer parameter representing the MultiFile index of the dataset under consideration.
\begin{enumerate}
\item The keyword {\tt target}\label{SIGSPEC_keyword.target} specifies a target dataset.
\item The keyword {\tt comp}\label{SIGSPEC_keyword.comp} specifies a comparison dataset.
\item The keyword {\tt skip}\label{SIGSPEC_keyword.skip} specifies a dataset to be ignored.
\end{enumerate}
To enhance the convenience for the user, not all files need to be specified. The keyword {\tt deftype}\label{SIGSPEC_keyword.deftype} may be used to assign a default dataset type.
\begin{enumerate}
\item Use {\tt deftype target} to assign the target attribute by default. If no {\tt deftype} keyword is provided, this setting is activated.
\item Use {\tt deftype comp} to assign the comp attribute by default.
\item Use {\tt deftype skip} to assign the skip attribute by default.
\end{enumerate}

Sampling profiles need to be computed for target datasets only. If the keyword {\tt profile} is given in the {\tt .ini} file, sampling profiles will only be generated for target datasets, and the file {\tt assign.log} will also contain target datasets only.

To make datasets comparable even if their quality is different, the DFT spectra of the comparison datasets are scaled according to the power integral over the entire frequency range under consideration.

Instead of the observables $c_k$, $k = 0,1,...,K$ of a comparison dataset, the transformed quantities
\begin{equation}
c^\prime _k := \frac{P\left( x_l\right)}{P\left( c_k\right)} c_k
\end{equation}
are used, where $x_l$, $l = 0,1,...,L$ denotes the observables of the target dataset under consideration and $P$ indicates the power integral of the quantity in parentheses. A DFT is calculated for each comparison dataset. There are two options to determine the resulting amplitude $A_T$ to be compared to the target amplitude $A$.

By default, the sig measures the probability of a peak generated by noise at the same variance as that of the given time series. In case of computing differential sigs, the normalisation has to be modified, since part of the power found in the target spectrum is assumed due to corresponding power in a comparison spectrum. To take this into account appropriately, a factor
\begin{equation}
\gamma := \frac{P\left( x_l\right)}{dP}
\end{equation}
is introduced, where $dP$ is the power integral of the difference between the target data and the transformed comparison data. Correspondingly, the differential sig is a measure of the additional power with respect to the comparison dataset to be due to noise.

\begin{enumerate}
\item If the keyword {\tt diff:comp}\label{SIGSPEC_keyword.diff:comp} is set in the {\tt .ini} file, a weighted arithmetic mean of the Fourier vectors, averaged over all comparison datasets is used to calculate $A_T$. The numbers of data points the comparison datasets consist of are used as weights. This option considers signal common among the comparison datasets only if the phases are aligned. Following the formalism by Reegen~(2007), the cartesian representation of the differential sig evaluates to
\[
\mathrm{sig}\left( a_{\mathrm{ZM}},b_{\mathrm{ZM}}\left|\right.\omega\right) = \gamma\frac{K\log\mathrm{e}}{\left< x^2\right>}\:\times
\]
\[
\left\lbrace\left[\frac{\left( a_{\mathrm{ZM}}-a_{T\,\mathrm{ZM}}\right)\cos\theta _0+\left(b_{\mathrm{ZM}}-b_{T\,\mathrm{ZM}}\right)\sin\theta _0}{\alpha _0}\right] ^2\right.
\]
\begin{equation}\label{SIGSPEC_EQ significance full cartesian}
\left. + \left[\frac{\left( a_{\mathrm{ZM}}-a_{T\,\mathrm{ZM}}\right)\sin\theta _0-\left(b_{\mathrm{ZM}}-b_{T\,\mathrm{ZM}}\right)\cos\theta _0}{\beta _0}\right] ^2\right\rbrace\: .
\end{equation}
\item If the keyword {\tt diff:compalign}\label{SIGSPEC_keyword.diff:compalign} is set in the {\tt .ini} file, a weighted arithmetic mean of the DFT amplitudes, averaged over all comparison datasets is considered as $A_T$. The numbers of data points the comparison datasets consist of are used as weights. This option considers signal common among the comparison datasets also if they lag in phase. The differential sig is obtained through
\begin{equation}
\mathrm{sig}\left( A\left|\right. A_T\right) = \gamma\frac{K\left( A-A_T\right)^2\log\mathrm{e}}{4\left< x^2\right>}\left[\frac{\cos ^2\left(\theta - \theta _0\right)}{\alpha _0^2} + \frac{\sin ^2\left(\theta - \theta _0\right)}{\beta _0^2}\right]\: ,
\end{equation}
following the annotation introduced by Reegen~(2007).
\end{enumerate}

The default setting is {\tt diff:off}\label{SIGSPEC_keyword.diff:off}, which switches off the computation of differential sigs.

Additional output is provided in the spectra (see p.\,\pageref{SIGSPEC_Spectra}), where columns 6 and 7 contain the DFT amplitudes and phases of the transformed comparison dataset, respectively.

\vspace{12pt}\noindent{\bf Example.}\label{SIGSPEC_EXdiffsig} \it The sample project {\tt diffsig} illustrates the analysis of target and comparison time series using differential significance spectra. There are nine time series input files available, indexed from {\tt 000038} through {\tt 000046}. The file {\tt diffsig.ini} contains the lines

\begin{scriptsize}\begin{verbatim}
mfstart 38
multifile -1
\end{verbatim}\end{scriptsize}

\noindent which forces {\sc SigSpec} to start with the file {\tt 000038.diffsig.dat} and compute all available datasets. In this case, {\sc SigSpec} takes into account all files from {\tt 000038} to {\tt 000046}. The two lines

\begin{scriptsize}\begin{verbatim}
deftype target
comp 38
\end{verbatim}\end{scriptsize}

\noindent in the file {\tt diffsig.ini} define the file {\tt 000038.diffsig.dat} as a comparison dataset and the rest as targets. Thus differential significance spectra are calculated for all time series from {\tt 000039} through {\tt 000045}, with respect to {\tt 000038} as comparison data. The calculation of differential sigs is activated by the line

\begin{scriptsize}\begin{verbatim}
diff:compalign
\end{verbatim}\end{scriptsize}

\noindent in the file {\tt diffsig.ini}, which produces differential sigs without respect to phase lags between comparison and target signals. The computations are made faster by the lines

\begin{scriptsize}\begin{verbatim}
ufreq 7
siglimit 0
iterations 1
\end{verbatim}\end{scriptsize}

The sampling of the input file {\tt 000038.diffsig.dat} represents the V photometry of IC\,4996\,\#\,89 (see Example {\tt SigSpecNative}, p.\,\pageref{SIGSPEC_EXnormalrun}), and the observable is a synthetically generated signal with unit amplitude at a frequency of 3.125 cycles per day, plus Gaussian noise with 5 units rms deviation. The corresponding significance spectrum, as obtained by typing

\begin{scriptsize}\begin{verbatim}
SigSpec 000038.diffsig
\end{verbatim}\end{scriptsize}

\noindent is displayed in the bottom panel of Fig.\,\ref{SIGSPEC_diffsig.s}\it . The five upper panels contain the differential significance spectra of the time series {\tt 000039} to {\tt 000046}. These datasets contain 11\,649 points and are based on the sampling used in the project {\tt harmonics} (p.\,\pageref{SIGSPEC_EXharmonics}). Gaussian noise with a standard deviation of 100 units is generated. Just as in case of the comparison data, a sinusoid at 3.125 cycles per day is synthesized, but the phase is not the same as for {\tt 000038.diffsig.dat}. The amplitudes of this signal are 5 units for {\tt 000039}, 6 units for {\tt 000040}, 7 units for {\tt 000041}, 8 units for {\tt 000042}, 9 units for {\tt 000043}, 10 units for {\tt 000044}, 11 units for {\tt 000045}, and 12 units for {\tt 000046}. With increasing signal amplitude in the target data, the differential sig of the main peak consistently increases. In Fig.\,\ref{SIGSPEC_diffsig.s} \it the datasets {\tt 000039} to {\tt 000046} are displayed from bottom to top.\sf

\figureDSSN{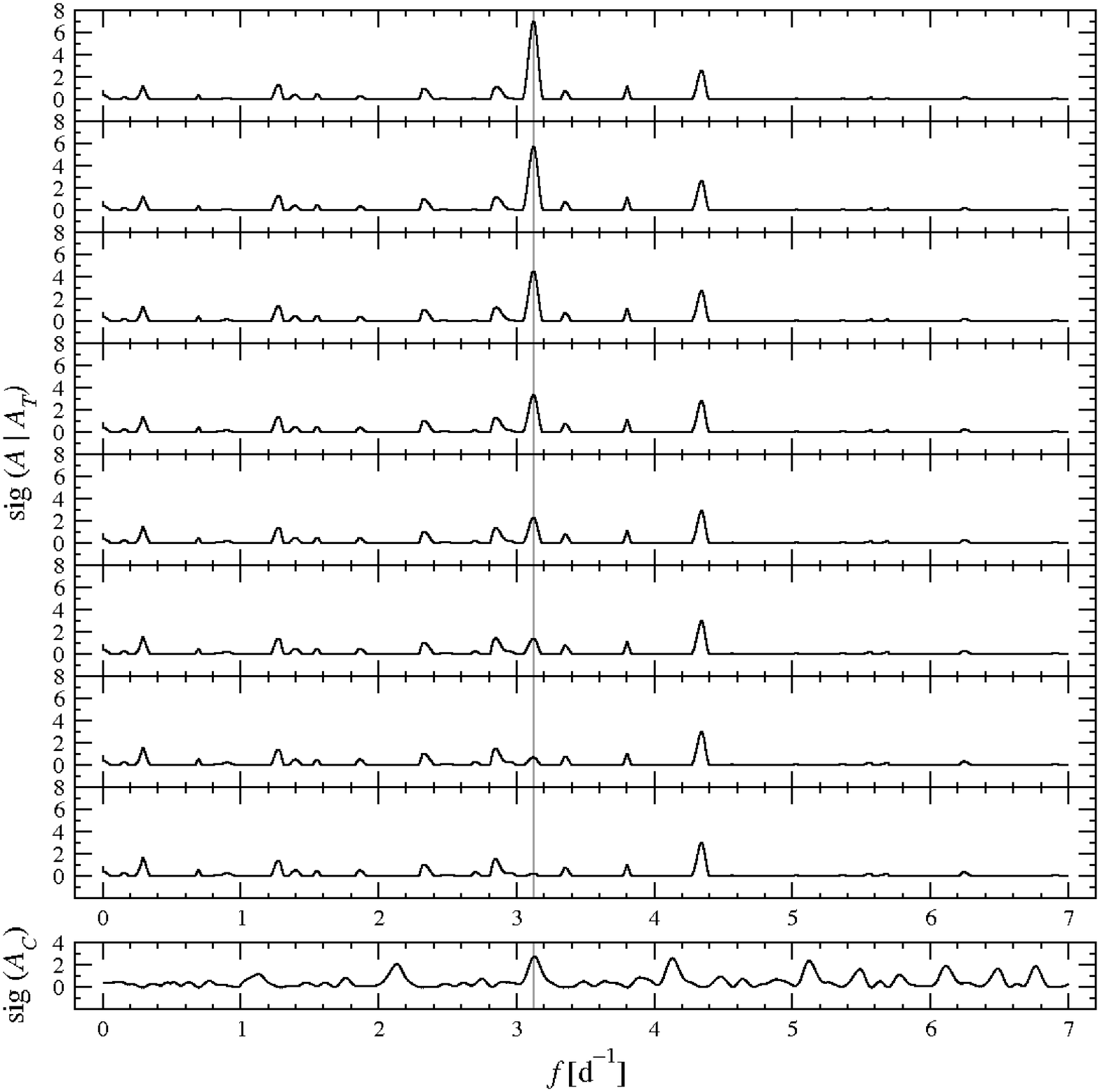}{Differential significance spectra for the sample project {\tt diffsig}. {\em Bottom:} significance spectrum of comparison data, representing a sinusoidal signal at $3.125$ cycles per day ({\em grey line}), plus Gaussian noise. {\em Top eight panels:} Differential significance spectra for target time series representing the Gaussian noise plus a sinusoidal signal at $3.125$ cycles per day. Both the time-domain sampling and the signal phase differ from the comparison data. From bottom to top, the amplitude of this signal increases.}{SIGSPEC_diffsig.s}{!htb}{clip,angle=0,width=110mm}

\section{The Built-in Simulator}\label{SIGSPEC_The Built-in Simulator}

{\sc SigSpec} contains a simulator to generate and analyse synthetic time series. To activate the simulator, a sequence of keywords may be given in the {\tt .ini} file to generate a variety of datasets. The sampling is taken from the time series input file.

The simulator activities specified by {\tt sim:signal}, {\tt sim:poly}, {\tt sim:exp}, {\tt sim:serial}, {\tt sim:temporal}, {\tt sim:rndsteps}, and {\tt sim:zeromean} are interpreted as a sequence and performed step by step, following their order in the {\tt .ini} file. {\sc SigSpec} generates the synthetic light curve by performing all specified actions following the order of occurrence in the {\tt .ini} file.

The synthetic time series is saved as a file with the same name as the input, but in the project directory, to avoid accidential overwriting of original data. If the time series input file is named {\tt <project>.dat}, then the synthetic time series is {\tt <project>/<project>.dat}. In MultiFile mode, if the time series input files are named {\tt \#multifile\#.<project>.dat}, the synthetic time series are {\tt <project/\#multifile\#.<project>.dat}.

\subsection{The simulator mode}\label{SIGSPEC_The simulator mode}

{\sc SigSpec} supports two different simulator modes.
\begin{enumerate}
\item The keyword {\tt sim:add}\label{SIGSPEC_keyword.sim:add} runs the simulator in additive mode. The program keeps the original observable values and adds the synthetic values. For example, this function is useful to add synthetic noise to a given time series.
\item The keyword {\tt sim:replace}\label{SIGSPEC_keyword.sim:replace} forces the simulator to overwrite the original observable values with the synthetic values.
\item The keyword {\tt sim:off}\label{SIGSPEC_keyword.sim:off} is used, if no simulator activity is desired. Since the simulator is deactivated by default, this keyword is redundant and only implemented for completeness.
\end{enumerate}

\subsection{Random numbers}\label{SIGSPEC_Random numbers}

The {\sc SigSpec} simulator is capable of modelling three different types of random processes:
\begin{itemize}
\item serially correlated noise (keyword {\tt sim:serial}, p.\,\pageref{SIGSPEC_Serially correlated noise}),
\item temporally correlated noise (keyword {\tt sim:temporial}, p.\,\pageref{SIGSPEC_Temporally correlated noise}),
\item random steps (keyword {\tt sim:rndsteps}, p.\,\pageref{SIGSPEC_Random steps}.
\end{itemize}

The random number generator employed for these models may be initialised in two different ways.
\begin{enumerate}
\item The user may pass an integer value to the program. This value has to be written into a file {\tt <project>.rnd}.
\item If the file {\tt <project>.rnd} is not present, the simulator initialises the random number generator using the system time.
\end{enumerate}

The last integer value in the sequence of random numbers is written to a file {\tt <project>/<project>.rnd}. This allows to embed {\sc SigSpec} into an external loop for numerical simulations. If the output file {\tt <project>/<project>.rnd} is moved to {\tt <project>.rnd} externally between consecutive {\sc SigSpec} runs, the program may used iteratively without breaking the random number sequence.

\vspace{12pt}\noindent{\bf Example.} \it The simulator is employed in the sample projects {\tt sim-serial}, {\tt sim-temporal} and {\tt sim-rndsteps}. To initialise the random number generator, a file {\tt sim-serial.rnd}, {\tt sim-temporal.rnd} and {\tt sim-rndsteps.rnd}, respectively, is used to make the output reproducible.

Consequently, the user has three options to explore the these samples.
\begin{enumerate}
\item If the samples are processed as they are, {\sc SigSpec} reproduces the given output exactly.
\item If the {\tt .rnd} file in the input directory is removed by the user, {\sc SigSpec} produces a new set of random numbers. The random number generator is initialised employing the system time.
\item If the content of the {\tt .rnd} file in the input directory is modified by the user, {\sc SigSpec} produces a new set of random numbers. The random number generator is initialised employing the new number in the {\tt .rnd} file.
\end{enumerate}\sf

\subsection{Sinusoidal signal}

The keyword {\tt sim:signal}\label{SIGSPEC_keyword.sim:signal} is given with five floating-point parameters. They specify
\begin{enumerate}
\item the lower time limit,
\item the upper time limit,
\item the amplitude,
\item the time zeropoint (a fixed time where the signal shall attain a maximum), and
\item the frequency [inverse time units].
\end{enumerate}
If the lower and upper time limits are both set zero, the signal is generated for the entire time base.

\figureDSSN{f27.eps}{Time series generated by the simulator in the sample project {\tt sim-signal}. {\em Open circles:} Original V photometry of IC\,4996\,\#\,89. {\em Dots:} Two sinusoidal signals added by the simulator.}{SIGSPEC_sim-sin.dat}{!htb}{clip,angle=0,width=110mm}

\vspace{12pt}\noindent{\bf Example.} \it The sample project {\tt sim-signal} contains the simulation and analysis of two sinusoidal signals, one over the entire time base, one on a restricted time interval. In this sample project, the V photometry of IC\,4996\,\#\,89 (see Example {\tt SigSpecNative}, p.\,\pageref{SIGSPEC_EXnormalrun}) is modified, according to the line

\begin{scriptsize}\begin{verbatim}
sim:add
\end{verbatim}\end{scriptsize}

\noindent in the file {\tt sim-signal.ini}. The line

\begin{scriptsize}\begin{verbatim}
sim:signal 0 0 0.00727 2521.4542 4.68573
\end{verbatim}\end{scriptsize}

\noindent produces a sinusoidal signal over the entire time base (corresponding to the first two arguments being zero). The amplitude is 7.27 mmag, and the frequency is 4.68573 cycles per day. At HJD\,2452521.4542 the sinusoid shall attain zero value. Correspondingly, the line

\begin{scriptsize}\begin{verbatim}
sim:signal 2521 2525 0.00543 2524.2356 6.24512
\end{verbatim}\end{scriptsize}

\noindent is associated to a sinusoid with amplitude 5.43 mmag, frequency 6.24512 cycles per day, and a zeropoint at HJD\,2452524.2356. This signal is not generated for the entire time base but only from HJD\,2452521 to HJD\,2452525. Fig.\,\ref{SIGSPEC_sim-sin.dat} \it displays the light curves of the original and the synthetic data.

The screen output contains the lines

\begin{scriptsize}\begin{verbatim}
*** simulator: add *****************************************

signal
signal
\end{verbatim}\end{scriptsize}

indicating that the simulator adds the synthetic values to the original observables, and that two sinusoids are generated.

\figureDSSN{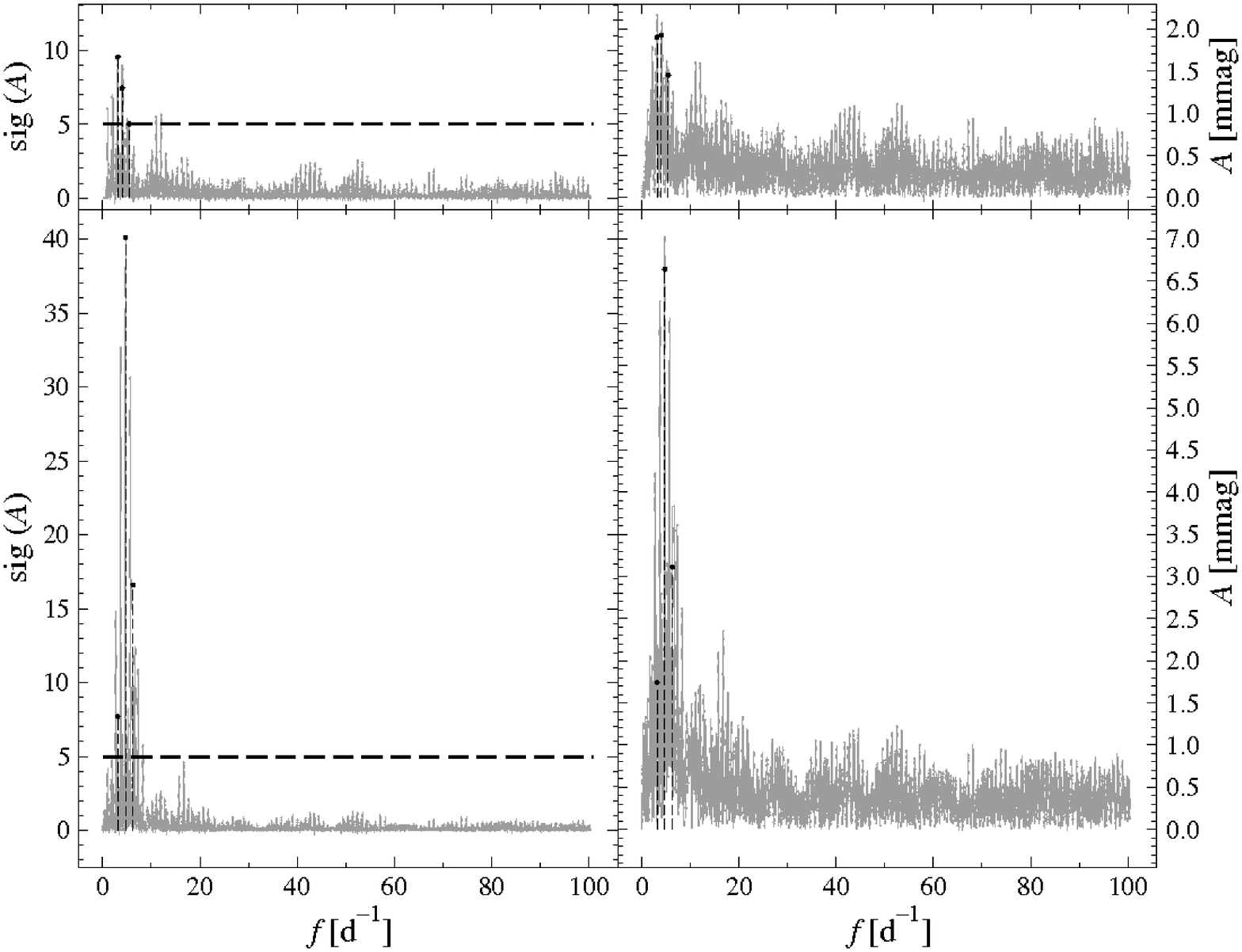}{Fourier spectra for the sample project {\tt sim-signal}. {\em Left:} significance spectra. {\em Right:} DFT amplitudes. {\em Top:} original spectra (file {\tt SigSpecNative/s000000.dat}). {\em Bottom:} spectra with two sinusoidal signals added. All spectra are plotted {\em grey}. The significant components are indicated by black {\em dots} with dashed drop lines (file {\tt SigSpecNative/result.dat} for the top panels, file {\tt sim-signal/s000000.dat} for the bottom panels). The default sig threshold of 5 is represented by a horizontal {\em dashed line} in the left panels.}{SIGSPEC_sim-sin.s}{!htb}{clip,angle=0,width=110mm}

Fig.\,\ref{SIGSPEC_sim-sin.s} 
compares the Fourier spectra of the synthetic time series to those of the original time series (as used in Example {\tt SigSpecNative}, p.\,\pageref{SIGSPEC_EXnormalrun}, and displayed in Fig.\,\ref{SIGSPEC_normalrun.s}\it , p.\,\pageref{SIGSPEC_normalrun.s}. Both signals introduced by the simulator are identified, but the prewhitening of the component at 6.25 cycles per day is performed over the whole time base, although the signal is present only in an interval. This introduces additional noise, which causes the signal at 3.99 cycles per day to drop below the significance limit of 5 and avoids the detection of the component at 5.41 cycles per day.\sf

\subsection{Polynomial trend}

The keyword {\tt sim:poly}\label{SIGSPEC_keyword.sim:poly} is given with five floating-point parameters. They specify
\begin{enumerate}
\item the lower time limit,
\item the upper time limit,
\item the coefficient $P_0$,
\item the time zeropoint $t_0$, and
\item the exponent $X$.
\end{enumerate}
If the exponent is a non-integer number, the simulator evaluates
\begin{equation}
P\left( t\right) := P_0\left| t-t_0\right| ^X
\end{equation}
instead and produces a power function.

For integer exponents, the trend is generated by the relation
\begin{equation}
P\left( t\right) := P_0\left( t-t_0\right) ^X\: .
\end{equation}
Thus a full polynomial may be constructed by multiple keywords {\tt sim:poly} with different parameters and integer exponents.

If the lower and upper time limits are both set zero, the polynomial trend is generated for the entire time base.

\figureDSSN{f29.eps}{Time series generated by the simulator in the sample project {\tt sim-poly}. The sampling represents the V photometry of IC\,4996\,\#\,89. The simulator replaces the origninal observable by 16 different power functions.}{SIGSPEC_sim-poly.dat}{!htb}{clip,angle=0,width=110mm}

\vspace{12pt}\noindent{\bf Example.}\label{SIGSPEC_EXsim-poly} \it The sample project {\tt sim-poly} contains the simulation and analysis of 16 individual power functions defined on different time intervals (Fig.\,\ref{SIGSPEC_sim-poly.dat}, \it p.\,\pageref{SIGSPEC_sim-poly.dat}). The sampling of the V photometry of IC\,4996\,\#\,89 is used, and the simulator replaces the original observable values, according to the line 

\begin{scriptsize}\begin{verbatim}
sim:replace
\end{verbatim}\end{scriptsize}

\noindent in the file {\tt sim-poly.ini}. The specifications for the power functions are contained in the lines

\begin{scriptsize}\begin{verbatim}
sim:poly   2520.215 2521.088 4.298 2520.626  0.581
sim:poly   2521.088 2521.679 2.932 2521.443  1.195
sim:poly   2521.679 2522.442 1.092 2522.067  1.063
sim:poly   2522.442 2522.595 5.372 2522.466  0.676
sim:poly   2522.595 2523.351 2.495 2522.682  2.042
sim:poly   2523.351 2523.924 2.839 2523.607  0.221
sim:poly   2523.924 2524.478 8.357 2525.412 -0.899
sim:poly   2524.478 2525.399 2.304 2524.576  1.432
sim:poly   2525.399 2526.107 2.573 2525.721  1.205
sim:poly   2526.107 2526.550 6.350 2526.493  0.031
sim:poly   2526.550 2526.847 4.192 2526.589  2.893
sim:poly   2526.847 2527.616 0.345 2527.652 -0.472
sim:poly   2527.616 2528.264 3.583 2527.783  0.725
sim:poly   2528.264 2528.777 1.246 2528.704  0.610
sim:poly   2528.777 2529.606 3.534 2529.535  1.752
sim:poly   2529.606 2530.242 9.002 2529.694  1.119
\end{verbatim}\end{scriptsize}

The screen output contains the lines

\begin{scriptsize}\begin{verbatim}
*** simulator: replace *************************************

polynomial trend
polynomial trend
polynomial trend
polynomial trend
polynomial trend
polynomial trend
polynomial trend
polynomial trend
polynomial trend
polynomial trend
polynomial trend
polynomial trend
polynomial trend
polynomial trend
polynomial trend
polynomial trend
\end{verbatim}\end{scriptsize}

to indicate that the simulator replaces the original observables by the synthetic values, and that 16 power functions are generated.

{\sc SigSpec} detects 19 significant signal components, which are not discussed here.\sf

\subsection{Exponential trend}

The keyword {\tt sim:exp}\label{SIGSPEC_keyword.sim:exp} is given with five floating-point parameters. They specify
\begin{enumerate}
\item the lower time limit,
\item the upper time limit,
\item the coefficient $E_0$,
\item the time zeropoint $t_0$, and
\item the exponent $X$.
\end{enumerate}
The polynomial trend is generated by the relation
\begin{equation}
E\left( t\right) := E_0\,\mathrm{e}^{X\left( t-t_0\right)}\: .
\end{equation}
If the lower and upper time limits are both set zero, the exponential trend is generated for the entire time base.

\figureDSSN{f30.eps}{Time series generated by the simulator in the sample project {\tt sim-exp}. The sampling represents the V photometry of IC\,4996\,\#\,89. The simulator replaces the origninal observable by two exponential functions, one over the entire time base, and the other one on an interval between HJD\,$2452521.4532$ and HJD\,$2452526.8832$.}{SIGSPEC_sim-exp.dat}{!htb}{clip,angle=0,width=110mm}

\vspace{12pt}\noindent{\bf Example.} \it The sample project {\tt sim-exp} contains the simulation and analysis of two exponential trends, one over the entire time base, one on a restricted time interval, corresponding to the lines

\begin{scriptsize}\begin{verbatim}
sim:exp 2521.4532 2526.8832 1.3256 2526.7384  0.65834
sim:exp    0         0      2.2841 2520.8562 -0.03425
\end{verbatim}\end{scriptsize}

\noindent in the file {\tt sim-exp.ini}. The sampling of the V photometry of IC\,4996\,\#\,89 is used, and the simulator replaces the original observable values, according to the line

\begin{scriptsize}\begin{verbatim}
sim:replace
\end{verbatim}\end{scriptsize}

\noindent The screen output contains the expression {\tt exponential trend} to indicate that such a trend is generated. In this example, the entry is found twice. The resulting light curve is displayed in Fig.\,\ref{SIGSPEC_sim-exp.dat}, \it p. \pageref{SIGSPEC_sim-exp.dat}. 

{\sc SigSpec} detects 54 significant signal components, which are not discussed here.\sf

\subsection{Serially correlated noise}\label{SIGSPEC_Serially correlated noise}

This simulator module produces Gaussian noise the standard deviation of which may vary in time according to a polynomial trend. A serial correlation coefficient between consecutive data points may be specified additionally.

The keyword {\tt sim:serial}\label{SIGSPEC_keyword.sim:serial} is given with six floating-point parameters. They specify
\begin{enumerate}
\item the lower time limit,
\item the upper time limit,
\item the coefficient $\sigma _0$ for the standard deviation of the Gaussian noise,
\item the time zeropoint $t_0$ for the polynomial trend of the standard deviation,
\item the exponent $X$ for the polynomial trend of the standard deviation, and
\item the serial correlation coefficient.
\end{enumerate}
The standard deviation of the Gaussian noise follows the relation
\begin{equation}
\sigma\left( t\right) := \sigma _0\,\left( t-t_0\right) ^X\: .
\end{equation}

\figureDSSN{f31.eps}{Time series generated by the simulator in the sample projects {\tt sim-serial} ({\em dots}) and {\tt sim-temporal} ({\em open circles}), respectively. The sampling represents the V photometry of IC\,4996\,\#\,89. In both samples, the original observable values are replaced by the simulator.}{SIGSPEC_sim-serial.dat}{!htb}{clip,angle=0,width=110mm}

A full polynomial may be constructed by multiple keywords {\tt sim:serial} with different parameters.

If the lower and upper time limits are both set zero, the noise is generated for the entire time base.

\vspace{12pt}\noindent{\bf Example.}\label{SIGSPEC_EXsim-serial} \it The sample project {\tt sim-serial} contains the simulation and analysis of serially correlated noise. The sampling of the V photometry of IC\,4996\,\#\,89 is used, and the simulator replaces the original observable values, according to the line

\begin{scriptsize}\begin{verbatim}
sim:replace
\end{verbatim}\end{scriptsize}

\noindent in the file {\tt sim-serial.ini}. The line

\begin{scriptsize}\begin{verbatim}
sim:serial   0 0 1 0 0 0.8
\end{verbatim}\end{scriptsize}

\figureDSSN{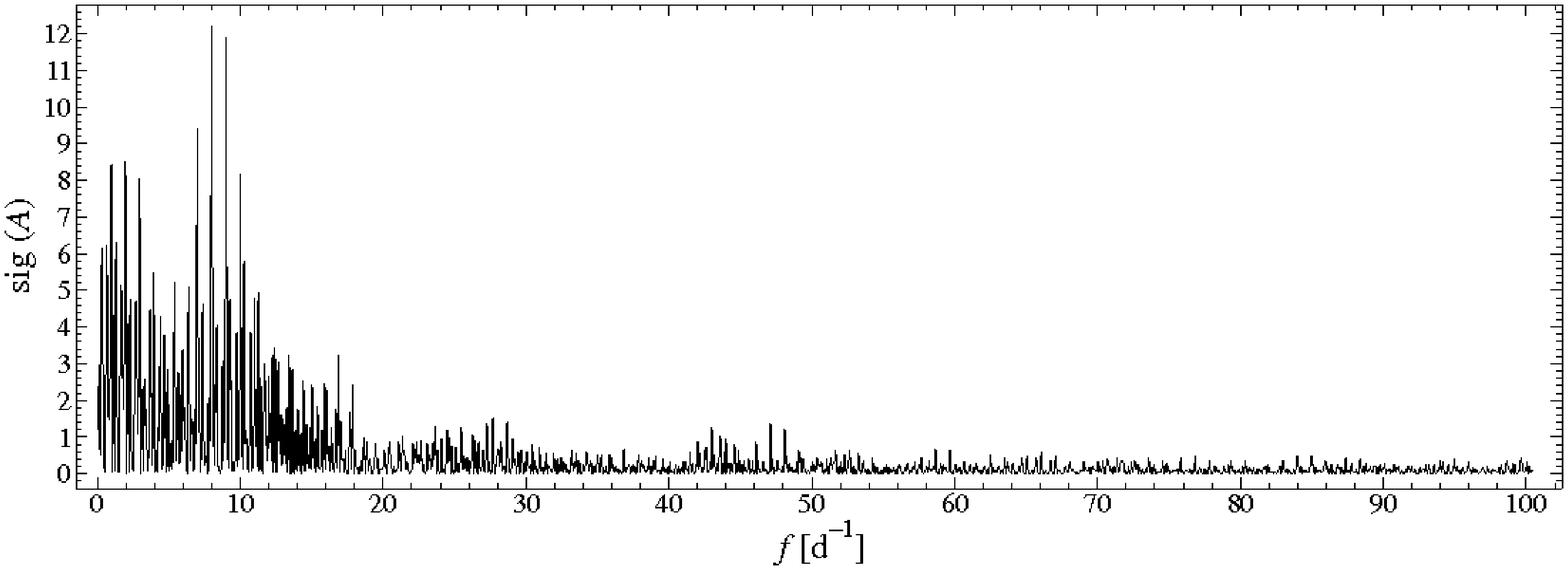}{Typical significance spectrum for serially correlated noise, based on the sampling of the V photometry of IC\,4996\,\#\,89. Serial correlation produces systematically higher sigs in the low frequency region.}{SIGSPEC_sim-serial.s}{!htb}{clip,angle=0,width=110mm}

\noindent specifies noise with a constant standard deviation of 1 and a serial correlation coefficient of 0.8. Setting the first two parameters zero provides synthetic data for the entire time series. The resulting light curve is displayed in Fig.\,\ref{SIGSPEC_sim-serial.dat}\it . The line

\begin{scriptsize}\begin{verbatim}
random number generator: file sim-serial.rnd
\end{verbatim}\end{scriptsize}

\noindent in the screen output indicates that a file {\tt sim-serial.rnd} is found and used to initialise the random number generator. If such a file were not present, the system time would be used:

\begin{scriptsize}\begin{verbatim}
random number generator: system time initialisation
\end{verbatim}\end{scriptsize}

A significance spectrum is displayed in Fig.\,\ref{SIGSPEC_sim-serial.s}\it . The overall shape of the spectrum is typical for serially correlated noise, characterised by higher amplitudes and sigs for low frequencies.\sf

\subsection{Temporally correlated noise}\label{SIGSPEC_Temporally correlated noise}

This simulator module produces Gaussian noise the standard deviation of which may vary in time according to a polynomial trend. A temporal correlation coefficient $R_T$ between consecutive data points $t_{n-1}$, $t_n$ may be specified. In contrary to the serial correlation, the temporal correlation takes into account the width of the time interval between pairs of data points, which has implications on the noise behaviour of non-equidistantly sampled data. The serial correlation $R_S$ drops exponentially with the distance in time according to
\begin{equation}\label{SIGSPEC_EQcorrcoefconversion}
R_S := R_T^{t_n - t_{n-1}}\: .
\end{equation}
In this context, the temporal correlation coefficient may be interpreted as the serial correlation coefficient of two data points separated by one unit of time.

The keyword {\tt sim:temporal}\label{SIGSPEC_keyword.sim:temporal} is given with six floating-point parameters. They specify
\begin{enumerate}
\item the lower time limit,
\item the upper time limit,
\item the coefficient $\sigma _0$ for the standard deviation of the Gaussian noise,
\item the time zeropoint $t_0$ for the polynomial trend of the standard deviation,
\item the exponent $X$ for the polynomial trend of the standard deviation, and
\item the temporal correlation coefficient $R_T$.
\end{enumerate}
The standard deviation of the Gaussian noise follows the relation
\begin{equation}
\sigma\left( t\right) := \sigma _0\,\left( t-t_0\right) ^X\: .
\end{equation}
A full polynomial may be constructed by multiple keywords {\tt sim:temporal} with different parameters.

If the lower and upper time limits are both set zero, the noise is generated for the entire time base.

\figureDSSN{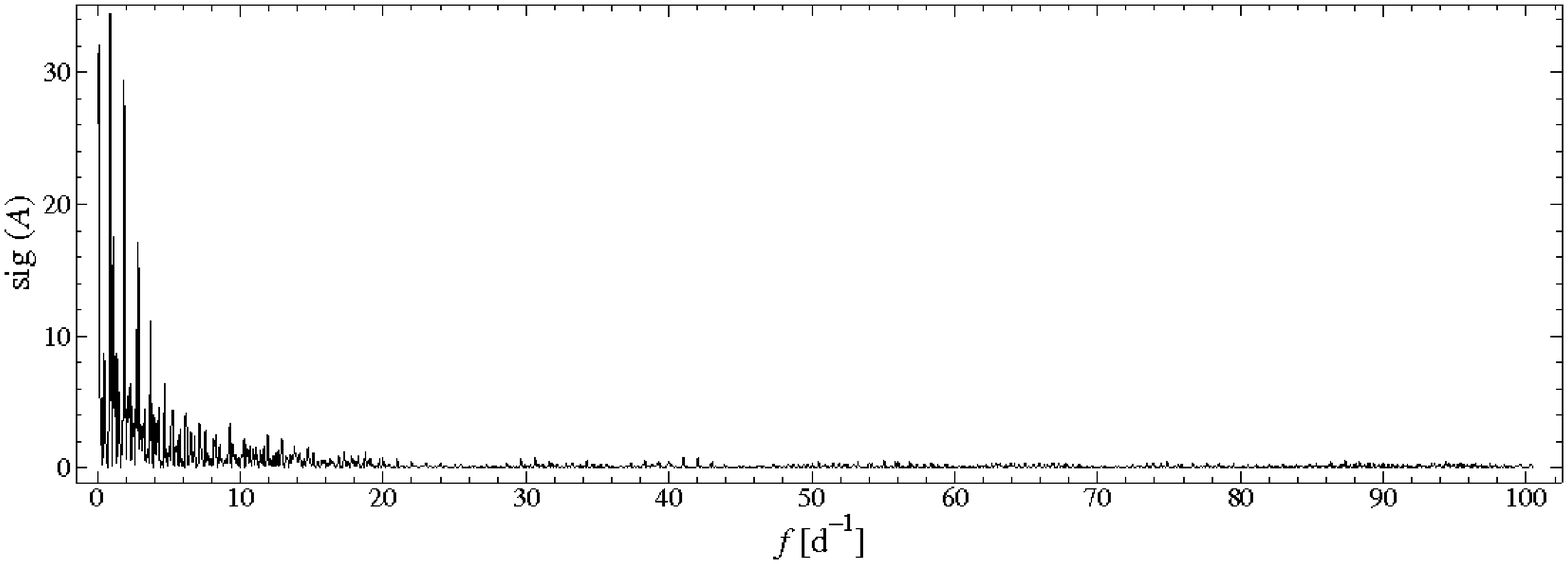}{Typical significance spectrum for temporally correlated noise, based on the sampling of the V photometry of IC\,4996\,\#\,89. Temporal correlation produces systematically higher sigs in the low frequency region, which is quite comparable to serial correlation (Fig.\,\ref{SIGSPEC_sim-serial.s}).}{SIGSPEC_sim-temporal.s}{!htb}{clip,angle=0,width=110mm}

\vspace{12pt}\noindent{\bf Example.} \it The sample project {\tt sim-temporal} contains the simulation and analysis of temporally correlated noise. The sampling of the V photometry of IC\,4996\,\#\,89 is used, and the simulator replaces the original observable values, according to the line

\begin{scriptsize}\begin{verbatim}
sim:replace
\end{verbatim}\end{scriptsize}

\noindent in the file {\tt sim-temporal.ini}. The line

\begin{scriptsize}\begin{verbatim}
sim:temporal   0 0 1 0 0 0.01
\end{verbatim}\end{scriptsize}

\noindent specifies noise with a constant standard deviation of 1 and a temporal correlation coefficient of 0.01. Setting the first two parameters zero provides synthetic data for the entire time series. The resulting light curve is displayed in Fig.\,\ref{SIGSPEC_sim-serial.dat}\it . Comparing this light curve to the dataset generated in the project {\tt sim-serial} (p.\,\pageref{SIGSPEC_EXsim-serial}), the correlation between consecutive data points is obviously much stronger in the present example. Using Eq.\,\ref{SIGSPEC_EQcorrcoefconversion}\it with a typical sampling interval width of 9 min for the dataset under consideration, the temporal correlation coefficient of 0.01 corresponds to a serial correlation coefficient of $\approx$ 0.97.

The significance spectrum displayed in Fig.\,\ref{SIGSPEC_sim-temporal.s} \it shows the same overall characteristics as the corresponding spectrum for serially correlated noise (Fig.\,\ref{SIGSPEC_sim-serial.s}\it , but the sigs at low frequencies are considerably higher, which is a consequence of the strong serial correlation associated to this setup.\sf

\subsection{Random steps}\label{SIGSPEC_Random steps}

This module generates steps following two random processes:
\begin{enumerate}
\item the constant attained by the synthetic observable throughout each step follows a Gaussian distribution with an expected value 0,
\item a Poisson process is used to define when a step has to be incorporated.
\end{enumerate}

The keyword {\tt sim:rndsteps}\label{SIGSPEC_keyword.sim:rndsteps} is given with four floating-point parameters. They specify
\begin{enumerate}
\item the lower time limit,
\item the upper time limit,
\item the standard deviation of the Gaussian distribution defining the constants attained throughout each step,
\item the expected time range for the Poisson distribution of steps.
\end{enumerate}
If the lower and upper time limits are both set zero, the steps are generated for the entire time base.

\figureDSSN{f34.eps}{Time series generated by the simulator in the sample project {\tt sim-rndsteps}. The sampling represents the V photometry of IC\,4996\,\#\,89. The original observable values are replaced by the simulator.}{SIGSPEC_sim-rndsteps.dat}{!htb}{clip,angle=0,width=110mm}

\vspace{12pt}\noindent{\bf Example.} \it The sample project {\tt sim-rndsteps} illustrates the simulation and analysis of random steps upon the sampling of the V photometry of IC\,4996\,\#\,89. The simulator replaces the original observable values, according to the line

\begin{scriptsize}\begin{verbatim}
sim:replace
\end{verbatim}\end{scriptsize}

\noindent in the file {\tt sim-rndsteps.ini}. The line

\begin{scriptsize}\begin{verbatim}
sim:rndsteps   0 0 0.5 0.07
\end{verbatim}\end{scriptsize}
\noindent in the file {\tt sim-rndsteps.ini} produces random steps the values of which are distributed according to a Gaussian with standard deviation 0.5. The expected distance in time of consecutive steps 0.07 days. The resulting light curve is displayed in Fig.\,\ref{SIGSPEC_sim-rndsteps.dat}\it .

\figureDSSN{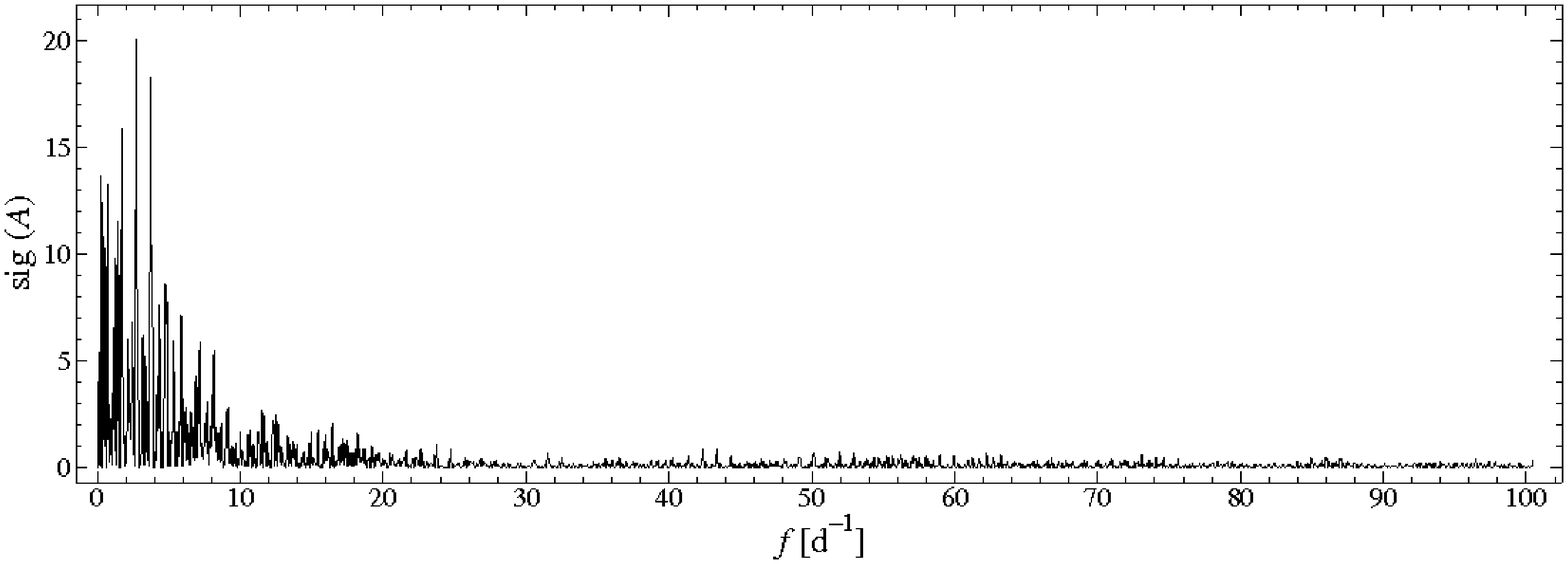}{Typical significance spectrum for random steps, based on the sampling of the V photometry of IC\,4996\,\#\,89. Each constant in the step function displayed in Fig.\,\ref{SIGSPEC_sim-rndsteps.dat} contributes a spectral window to this DFT.}{SIGSPEC_sim-rndsteps.s}{!htb}{clip,angle=0,width=110mm}

Since the observables are constant between the steps, one may consider each of the corresponding time intervals to contribute a spectral window to the DFT, or significance spectrum, correspondingly. The significance spectrum associated to the light curve in Fig.\,\ref{SIGSPEC_sim-rndsteps.dat} \it is displayed in Fig.\,\ref{SIGSPEC_sim-rndsteps.s} \it and respresents such a superposition of spectral windows.\sf

\subsection{Zero-mean adjustment}

The keyword {\tt sim:zeromean}\label{SIGSPEC_keyword.sim:zeromean} may be used to adjust the mean value of the time series (or a subset) to zero. It is given with two floating-point parameters,
\begin{enumerate}
\item the lower time limit, and
\item the upper time limit.
\end{enumerate}
If the lower and upper time limits are both set zero, the mean value of the entire synthetic time series is adjusted to zero. This option was adopted for consistency, but does not provide additional functionality, because a zero-mean correction of the whole data set is performed at every step of the prewhitening cascade by default.

\vspace{12pt}\noindent{\bf Example.} \it In the sample project {\tt sim-zeromean}, {\sc SigSpec} models the same time series as in the project {\tt sim-poly} (p.\,\pageref{SIGSPEC_EXsim-poly}), according to the first part of the file {\tt sim-zeromean.ini}:

\begin{scriptsize}\begin{verbatim}
sim:poly   2520.215 2521.088 4.298 2520.626  0.581
sim:poly   2521.088 2521.679 2.932 2521.443  1.195
sim:poly   2521.679 2522.442 1.092 2522.067  1.063
sim:poly   2522.442 2522.595 5.372 2522.466  0.676
sim:poly   2522.595 2523.351 2.495 2522.682  2.042
sim:poly   2523.351 2523.924 2.839 2523.607  0.221
sim:poly   2523.924 2524.478 8.357 2525.412 -0.899
sim:poly   2524.478 2525.399 2.304 2524.576  1.432
sim:poly   2525.399 2526.107 2.573 2525.721  1.205
sim:poly   2526.107 2526.550 6.350 2526.493  0.031
sim:poly   2526.550 2526.847 4.192 2526.589  2.893
sim:poly   2526.847 2527.616 0.345 2527.652 -0.472
sim:poly   2527.616 2528.264 3.583 2527.783  0.725
sim:poly   2528.264 2528.777 1.246 2528.704  0.610
sim:poly   2528.777 2529.606 3.534 2529.535  1.752
sim:poly   2529.606 2530.242 9.002 2529.694  1.119
\end{verbatim}\end{scriptsize}

\figureDSSN{f36.eps}{Time series generated by the simulator in the sample project {\tt sim-zeromean}. The sampling represents the V photometry of IC\,4996\,\#\,89. First the simulator generates a set of power functions over intervals within the time series ({\em grey}), then the actual light curve ({\em black}) is produced by shifting the mean observable for each power function to zero individually.}{SIGSPEC_sim-zeromean.dat}{!htb}{clip,angle=0,width=110mm}

\noindent This block of {\tt sim:poly} keywords is followed by a corresponding block of {\tt sim:zeromean} keywords:

\begin{scriptsize}\begin{verbatim}
sim:zeromean   2520.215 2521.088
sim:zeromean   2521.088 2521.679
sim:zeromean   2521.679 2522.442
sim:zeromean   2522.442 2522.595
sim:zeromean   2522.595 2523.351
sim:zeromean   2523.351 2523.924
sim:zeromean   2523.924 2524.478
sim:zeromean   2524.478 2525.399
sim:zeromean   2525.399 2526.107
sim:zeromean   2526.107 2526.550
sim:zeromean   2526.550 2526.847
sim:zeromean   2526.847 2527.616
sim:zeromean   2527.616 2528.264
sim:zeromean   2528.264 2528.777
sim:zeromean   2528.777 2529.606
sim:zeromean   2529.606 2530.242
\end{verbatim}\end{scriptsize}

\noindent This block is responsible for shifting the mean observable to zero for each synthesized power function.

Fig.\,\ref{SIGSPEC_sim-zeromean.dat} \it compares the corresponding light curve with the light curve generated in the project {\tt sim-poly}. (See also Fig.\,\ref{SIGSPEC_sim-poly.dat}\it ) The 16 significant signal components detected by {\sc SigSpec} are of minor interest and not discussed here.\sf

\section{Signal-to-Noise Ratio and Lomb-Scargle Periodogram}\label{SIGSPEC_Signal-to-Noise Ratio and Lomb-Scargle Periodogram}

As pointed out by Reegen (2007), the {\sc SigSpec} method represents a tool for an iterative frequency analysis of a zero-mean corrected time series superior to signal-to-noise ratio estimation (Breger et al.~1993) and Lomb-Scargle periodogram (Lomb~1976; Scargle~1982). However, in some situations these alternative methods may be desired or even more reasonable. Namely the Lomb-Scargle periodogram represents the optimum statistical approach to the problem if the mean observable is meaningful rather than set zero arbitrarily. The relations between sig and signal-to-noise ratio or Lomb-Scargle periodogram, respectively, are introduced and discussed by Reegen (2007).

In order to meet a user's requirement of signal-to-noise ratio-based DFT analysis or Lomb-Scargle periodograms as well, the {\sc SigSpec} software offers the option to perform an analysis relying on amplitude signal-to-noise ratios by providing the keyword {\tt DFT}\label{SIGSPEC_keyword.DFT} in the {\tt .ini} file. If this keyword is specified, all {\sc SigSpec} computations rely on the approximation of sig by the amplitude signal-to-noise ratio according to
\begin{equation}\label{EQ significance SNR approx}
\mathrm{sig}\left( A\right)\approx\frac{K\log\mathrm{e}}{4}\,\frac{A^2}{\left< x^2\right>}\: ,
\end{equation}
where $K$ represents the number of time series data, $A$ denotes the Fourier amplitude, and $\left< x^2\right>$ refers to the variance of the observable.

Second, the keyword {\tt Lomb}\label{SIGSPEC_keyword.Lomb} forces {\sc SigSpec} to evaluate Lomb-Scargle periodograms rather than significance spectra. In this case, the sig is approximated by
\begin{equation}\label{EQ significance LS approx}
\mathrm{sig}\left( A\right)\approx\frac{K\log\mathrm{e}}{4}\,\frac{P_{\mathrm{LS}}}{\left< x^2\right>}\: ,
\end{equation}
where $P_{\mathrm{LS}}$ denotes the power level in terms of the Lomb-Scargle periodogram.

\vspace{12pt}\noindent{\bf Example.} \it In the sample projects {\tt DFT} and {\tt L-S}, the input time series represents the V photometry of IC\,4996\,\#\,89.

The file {\tt DFT.ini} contains a single entry

\begin{scriptsize}\begin{verbatim}
DFT
\end{verbatim}\end{scriptsize}

\noindent which forces {\sc SigSpec} to rely on the signal-to-noise ratio of DFT amplitudes. The screen output is:

\begin{scriptsize}\begin{verbatim}
   1 freq 3.13205  sig 9.75026  rms 0.00449592  csig 9.75026
   2 freq 3.98473  sig 6.80132  rms 0.00422861  csig 6.80083
   3 freq 5.40684  sig 5.31609  rms 0.0040257  csig 5.30209
   4 freq 17.3677  sig 4.1816  rms 0.00388775  csig 4.14988
\end{verbatim}\end{scriptsize}

\figureDSSN{f37.eps}{Significance spectrum of the V photometry of IC\,4996\,\#\,89 ({\em grey}) and approximation by the signal-to-noise ratio of DFT amplitudes ({\em black}).}{SIGSPEC_DFT.s}{!htb}{clip,angle=0,width=110mm}

\figureDSSN{f38.eps}{Significance spectrum of the V photometry of IC\,4996\,\#\,89 ({\em grey}) and approximation by the Lomb-Scargle periodogram ({\em black}).}{SIGSPEC_L-S.s}{!htb}{clip,angle=0,width=110mm}

The file {\tt L-S.ini} contains a single keyword

\begin{scriptsize}\begin{verbatim}
Lomb
\end{verbatim}\end{scriptsize}

\noindent and {\sc SigSpec} uses the Lomb-Scargle periodogram rather than sig for all computations. The screen output is:

\begin{scriptsize}\begin{verbatim}
   1 freq 3.13205  sig 9.75026  rms 0.00449592  csig 9.75026
   2 freq 3.98472  sig 6.79398  rms 0.00422861  csig 6.7935
   3 freq 5.40684  sig 5.31451  rms 0.0040257  csig 5.30033
   4 freq 17.3677  sig 4.18161  rms 0.00388775  csig 4.14977
\end{verbatim}\end{scriptsize}

The significance spectrum of the input time series is compared to the approximations by DFT amplitude signal-to-noise ratio and Lomb-Scargle periodogram in Figs.\,\ref{SIGSPEC_DFT.s} \it and \ref{SIGSPEC_L-S.s}\it , respectively.

A comparison of the two outputs and the screen output of the corresponding sig-based application (Example {\tt SigSpecNative}, p.\,\pageref{SIGSPEC_EXnormalrunfreqs}) reveals slightly different signal components. Especially for the second component the frequency of which is close to an integer multiple of 1 cycle per day and therefore susceptible to alias, the results are different for all three methods. However, the frequencies, amplitudes and phases in the files {\tt result.dat} are in good agreement and reflect the numerical uncertainties of the MultiSine fitting procedure only.\sf

\section{Frequently Asked Questions}\label{SIGSPEC_FAQ}

This section contains questions frequently asked by users familiar to common methods of astronomical time series analysis involving signal-to-noise ratio estimation in power spectra and consecutive prewhitenings. The intention is to clarify the differences between these classical techniques and {\sc SigSpec} from the user's perspective.

\subsection{Changing sig in a prewhitening sequence}

{\bf Given a time series showing two different peaks in the power spectrum, prewhitening of the dominant signal usually does not cause a major change in the height of the secondary peak in the spectrum of the residuals. Why does the corresponding sig change?}

\figureDSSN{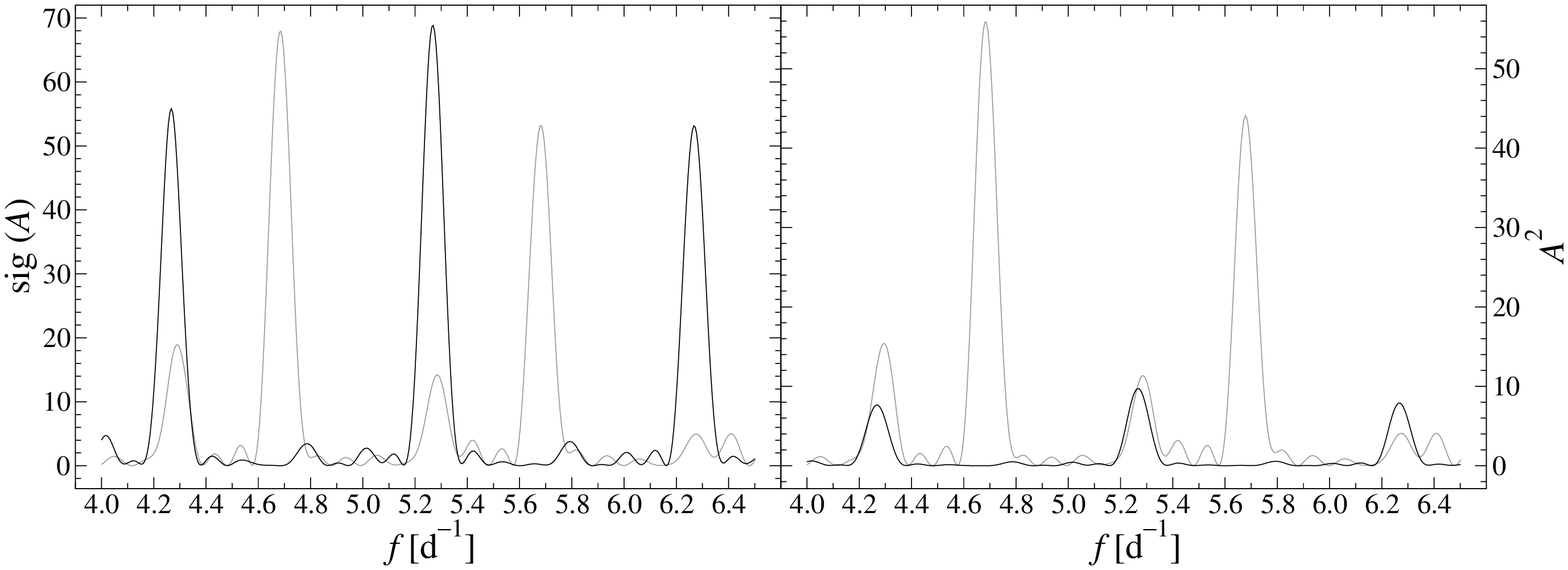}{{\em Grey} graphs: sig ({\em left}) and power (squared amplitude) spectra ({\em right}) of a synthetic time series containing two signals plus noise. The sampling represents the V photometry of IC\,4996\,\#\,89. {\em Black} graphs: spectra after subtracting the dominant signal ($f = 4.68573\,\mathrm{d}^{-1}$).}{SIGSPEC_change.s}{!htb}{clip,angle=0,width=110mm}

The situation is illustrated in Fig.\,\ref{SIGSPEC_change.s} displaying the sig ({\em left} panel) and power ({\em right} panel) spectra (in this sample just squared amplitude) generated by a synthetic time series. It consists of two sinusoidal signals, $f_1 = 4.68573\,\mathrm{d}^{-1}$, $A_1 = 7.27$, and $f_2 = 5.26934\,\mathrm{d}^{-1}$, $A_2 = 3.31$, plus noise with unit rms error. The plots contain a comparison of the initial spectra ({\em grey}) and the spectra after subtraction of the first signal component. In the right panel, the power associated to the peak at $5.27\,\mathrm{d}^{-1}$ differs only slightly between the two iterations, whereas the corresponding sig in the left panel increases dramatically in the second iteration.

The reason for this behaviour is that the sig refers to the probability of a random time series with the same rms error as the given one to produce a peak like the given one. In the first iteration, the sig calculation is based on the initial time series (rms error $5.84$), and in the second iteration, it relies on the residual time series after prewhitening of the peak at $f_1$, the rms error of which is $2.46$. The ratio of rms errors ($\approx 2.4$) is in agreement with the root ratio of sigs at $f_2$ in the two iterations ($\approx 2.2$).

This effect is more prominent for high sigs, because in this case prewhitening causes a major change in the statistical properties of the time series. If a peak with a low sig is prewhitened, the time series is affected marginally, and correspondingly, the sigs of other signals do not change much.

\subsection{The effect of binning}\label{SIGSPEC_The effect of binning}

{\bf Consider a time series representing a signal plus noise. If the data points are grouped into bins, the noise of the binned observables will reduce by the square root of the number of points in each bin. On the other hand, the number of data points the time series consists of reduces by the same amount. Since these two effects cancel each other, the noise level in the power spectrum will be the same for unbinned and binned data. What is the corresponding situation in terms of significance?}

\figureDSSN{f40.eps}{{\em Grey} graphs: sig ({\em left}) and power (squared amplitude) spectra ({\em right}) of a synthetic time series ($100$ equidistant data points) containing a sinusoidal signal plus noise with a standard deviation of $1$. The signal amplitude is $0.5$. {\em Black} graphs: same for time series data grouped into bins of two points. The resulting time series consists of $50$ data points.}{SIGSPEC_binweak.s}{!htb}{clip,angle=0,width=110mm}

Fig.\,\ref{SIGSPEC_binweak.s} contains the significance ({\em left}) and power (squared amplitude) spectra ({\em right}) of a synthetic time series containing a sinusoidal signal with a frequency of $0.075832\,\mathrm{d}^{-1}$ plus Gaussian noise with a standard deviation of $1$. The signal amplitude is $0.5$, providing an amplitude signal-to-noise ratio of $5.64$. All corresponding plots are displayed in {\em grey} colour. The {\em black} graphs represent the spectra generated by a binned version of the time series: each bin contains two data points, and the observable is the arithmetic mean.

In terms of sig as well as amplitude, binning affects neither the peak nor the mean amplitude remarkably: the reduced number of data points would increase the amplitude noise, but this effect is mitigated by the fact that binning reduces the rms residual in the time domain. For a multi-sine signal plus white noise, the number of significant peaks in a given frequency range will hardly be modified by data binning. A considerable change of these sigs by binning is an indication of the noise not being white. A correlation between consecutive measurements in the time series would be a reasonable explanation for such a behaviour.

\subsection{Binning of extremely strong signals}\label{SIGSPEC_Binning of extremely strong signals}

{\bf If an extremely strong signal is binned, the sig changes, whereas the signal amplitude and the noise level do not. Why?}

\figureDSSN{f41.eps}{{\em Grey} graphs: sig ({\em left}) and power (squared amplitude) spectra ({\em right}) of a synthetic time series ($100$ equidistant data points) containing a sinusoidal signal plus noise with a standard deviation of $1$. The signal amplitude is $10$. {\em Black} graphs: same for time series data grouped into bins of two points. The resulting time series consists of $50$ data points.}{SIGSPEC_binstrong.s}{!htb}{clip,angle=0,width=110mm}

Fig.\,\ref{SIGSPEC_binstrong.s} contains the significance ({\em left}) and power (squared amplitude) spectra ({\em right}) of a synthetic time series containing a sinusoidal signal with a frequency of $0.075832\,\mathrm{d}^{-1}$ plus Gaussian noise with a standard deviation of $1$. The signal amplitude of $10$ is associated to an ampltiude signal-to-noise ratio of more than $100$. All corresponding plots are displayed in {\em grey} colour. The {\em black} graphs represent the spectra generated by a binned version of the time series: each bin contains two data points, and the observable is the arithmetic mean.

For both strong and weak signals, binning affects neither the peak nor the mean amplitude remarkably: the reduced number of data points would increase the amplitude noise, but this effect is mitigated by the fact that binning reduces the rms residual in the time domain.

In terms of sig the situation is different: for very strong signals, the peak sig is reduced by binning. Classical techniques prewhiten a peak under consideration and employ the residuals to estimate a noise level. {\sc SigSpec} does not imply any prewhitening. In the case of a dominant signal plus a tiny scatter, the unbinned and binned data have comparable rms deviations, which are mainly determined by the signal. In the frequency domain, only the reduced number of binned data points comes into play.

Very strong signals let the sig drop to $\approx\frac{1}{N}$ by forming groups of $N$ data points: in Fig.\,\ref{SIGSPEC_binstrong.s}, {\em left} panel, the grey peak is about twice as high as the black peak.

\subsection{Linear interpolation: more information?}

{\bf Consider a time series representing a signal plus noise. Generating additional data points through linear interpolation increases the sig of the signal peak, although the power spectrum remains practically unchanged. This provides the possibility to boost signal sigs artificially, although the amount of information contained by the time series does not increase. Does this make sense?}

\figureDSSN{f42.eps}{{\em Grey} graphs: sig ({\em left}) and power (squared amplitude, logarithmic scale) spectrum ({\em right}) of a synthetic time series ($100$ equidistant data points) containing a sinusoidal signal without noise. {\em Black} graphs: same for a new time series generated by inserting $9$ additional linearly interpolated points such that the result is an equidistantly sampled dataset consisting of $991$ points.}{SIGSPEC_lint.s}{!htb}{clip,angle=0,width=110mm}

Fig.\,\ref{SIGSPEC_lint.s} displays the sig ({\em left}) and power (squared amplitude, {\em right}) spectrum of an equidistantly sampled time series consisting of $100$ data points and representing a sinusoidal signal with a frequency of $0.075832\,\mathrm{d}^{-1}$ and an amplitude of $1$ in {\em black} colour. No noise is added. Based on this time series, a new dataset is generated: between each pair of data points, $9$ additional, equidistant data points are inserted. The observables are assigned by linear interpolation. The number of data points in this new time series is thus $991$. The corresponding spectra are shown in {\em grey} colour. The longer dataset generates a peak significance that is roughly ten times higher than the initial one, whereas the power spectrum remains practically unchanged. Only the fact that the linear interpolation does not reveal the ``true'' observables that would be generated by the signal exactly is responsible for a small deviation of the black graph from the grey one.

The explanation for this behaviour is quite similar to the previous section ``The effect of binning'', p.\,\pageref{SIGSPEC_The effect of binning}, and correspondingly, the effect is mitigated for very noisy signals. Therefore in practical applications, it will be impossible to enhance the capability of a frequency analysis by artificially introducing new data points.

\subsection{Which sig threshold is reasonable?}

{\bf Occasionally, sigs or sig limits are shifted by $\log\frac{K}{2}$, $K$ denoting the number of time series data points. Which sig threshold is the true one?}

In fact both versions are correct, but they apply to different
questions. The version without $\log\frac{K}{2}$ refers to the probability that
an amplitude level (a peak) {\em at a given frequency and phase} occurs by
chance. The version including $\log\frac{K}{2}$ corresponds to the probability
that the highest out of $\frac{K}{2}$ independent peaks occurs by chance.
According to the sampling theorem, the DFT of $K$ data points (a system
with $K$ degrees of freedom) produces $\approx\frac{K}{2}$ independent frequencies in
Fourier space, if the sampling is equidistant. Although there is no
explicit prescription where to find a set of independent frequencies for
non-equidistant sampling, the system will still have $K$ degrees of
freedom, and the statistical considerations will be essentially the same.

A simple experiment makes the situation clearer: we roll
a dice and obtain the result ``4''. The probability that such an
experiment returns at least ``4'' (i.\,e. ``4'', ``5'' or ``6'') is, of course, $50$\,\%. This refers to the
examination of an individual peak without respect to all the others in
the spectrum. If we roll $10$ dices, the probability for at least one
showing ``4'' or more is dramatically higher, namely $>99.9$\,\%. This refers
to examining the highest out of $10$ peaks. The increasing probability of
obtaining such a result by chance corresponds to a decreasing
significance of the result.

\section{Keywords Reference}\label{SIGSPEC_Keywords Reference}

This section is a compilation of all keywords accepted by {\sc SigSpec}. A brief description of arguments and default values is given. The type of argument is provided by either {\tt <int>} or {\tt <double>}, and default values are given in parentheses, e.\,g.~{\tt (2)}. Empty parentheses indicate that there is no default setting.

\subsubsection{\tt antialc:adopt <int> (1)}\label{SIGSPEC_antialc:adopt}

number of AntiAlC test iterations adopted for the main prewhitening cascade, p.\,\pageref{SIGSPEC_keyword.antialc:adopt}

\subsubsection{\tt antialc:depth <int> (automatic)}\label{SIGSPEC_antialc:depth}

AntiAlC computation depth, p.\,\pageref{SIGSPEC_keyword.antialc:depth}

parameter: number of iterations used for peak combination testing

default: $\frac{1}{\sqrt{p_\mathrm{al}}}$, where $p_\mathrm{al}$ is the AntiAlC parameter, rounded to the successive integer value

\subsubsection{\tt antialc:par <double> ()}\label{SIGSPEC_antialc:par}

AntiAlC parameter $p_\mathrm{al}$: sig limit relative to maximum for the selection of candidate peaks ($0$ ... use the sig limit {\tt siglimit} instead), p.\,\pageref{SIGSPEC_keyword.antialc:par}

\subsubsection{\tt antialc:siglimit <double> ()}\label{SIGSPEC_antialc:siglimit}

significance limit for the AntiAlC candidate peak selection (no significance limit by default; the limit defined by the keyword {\tt siglimit} is used instead), p.\,\pageref{SIGSPEC_keyword.antialc:siglimit}

\subsubsection{\tt col:obs <int> (2)}\label{SIGSPEC_col:obs}

observable column index (unique), starting with $1$, p.\,\pageref{SIGSPEC_keyword.col:obs}

\subsubsection{\tt col:ssid <int> ()}\label{SIGSPEC_col:ssid}

subset identifier column index (also multiple), starting with $1$, p.\,\pageref{SIGSPEC_keyword.col:ssid}

\subsubsection{\tt col:time <int> (1)}\label{SIGSPEC_col:time}

time column index (unique), starting with $1$, p.\,\pageref{SIGSPEC_keyword.col:time}

\subsubsection{\tt col:weights <int> ()}\label{SIGSPEC_col:weights}

weights column index (also multiple), starting with $1$, p.\,\pageref{SIGSPEC_keyword.col:weights}

\subsubsection{\tt comp <int> ()}\label{SIGSPEC_comp}

specifies the file indicated by the parameter as comparison dataset, p.\,\pageref{SIGSPEC_keyword.comp}

\subsubsection{\tt correlograms <int> <int> <int> () ()}\label{SIGSPEC_correlograms}

specifications for correlogram files {\tt c\#iteration\#.dat}, p.\,\pageref{SIGSPEC_keyword.correlograms}

parameters:
\begin{itemize}
\item correlogram order (maximum index lag), default: half of the number of time series data points,
\item number of files to generate ($< 0$ for all correlogram files, default: no correlogram computation),
\item step width (number of iterations) for output.
\end{itemize}

\subsubsection{\tt csiglimit <double> ()}\label{SIGSPEC_csiglimit}

lower cumulative sig limit, p.\,\pageref{SIGSPEC_keyword.csiglimit}

\subsubsection{\tt deftype <target/comp/skip> (target)}\label{SIGSPEC_deftype}

specifies the type of dataset to be assigned to a time series by default, p.\,\pageref{SIGSPEC_keyword.deftype}

\subsubsection{\tt DFT}\label{SIGSPEC_DFT}

forces {\sc SigSpec} to approximate all sigs by signal-to-noise ratios of DFT amplitudes, p.\,\pageref{SIGSPEC_keyword.DFT}

\subsubsection{\tt diff:comp}\label{SIGSPEC_diff:comp}

specifies the DFT amplitude spectrum of the comparison datasets to be calculated through a weighted mean of Fourier vectors, p.\,\pageref{SIGSPEC_keyword.diff:comp}

\subsubsection{\tt diff:compalign}\label{SIGSPEC_diff:compalign}

specifies the DFT amplitude spectrum of the comparison datasets to be calculated as a weighted mean of Fourier amplitudes, p.\,\pageref{SIGSPEC_keyword.diff:compalign}

This setting forces {\sc SigSpec} to take into account also correlated signal components that lag in phase with respect to each other and the target dataset, respectively.

\subsubsection{\tt diff:off}\label{SIGSPEC_diff:off}

switches off the differential significance computation (default), p.\,\pageref{SIGSPEC_keyword.diff:off}

\subsubsection{\tt freqspacing <double> ()}\label{SIGSPEC_freqspacing}

spacing between consecutive frequencies [inverse time units], p.\,\pageref{SIGSPEC_keyword.freqspacing}

\subsubsection{\tt harmonics <int> ()}\label{SIGSPEC_harmonics}

activates the simultaneous analysis of a fundamental plus harmonics (the frequencies of which are integer multiples of the fundamental) the number of which is specified by the parameter, p.\,\pageref{SIGSPEC_keyword.harmonics}

\subsubsection{\tt iterations <int> ()}\label{SIGSPEC_iterations}

number of prewhitening iterations, p.\,\pageref{SIGSPEC_keyword.iterations}

\subsubsection{\tt lfreq <double> (0)}\label{SIGSPEC_lfreq}

lower frequency limit [inverse time units], p.\,\pageref{SIGSPEC_keyword.lfreq}

\subsubsection{\tt Lomb}\label{SIGSPEC_Lomb}

forces {\sc SigSpec} to approximate all sigs by the Lomb-Scargle periodogram, p.\,\pageref{SIGSPEC_keyword.Lomb}

\subsubsection{\tt mfstart <int> (0)}\label{SIGSPEC_mfstart}

index of the first time series input file to apply the MultiFile mode to, p.\,\pageref{SIGSPEC_keyword.mfstart}

\subsubsection{\tt mstracks <int> <int> ()}\label{SIGSPEC_mstracks}

MultiSine tracks are written to files {\tt m\#index\#.dat}, where {\#index\#} refers to the signal component in the result files. The parameters are
\begin{enumerate}
\item the maximum number of iterations for which to write entries to the MultiSine track files, and
\item the step width (number of iterations) for output, p.\,\pageref{SIGSPEC_keyword.mstracks}.
\end{enumerate}

The file name may be assigned additional indices for Time-Resolved analysis (p.\,\pageref{SIGSPEC_Time-resolved Analysis}) and/or MultiFile mode (p.\,\pageref{SIGSPEC_MultiFile Mode}).

\subsubsection{\tt multifile <int> ()}\label{SIGSPEC_multifile}

activates MultiFile mode, p.\,\pageref{SIGSPEC_keyword.multifile}

parameter: maximum index of time series input files ($\le 0$ ... infinite)

\subsubsection{\tt multisine:lock}\label{SIGSPEC_multisine:lock}

forces {\sc SigSpec} to use the ``raw'' frequencies, amplitudes, and phases (without MultiSine fitting) for the subsequent analysis, p.\,\pageref{SIGSPEC_keyword.multisine:lock}.

\subsubsection{\tt multisine:newton <double> <double> <double> (0.000001 1 0.000001)}\label{SIGSPEC_multisine:newton}

accuracy parameters for the MultiSine least-squares fits
\begin{enumerate}
\item scaling factor for the overall precision of resulting frequencies,
\item degree of dependence of the frequency accuracy on the peak sig,
\item the minimum relative improvement of rms residual between consecutive iterations to continue the fitting process, p.\,\pageref{SIGSPEC_keyword.multisine:newton}.
\end{enumerate}

\subsubsection{\tt multisine:unlock}\label{SIGSPEC_multisine:unlock}

forces {\sc SigSpec} to use the frequencies, amplitudes, and phases improved by MultiSine least-squared fits for the subsequent analysis (default), p.\,\pageref{SIGSPEC_keyword.multisine:unlock}.

\subsubsection{\tt nycoef <double> (0.5)}\label{SIGSPEC_nycoef}

Nyquist Coefficient (between 0 and 1), p.\,\pageref{SIGSPEC_keyword.nycoef}

\subsubsection{\tt nyscan}\label{SIGSPEC_nyscan}

Nyquist Coefficients for the specified frequency range (file {\tt nycoef.dat} or {\tt <\#multifile\#>.nycoef.dat}), p.\,\pageref{SIGSPEC_keyword.nyscan}

\subsubsection{\tt osratio <double> (20)}\label{SIGSPEC_osratio}

oversampling ratio, p.\,\pageref{SIGSPEC_keyword.osratio}

\subsubsection{\tt phdist:cart}\label{SIGSPEC_phdist:cart}

generates a Phase Distribution Diagram in three-dimensional cartesian coordinates, p.\,\pageref{SIGSPEC_keyword.phdist:cart}

\subsubsection{\tt phdist:colmodel:lin}\label{SIGSPEC_phdist:colmodel:lin}

specifies the linear colour model, i.\,e., phase probability density is used as a colour scale, p.\,\pageref{SIGSPEC_keyword.phdist:colmodel:lin}

\subsubsection{\tt phdist:colmodel:rank}\label{SIGSPEC_phdist:colmodel:rank}

specifies the rank colour model, i.\,e., the rank in an ascending sequence of sock significances is used as a colour scale, p.\,\pageref{SIGSPEC_keyword.phdist:colmodel:rank}

\subsubsection{\tt phdist:colour <double> <double> <double> <double>}\label{SIGSPEC_phdist:colour}

A set of {\tt phdist:colour} lines defines an RGB path for colourising the Phase Distribution Diagram, p.\,\pageref{SIGSPEC_keyword.phdist:colour}.

parameters:
\begin{itemize}
\item red channel ($0...255$)
\item green channel ($0...255$)
\item blue channel ($0...255$)
\item scale
\end{itemize}

The scale parameter refers directly to probability density of phases in case of {\tt phdist:colmodel:lin}, or to a fractile of probability density on the interval $\left[ 0,1\right]$ in case of {\tt phdist:colmodel:rank}.

\subsubsection{\tt phdist:cyl}\label{SIGSPEC_phdist:cyl}

generates a Phase Distribution Diagram in three-dimensional cylindrical coordinates (default setting), p.\,\pageref{SIGSPEC_keyword.phdist:cyl}. The frequency is the height axis, the phase is the azimuth angle, and the radial coordinate refers to the probability density of phase.

\subsubsection{\tt phdist:fill <double> (0)}\label{SIGSPEC_phdist:fill}

specifies a filling factor to compute extra frequencies if the difference of phase PDFs between two adjacent frequencies is too high, p.\,\pageref{SIGSPEC_keyword.phdist:fill}.

parameter: number of additional frequencies per unit probability density (difference between two adjacent frequencies)

\subsubsection{\tt phdist:phases <int> ()}\label{SIGSPEC_phdist:phases}

generates a Phase Distribution Diagram for the sampling of the given time series, p.\,\pageref{SIGSPEC_keyword.phdist:phases}. By default, no Phase Distribution Diagram is computed.

parameter: number of phase angles in the interval $\left[ 0,\pi\right[$, if the maximum probability density is $\le 1$. Between $1$ and $2$, twice this number is used, and so on. This enhances the visibility of the Phase Distribution Diagram also in frequency and phase regions associated with a very eccentric phase distribution.

\subsubsection{\tt preview <double> ()}\label{SIGSPEC_preview}

generates a preview, p.\,\pageref{SIGSPEC_keyword.preview}. Instead of a prewhitening cascade, only one significance spectrum is computed. All local maxima above the specified significance limit are written to a file {\tt preview.dat}. By default, no preview is computed.

parameter: significance limit

\subsubsection{\tt profile}\label{SIGSPEC_profile}

{\sc SigSpec} generates a file {\tt profile.dat} containing the sampling profile for the given time series, p.\,\pageref{SIGSPEC_keyword.profile}. By default, the file {\tt profile.dat} is not generated.

This keyword is ignored in MultiFile mode, where sampling profiles are calculated and written to files whenever required by the program. See ``MultiFile Mode'', p.\,\pageref{SIGSPEC_MultiFile Mode} for further information.

\subsubsection{\tt residuals <int> <int> ()}\label{SIGSPEC_residuals}

output files containing residual time series (only {\tt residuals.dat} for the residuals after prewhitening all significant compontents by default). The parameters are
\begin{enumerate}
\item the maximum number of iterations (files {\tt t\#iteration\#.dat}), and
\item the step width (number of iterations) for output, p.\,\pageref{SIGSPEC_keyword.residuals}.
\end{enumerate}

The file name may be assigned additional indices for Time-Resolved analysis (p.\,\pageref{SIGSPEC_Time-resolved Analysis}) and/or MultiFile mode (p.\,\pageref{SIGSPEC_MultiFile Mode}).

\subsubsection{\tt results <int> <int> ()}\label{SIGSPEC_results}

output files containing a list of significant signal components. The default setting is to produce only a file {\tt result.dat} for the final list. The parameters are
\begin{enumerate}
\item the maximum number of iterations for which to write additional result files {\tt r\#iteration\#.dat}, and
\item the step width (number of iterations) for output, p.\,\pageref{SIGSPEC_keyword.results}.
\end{enumerate}

The file name may be assigned additional indices for Time-Resolved analysis (p.\,\pageref{SIGSPEC_Time-resolved Analysis}) and/or MultiFile mode (p.\,\pageref{SIGSPEC_MultiFile Mode}).

\subsubsection{\tt siglimit <double> ($5$)}\label{SIGSPEC_siglimit}

lower sig limit (0 to deactivate), p.\,\pageref{SIGSPEC_keyword.siglimit}

\subsubsection{\tt sim:add}\label{SIGSPEC_sim:add}

add synthetic data to given observable, p.\,\pageref{SIGSPEC_keyword.sim:add}

\subsubsection{\tt sim:exp <double> <double> <double> <double> <double> ()}\label{SIGSPEC_sim:exp}

exponential trend, p.\,\pageref{SIGSPEC_keyword.sim:exp}

parameters:
\begin{itemize}
\item lower time limit [time units]
\item upper time limit [time units]
\item scale
\item time zeropoint [time units]
\item exponent
\end{itemize}

\subsubsection{\tt sim:off}\label{SIGSPEC_sim:off}

deactivate simulator (default), p.\,\pageref{SIGSPEC_keyword.sim:off}

\subsubsection{\tt sim:poly <double> <double> <double> <double> <double> ()}\label{SIGSPEC_sim:poly}

polynomial trend, p.\,\pageref{SIGSPEC_keyword.sim:poly}

parameters:
\begin{itemize}
\item lower time limit [time units]
\item upper time limit [time units]
\item scale
\item time zeropoint [time units]
\item exponent
\end{itemize}

full polynomial by multiple declaration with different scales, time zeropoints, and exponents

\subsubsection{\tt sim:replace}\label{SIGSPEC_sim:replace}

replace given observable by synthetic data, p.\,\pageref{SIGSPEC_keyword.sim:replace}

\subsubsection{\tt sim:rndsteps <double> <double> <double> <double> ()}\label{SIGSPEC_sim:rndsteps}

random steps, p.\,\pageref{SIGSPEC_keyword.sim:rndsteps}

parameters:
\begin{itemize}
\item lower time limit [time units]
\item upper time limit [time units]
\item standard deviation for Gaussian distribution of (constant) step values
\item expected time range for Poisson distribution of steps [time units]
\end{itemize}

\subsubsection{\tt sim:serial <double> <double> <double> <double> <double> <double> ()}\label{SIGSPEC_sim:serial}

serially correlated noise, p.\,\pageref{SIGSPEC_keyword.sim:serial}

parameters:
\begin{itemize}
\item lower time limit [time units]
\item upper time limit [time units]
\item scale for standard deviation
\item time zeropoint for polynomial trend of standard deviation [time units]
\item exponent for polynomial trend of standard deviation
\item serial correlation coefficient
\end{itemize}

full polynomial by multiple declaration with different scales, time zeropoints, and exponents

\subsubsection{\tt sim:signal <double> <double> <double> <double> <double> ()}\label{SIGSPEC_sim:signal}

sinusoidal signal, p.\,\pageref{SIGSPEC_keyword.sim:signal}

parameters:
\begin{itemize}
\item lower time limit [time units]
\item upper time limit [time units]
\item amplitude
\item time zeropoint [time units]
\item frequency [inverse time units]
\end{itemize}

\subsubsection{\tt sim:temporal <double> <double> <double> <double> <double> <double> ()}\label{SIGSPEC_sim:temporal}

temporally correlated noise, p.\,\pageref{SIGSPEC_keyword.sim:temporal}

parameters:
\begin{itemize}
\item lower time limit [time units]
\item upper time limit [time units]
\item scale for standard deviation
\item time zeropoint for polynomial trend of standard deviation [time units]
\item exponent for polynomial trend of standard deviation
\item temporal correlation coefficient
\end{itemize}

full polynomial by multiple declaration with different scales, time zeropoints, and exponents

\subsubsection{\tt sim:zeromean <double> <double> ()}\label{SIGSPEC_sim:zeromean}

zero-mean adjustment, p.\,\pageref{SIGSPEC_keyword.sim:zeromean}

parameters:
\begin{itemize}
\item lower time limit [time units]
\item upper time limit [time units]
\end{itemize}

\subsubsection{\tt skip <int> ()}\label{SIGSPEC_skip}

forces {\sc SigSpec} to skip the file indicated by the parameter, p.\,\pageref{SIGSPEC_keyword.skip}

\subsubsection{\tt sock:cart}\label{SIGSPEC_sock:cart}

generates a Sock Diagram in three-dimensional cartesian coordinates, p.\,\pageref{SIGSPEC_keyword.sock:cart}

\subsubsection{\tt sock:colmodel:lin}\label{SIGSPEC_sock:colmodel:lin}

specifies the linear colour model, i.\,e., sock significance is used as a colour scale, p.\,\pageref{SIGSPEC_keyword.sock:colmodel:lin}

\subsubsection{\tt sock:colmodel:rank}\label{SIGSPEC_sock:colmodel:rank}

specifies the rank colour model, i.\,e., the rank in an ascending sequence of sock significances is used as a colour scale, p.\,\pageref{SIGSPEC_keyword.sock:colmodel:rank}

\subsubsection{\tt sock:colour <double> <double> <double> <double>}\label{SIGSPEC_sock:colour}

A set of {\tt sock:colour} lines defines an RGB path for colourising the Sock Diagram, p.\,\pageref{SIGSPEC_keyword.sock:colour}.

parameters:
\begin{itemize}
\item red channel ($0...255$)
\item green channel ($0...255$)
\item blue channel ($0...255$)
\item scale
\end{itemize}

For the linear colour model selected by the keyword {\tt sock:colmodel:lin}, the scale parameter refers directly to sock significance. If the rank colour model is selected ({\tt sock:colmodel:rank}), it refers to a fractile of sock significance on the interval $\left[ 0,1\right]$.

\subsubsection{\tt sock:cyl}\label{SIGSPEC_sock:cyl}

generates a Sock Diagram in three-dimensional cylindrical coordinates (default setting), p.\,\pageref{SIGSPEC_keyword.sock:cyl}. The frequency is the height axis, the phase is the azimuth angle, and the radial coordinate refers to the sock significance.

\subsubsection{\tt sock:fill <double> (0)}\label{SIGSPEC_sock:fill}

specifies a filling factor to compute extra frequencies if the sock significance difference between two adjacent frequencies is too high, p.\,\pageref{SIGSPEC_keyword.sock:fill}.

parameter: number of additional frequencies per unit sig (difference between two adjacent frequencies)

\subsubsection{\tt sock:phases <int> ()}\label{SIGSPEC_sock:phases}

generates a Sock Diagram for the sampling of the given time series, p.\,\pageref{SIGSPEC_keyword.sock:phases}. By default, no Sock Diagram is computed.

parameter: number of phase angles in the interval $\left[ 0,\pi\right[$, if the maximum sock significance is $\le 1$. Between $1$ and $2$, twice this number is used, and so on. This enhances the visibility of the Sock Diagram also in frequency and phase regions associated with a high sock significance.

\subsubsection{\tt spectra <int> <int> ()}\label{SIGSPEC_spectra}

output files containing spectra (only {\tt s000000.dat} for the spectrum of the initial time series and {\tt resspec.dat} for the spectrum of the residuals after prewhitening all significant compontents by default). The parameters are
\begin{enumerate}
\item the maximum number of iterations (files {\tt s\#iteration\#.dat}), and
\item the step width (number of iterations) for output, p.\,\pageref{SIGSPEC_keyword.spectra}.
\end{enumerate}

The file name may be assigned additional indices for Time-Resolved analysis (p.\,\pageref{SIGSPEC_Time-resolved Analysis}) and/or MultiFile mode (p.\,\pageref{SIGSPEC_MultiFile Mode}).

\subsubsection{\tt target <int> ()}\label{SIGSPEC_target}

specifies the file indicated by the parameter as target dataset, p.\,\pageref{SIGSPEC_keyword.target}

\subsubsection{\tt timeres:range <double> ()}\label{SIGSPEC_timeres:range}

subset interval width [time units], p.\,\pageref{SIGSPEC_keyword.timeres:range}

\subsubsection{\tt timeres:step <double> ()}\label{SIGSPEC_timeres:step}

step width between subset centres [time units], p.\,\pageref{SIGSPEC_keyword.timeres:step}

\subsubsection{\tt timeres:w:cos <double> <double> ()}\label{SIGSPEC_timeres:w:cos}

cosine weights, p.\,\pageref{SIGSPEC_TABwts}

parameters:
\begin{itemize}
\item frequency [inverse time units]
\item phase [rad]
\end{itemize}

\subsubsection{\tt timeres:w:cosp <double> <double> <double> ()}\label{SIGSPEC_timeres:w:cosp}

weights according to the power of a cosine, p.\,\pageref{SIGSPEC_TABwts}

parameters:
\begin{itemize}
\item frequency [inverse time units]
\item phase [rad]
\item exponent
\end{itemize}

\subsubsection{\tt timeres:w:damp <double> ()}\label{SIGSPEC_timeres:w:damp}

exponential damping, p.\,\pageref{SIGSPEC_TABwts}

parameter: width [time units]

\subsubsection{\tt timeres:w:exp <double> ()}\label{SIGSPEC_timeres:w:exp}

exponential weights, p.\,\pageref{SIGSPEC_TABwts}

parameter: width [time units]

\subsubsection{\tt timeres:w:gauss <double> ()}\label{SIGSPEC_timeres:w:gauss}

Gaussian weights, p.\,\pageref{SIGSPEC_TABwts}

parameter: standard deviation [time units]

\subsubsection{\tt timeres:w:ipow <double> ()}\label{SIGSPEC_timeres:w:ipow}

inverse power weights, p.\,\pageref{SIGSPEC_TABwts}

parameter: exponent

\subsubsection{\tt timeres:w:none}\label{SIGSPEC_timeres:w:none}

unweighted moving averages, i.\,e.~a rectangular filter, p.\,\pageref{SIGSPEC_TABwts}

\subsubsection{\tt ufreq <double> ()}\label{SIGSPEC_ufreq}

upper frequency limit [inverse time units], p.\,\pageref{SIGSPEC_keyword.ufreq}

\subsubsection{\tt win}\label{SIGSPEC_win}

{\sc SigSpec} generates a file {\tt win.dat} containing the spectral window for the given time series. By default, the file {\tt win.dat} is not generated, p.\,\pageref{SIGSPEC_keyword.win}.

\section{Online availability}

The ANSI-C code is available online at {\tt http://www.sigspec.org}. For further information, please contact P.~Reegen, {\tt peter.reegen@univie.ac.at}.

\acknowledgments{
PR received financial support from the Fonds zur F\"or\-de\-rung der wis\-sen\-schaft\-li\-chen Forschung (FWF, projects P14546-PHY, P17580-N2) and the BM:BWK (project COROT). Furthermore, it is a pleasure to thank T.~Appourchaux (IAS, Orsay), A.~Baglin (Obs.~de Paris, Meudon), T.~Boehm (Obs.~M.-P., Tou\-lou\-se), M.~Breger, R.~Dvorak, M.\,G.~Firneis, D.~Frast (Univ.~of Vienna), R.~Garrido (Inst.~Astrof.~Andalucia, Granada), M.~Gruberbauer (Univ.~of Vienna), D.\,B.~Guenther (St.~Mary's Univ., Halifax), M.~Hareter, D.~Huber, T.~Kallinger (Univ.~of Vienna), R.~Kusch\-nig (UBC, Vancouver), S.~Mar\-chen\-ko (Western Kentucky Univ., Bowling Green, KY), M.~Masser (Univ. of Vienna), J.\,M. Matthews (UBC, Vancouver), E.~Michel (Obs. de Paris, Meudon), A.\,F.\,J.~Moffat (Univ.~de Montreal), E.~Paunzen, D.~Punz (Univ.~of Vienna), V.~Ripepi (INAF, Naples), S.\,M. Rucinski (D.~Dunlap Obs., Toronto), T.\,A.~Ryab\-chik\-ova (Inst.~Astrpn. RAS, Moscow), D.~Sasselov (Harvard-Smith\-sonian Center, Cambridge, MA), S.~Schraml (Univ.~of Technology, Vienna), G.\,A.~Wade (Royal Military College, Kingston), G.\,A.\,H.~Walker (UBC, Vancouver), W.\,W. Weiss, and K.~Zwintz (Univ.~of Vienna) for valuable discussion and support with extensive software tests. I acknowledge the anonymous referee for a detailed examination of both this publication and the corresponding software, as well as for the constructive feedback that helped to improve the overall quality a lot. Finally, I address my very special thanks to J.\,D.~Scargle for his valuable support.
}

\References{
Breger, M., Stich, J., Garrido, R., et al.~1993, A\&A, 271, 482\\

Breger, M., Rucinski, S.\,M., Reegen, P.~2007, AJ, 134, 1994\\

Kallinger, T., Reegen, P., Weiss, W.\,W.~2008, A\&A, 481, 571\\

Kjeldsen, H., Bedding, T.\,R., Viskum, M., Frandsen, S.~1995, AJ, 109, 1313

Lomb, N.\,R.~1976, ApSS, 39, 447\\

Reegen, P.~2005, in {\it The A-Star Puzzle}, Proceedings of IAUS 224, eds. J. Zverko, J. Ziznovsky, S.J. Adelman, W.W. Weiss (Cambridge: Cambridge Univ.~Press), p.~791\\

Reegen, P.~2007, A\&A, 467, 1353\\

Scargle, J.\,D.~1982, ApJ, 263, 835\\

Strassmeier, K.\,G., Boyd, L.\,J., Epand, D.\,H., Granzer, T.~1997, PASP, 109, 697\\

Zwintz, K., Marconi, M., Kallinger, T., Weiss, W.\,W.~2004, in {\it The A-Star Puzzle}, Proceedings of IAUS 224, eds. J.~Zverko, J.~Ziznovsky, S.\,J.~Adelman, W.\,W.~Weiss (Cambridge: Cambridge Univ.~Press), p.\,353\\

Zwintz, K., Weiss, W.\,W.~2006, A\&A, 457, 237\\

}


\setlength{\columnseprule}{0pt}\clearpage
\thispagestyle{empty}
\cleardoublepage
\thispagestyle{empty}

\end{document}